\documentclass[aps,prb,amsmath,amssymb,superscriptaddress,10pt,a4paper,floatfix,twocolumn]{revtex4-1}

\usepackage{braket}
\usepackage{graphicx}
\usepackage[english]{babel}
\usepackage{color}   
\usepackage{bbm}
\usepackage{tabularx}
\usepackage{algorithm}
\usepackage{algpseudocode}
\usepackage{bm}
\usepackage{tikz}
\usepackage[pdfpagelabels,plainpages=false,bookmarks=true,colorlinks,linkcolor=red,urlcolor=blue,citecolor=blue]{hyperref}

\newcommand{\unity}{\openone}

\newcommand{\Fig}[1]{Fig.~\ref{#1}}
\newcommand{\Sec}[1]{Sec.~\ref{#1}}
\newcommand{\App}[1]{Appendix~\ref{#1}}
\newcommand{\Tab}[1]{Table~\ref{#1}}
\DeclareMathOperator{\Tr}{Tr}

\newcommand{\AL}{A^{s}_{L}}
\newcommand{\AR}{A^{s}_{R}}
\newcommand{\AC}{A^{s}_{C}}
\newcommand{\ints}{\mathbb{Z}}
\renewcommand{\P}{\mathcal{P}}
\renewcommand{\O}{\mathcal{O}}

\newcommand{\rbra}[1]{(#1|}
\newcommand{\rket}[1]{|#1)}
\newcommand{\T}{\mathcal{T}}
\newcommand{\rproj}[1]{\rket{#1}\rbra{#1}}
\newcommand{\rbraket}[2]{\left(#1\middle|#2\right)}
\newcommand{\PDS}{\rproj{0}}

\newcommand{\algorithmname}{VUMPS}
\newcommand{\gradnorm}{\lVert B\rVert}

\def\figpath{.}

\newcommand{\lineH}[3]{\draw (#1,#3) -- (#2,#3);}
\newcommand{\lineV}[3]{\draw (#3,#1) -- (#3,#2);}
\newcommand{\mpsT}[3]{\draw[rounded corners] (0.5+#1,0.5+#2) rectangle (-0.5+#1,-0.5+#2); \draw (#1,#2) node {$#3$};}
\newcommand{\dobase}[1]{\draw (0,#1) node (X) {$\phantom{X}$};}

\newcommand{\drawMatrixLeft}[1]{
\begin{tikzpicture}[baseline = (X.base),every node/.style={scale=0.750},scale=.55]
\draw (0,0) circle (.5);
\draw (0,0) node (X) {#1};
\draw (0,0.5) edge[out=90,in=180] (1,1.5);
\draw (0,-0.5) edge[out=270,in=180] (1,-1.5);
\end{tikzpicture}
}
\newcommand{\drawMatrixRight}[1]{
\begin{tikzpicture}[baseline = (X.base),every node/.style={scale=0.750},scale=.55]
\draw (2,1.5) edge[out=0,in=90] (3,0.5);
\draw (3,0) circle (.5);
\draw (3,-.5) edge[out = -90, in = 0] (2, -1.5);
\draw (3,0) node (X) {#1};
\end{tikzpicture}
}

\begin{document}
\title{Variational optimization algorithms for uniform matrix product states}

\author{V. \surname{Zauner-Stauber}}
\affiliation{Vienna Center for Quantum Technology, University of Vienna, Boltzmanngasse 5, 1090 Wien, Austria}
\author{L. \surname{Vanderstraeten}}
\affiliation{Ghent University, Faculty of Physics, Krijgslaan 281, 9000 Gent, Belgium}
\author{M.T. \surname{Fishman}}
\affiliation{Institute for Quantum Information and Matter, California Institute of Technology, Pasadena, California 91125, USA}
\author{F. \surname{Verstraete}}
\affiliation{Vienna Center for Quantum Technology, University of Vienna, Boltzmanngasse 5, 1090 Wien, Austria}
\affiliation{Ghent University, Faculty of Physics, Krijgslaan 281, 9000 Gent, Belgium}
\author{J. \surname{Haegeman}}
\affiliation{Ghent University, Faculty of Physics, Krijgslaan 281, 9000 Gent, Belgium}
 
\begin{abstract}
We combine the Density Matrix Renormalization Group (DMRG) with Matrix Product State tangent space concepts to construct a variational algorithm for finding ground states of one dimensional quantum lattices in the thermodynamic limit. A careful comparison of this variational uniform Matrix Product State algorithm (VUMPS) with infinite Density Matrix Renormalization Group (IDMRG) and with infinite Time Evolving Block Decimation (ITEBD) reveals substantial gains in convergence speed and precision. We also demonstrate that VUMPS works very efficiently for Hamiltonians with long range interactions and also for the simulation of two dimensional models on infinite cylinders. The new algorithm can be conveniently implemented as an extension of an already existing DMRG implementation.
\end{abstract}
\maketitle

\section{Introduction}
\label{sec:intro}
The strategy of renormalization group (RG) techniques to successively reduce a large number of microscopic degrees of freedom to a smaller set of effective degrees of freedom has led to powerful numerical and analytical methods to probe and understand the effective macroscopic behavior of both classical and quantum many body systems.\cite{RG_Wilson, RG_Fisher1, RG_Fisher2, RG_ZJ} However, it was not until the advent of White's celebrated \textit{Density Matrix Renormalization Group} (DMRG) \cite{DMRG1,DMRG2} that variational RG methods reached unprecedented accuracy in numerically studying strongly correlated one dimensional quantum lattice systems at low temperature. The underlying variational ansatz of \textit{Matrix Product States} (MPS) \cite{MPS1_FNW,MPS2_KSZ,MPS3_OR,MPS4_PVWC,MPS5_VMC,MPS6_S,MPSTP} belongs to a class of ansatzes known as \textit{Tensor Network States}.\cite{MPS5_VMC,TN_Orus,TN_BC} These variational classes encode the many body wavefunction in terms of virtual entanglement degrees of freedom living on the boundary and thus satisfy an area law scaling of entanglement entropy per construction. As such, they provide a natural parameterization for the physical corner of Hilbert space, where low energy states of quantum many body systems ought to live in.\cite{AreaLaw_Hastings,AreaLaw_Eisert} MPS in particular are especially fit for studying ground states of strongly correlated one dimensional quantum systems with local interactions.\cite{MPS_faithful,MPS_Hastings,MPS_excitations}

The variational parameters in MPS are contained within local tensors associated with the individual sites of the lattice system. For homogeneous systems, the global wave function can then be captured using just a single (or a small number of) such tensors, independent of the system size. 
They consequently offer very natural access to the thermodynamic limit, providing a clear advantage over other numerical approaches such as  Exact Diagonalization or Quantum Monte Carlo.

On finite lattices, (one-site) DMRG implements the variational principle (energy minimization) by exploiting that the quantum state is a multilinear function of the local tensors. By fixing all but one tensors, the global eigenvalue problem is transformed into an effective eigenvalue problem for the local tensor.\cite{DMRG1,DMRG2,DMRG_variational,DMRG_SS,DMRG_MC,MPS6_S}
Using a translation invariant parameterization gives rise to an energy expectation value with a highly non-linear dependence on the tensor(s). Two different algorithms are widely used to obtain such an MPS in the thermodynamic limit. \textit{Infinite system DMRG} (IDMRG)\cite{DMRG1,DMRG2,IDMRG} proceeds by performing regular DMRG on a successively growing lattice, inserting and optimizing over new tensors in the center of the lattice in each step only, effectively mimicking an infinite lattice by using a finite, albeit very large lattice. After convergence the most recently inserted tensors in the center are taken as a unit cell for an infinite MPS approximation of the ground state. An alternative approach is known as \textit{infinite time evolving block decimation} (ITEBD).\cite{TEBD,ITEBD} It works directly in the thermodynamic limit and is based on evolving an initial state in imaginary time by using a Trotter decomposition of the evolution operator. 

We present a new variational algorithm, inspired by tangent space ideas,\cite{TDVP,MPSTP,TDVP_Uni} that combines the advantages of IDMRG and ITEBD and addresses some of their shortcomings. As such it is directly formulated in the thermodynamic limit, but at the same time optimizes the state by solving effective eigenvalue problems, rather than employing imaginary time evolution.
We find that it leads to a significant increase in efficiency in all of our test cases.  The following section introduces MPS notations and definitions and presents our variational algorithm, heuristically motivated from the perspective of finite size DMRG. \Sec{sec:numerics} illustrates the performance of our algorithm on various test cases, and compares to conventional IDMRG and ITEBD results. After the conclusion in \Sec{sec:conclusion}, we provide further technical details in the appendices. \App{sec:theory} contains additional theoretical background: we derive the self-consistent conditions that characterize the variational minimum and provide additional motivation for our algorithm from the perspective of the MPS tangent space. \App{sec:bonddim} presents a suitable strategy to expand the bond dimension of translation invariant MPS. \App{sec:effH_construct} explains how to construct effective Hamiltonians in the thermodynamic limit. These involve infinite geometric sums of the transfer matrix, which are further studied in \App{sec:invertE}.

\section{A Variational Algorithm for Matrix Product States in the Thermodynamic Limit}
\label{sec:algorithm}
In this section we introduce a variational algorithm for optimizing MPS directly in the thermodynamic limit. Because the algorithm strongly resembles conventional DMRG, we explain it by describing a single iteration step from the viewpoint of DMRG and show that only a few additional ingredients are needed to arrive at our variational algorithm. We only briefly motivate these extra ingredients for the sake of readability, and refer to \App{sec:theory} for additional explanations and more rigorous theoretical motivations. As such, the new algorithm can easily be implemented as an extension to an already existing (I)DMRG implementation.

We start by considering a setting familiar from conventional DMRG: a finite homogeneous one dimensional quantum lattice system, where every site corresponds to a $d$ level system. We label the sites by an integer $n$ and thus have a basis $\{\ket{s}_n,s=1,\ldots,d\}$ for the local Hilbert space on site $n$. The total Hilbert space is spanned by the product basis $\ket{\bm{s}}= \bigotimes_{n} \ket{s}_n$. We assume the dynamics of the system to be governed by a translation invariant Hamiltonian $H$. 

We further consider a variational parameterization of a ground state approximation of the system, for now in terms of a finite size (site dependent) MPS, but we will ultimately be interested in the thermodynamic limit. DMRG proceeds to find the best variational ground state approximation by employing an alternating least squares minimization: It starts from some initial state and successively optimizes each of the individual MPS tensors site by site by solving effective (Hamiltonian) eigenvalue problems, in a sweeping process through the lattice until convergence, where each iteration depends on already optimized tensors from previous iterations (see e.g.\ Refs~\onlinecite{DMRG1,DMRG2,DMRG_variational,DMRG_MC,MPS6_S}).

We are now however interested in the thermodynamic limit $n \in \mathbb{Z}$ (but will ignore the technical complications involving a rigorous definition of a Hilbert space in that limit). In that case the MPS ground state approximation will be given in terms of a translation invariant uniform MPS, described by a single MPS tensor (or a unit cell of $N$ tensors), repeated on all sites. Two immediate difficulties arise: Firstly, conventional DMRG updates the variational state site by site, thus breaking translation invariance. Secondly, the effective Hamiltonian for a single-site optimization has to be constructed from an infinite environment. 

After briefly introducing the variational class of uniform MPS and introducing necessary notation and conventions (for further details see \Sec{sec:umps}), we describe how the new algorithm modifies DMRG accordingly to exactly account for these two issues in order to arrive at a variational ground state algorithm directly formulated in the thermodynamic limit. 

\subsection{Uniform MPS}
\label{sec:umps_short}
A uniform MPS (uMPS) of bond dimension $D$ defined on an infinite translation invariant lattice is parameterized by a single collection of $d$ matrices $A^{s}\in\mathbb{C}^{D\times D}$ for $s=1,\ldots,d$. 
The overall translation invariant variational state is then given by
\begin{equation}
\label{eq:MPS1}
\ket{\Psi(A)}=\sum_{\bm{s}}\Big(\ldots A^{s_{n-1}}A^{s_{n}}A^{s_{n+1}}\ldots\Big)\ket{\bm{s}}
\end{equation} 
and can be represented diagrammatically as
\begin{equation*}
\ket{\Psi(A)} =  \dots
\begin{tikzpicture}[baseline = (X.base),every node/.style={scale=0.75},scale=.55]
\draw (0.5,1.5) -- (1,1.5); 
\draw[rounded corners] (1,2) rectangle (2,1);
\draw (1.5,1.5) node (X) {$A$};
\draw (2,1.5) -- (3,1.5); 
\draw[rounded corners] (3,2) rectangle (4,1);
\draw (3.5,1.5) node {$A$};
\draw (4,1.5) -- (5,1.5);
\draw[rounded corners] (5,2) rectangle (6,1);
\draw (5.5,1.5) node {$A$};
\draw (6,1.5) -- (7,1.5); 
\draw[rounded corners] (7,2) rectangle (8,1);
\draw (7.5,1.5) node {$A$};
\draw (8,1.5) -- (9,1.5); 
\draw[rounded corners] (9,2) rectangle (10,1);
\draw (9.5,1.5) node {$A$};
\draw (10,1.5) -- (10.5,1.5);
\draw (1.5,1) -- (1.5,.5); \draw (3.5,1) -- (3.5,.5); \draw (5.5,1) -- (5.5,.5);
\draw (7.5,1) -- (7.5,.5); \draw (9.5,1) -- (9.5,.5);
\end{tikzpicture} \dots
\end{equation*}

Exploiting the invariance of \eqref{eq:MPS1} under local gauge transformations $A^{s}\to XA^{s}X^{-1}$, with $X\in\mathbb{C}^{D\times D}$ invertible, the state can be cast into certain favorable representations, among them the \textit{left and right canonical representation}
\begin{subequations}
\label{eq:gauges}
\begin{align}
 \sum_{s}{A_{L}^{s}}^{\dagger}A_{L}^{s}&=\unity &   \sum_{s}A_{L}^{s}\,R\, {A_{L}^{s}}^{\dagger}&=R\label{eq:leftgauge}\\
 \sum_{s}A_{R}^{s}{A_{R}^{s}}^{\dagger}&=\unity &   \sum_{s}{A_{R}^{s}}^{\dagger}\,L\, A_{R}^{s}&=L,\label{eq:rightgauge}
\end{align}
\end{subequations}
or diagrammatically
\begin{align*}
\begin{tikzpicture}[baseline = (X.base),every node/.style={scale=0.75},scale=.55]
\draw (1,-1.5) edge[out=180,in=180] (1,1.5);
\draw[rounded corners] (1,2) rectangle (2,1);
\draw[rounded corners] (1,-1) rectangle (2,-2);
\draw (1.5,1) -- (1.5,-1);
\draw (1.5,1.5) node {$A_L$};
\draw (1.5,-1.5) node {$\bar{A}_L$};
\draw (2,1.5) -- (2.5,1.5); \draw (2,-1.5) -- (2.5,-1.5);
\end{tikzpicture} 
&= 
\begin{tikzpicture}[baseline = (X.base),every node/.style={scale=0.750},scale=.55]
\draw (1,-1.5) edge[out=180,in=180] (1,1.5);
\end{tikzpicture}
& 
\begin{tikzpicture}[baseline = (X.base),every node/.style={scale=0.750},scale=.55]
\draw (.5,1.5) -- (1,1.5); \draw (.5,-1.5) -- (1,-1.5);
\draw[rounded corners] (1,2) rectangle (2,1);
\draw[rounded corners] (1,-1) rectangle (2,-2);
\draw (1.5,1) -- (1.5,-1);
\draw (1.5,1.5) node {$A_L$};
\draw (1.5,-1.5) node {$\bar{A}_L$};
\draw (3,0) circle (.5);
\draw (3,0) node (X) {$R$};
\draw (3,0.5) edge[out=90,in=0] (2,1.5);
\draw (3,-0.5) edge[out=270,in=0] (2,-1.5);
\end{tikzpicture}  
&= 
\begin{tikzpicture}[baseline = (X.base),every node/.style={scale=0.750},scale=.55]
\draw (3,0) circle (.5);
\draw (3,0) node (X) {$R$};
\draw (3,0.5) edge[out=90,in=0] (2,1.5);
\draw (3,-0.5) edge[out=270,in=0] (2,-1.5);
\end{tikzpicture}
\\
\begin{tikzpicture}[baseline = (X.base),every node/.style={scale=0.750},scale=.55]
\draw (0.5,1.5) -- (1,1.5); \draw (0.5,-1.5) -- (1,-1.5);
\draw[rounded corners] (1,2) rectangle (2,1);
\draw[rounded corners] (1,-1) rectangle (2,-2);
\draw (1.5,1) -- (1.5,-1);
\draw (1.5,1.5) node {$A_R$};
\draw (1.5,-1.5) node {$\bar{A}_R$};
\draw (2,-1.5) edge[out=0,in=0] (2,1.5);
\end{tikzpicture} 
&= 
\begin{tikzpicture}[baseline = (X.base),every node/.style={scale=0.750},scale=.55]
\draw (1,-1.5) edge[out=0,in=0] (1,1.5);
\end{tikzpicture}
& 
\begin{tikzpicture}[baseline = (X.base),every node/.style={scale=0.750},scale=.55]
\draw (0,0) circle (.5);
\draw (0,0) node {$L$};
\draw (1,1.5) edge[out=180,in=90] (0,0.5);
\draw (1,-1.5) edge[out=180,in=270] (0,-0.5);
\draw[rounded corners] (1,2) rectangle (2,1);
\draw[rounded corners] (1,-1) rectangle (2,-2);
\draw (1.5,1) -- (1.5,-1);
\draw (1.5,1.5) node {$A_R$};
\draw (1.5,-1.5) node {$\bar{A}_R$};
\draw (2,1.5) -- (2.5,1.5); \draw (2,-1.5) -- (2.5,-1.5);
\end{tikzpicture} 
&= 
\begin{tikzpicture}[baseline = (X.base),every node/.style={scale=0.750},scale=.55]
\draw (0,0) circle (.5);
\draw (0,0) node {$L$};
\draw (1,1.5) edge[out=180,in=90] (0,0.5);
\draw (1,-1.5) edge[out=180,in=270] (0,-0.5);
\end{tikzpicture}
\end{align*}
Here $L$ and $R$ correspond to the left and right reduced density matrices of a bipartition of the state respectively. We henceforth refer to $A_L$ ($A_R$) as a left (right) isometric tensor, or just a left (right) isometry.

Defining the left and right transfer matrices 
\begin{equation}
T_{L/R}=\sum_{s}\bar{A}^{s}_{L/R}\otimes A^{s}_{L/R}
\label{eq:TM_LR}
\end{equation}
and using the notation $\rbra{x}$ and $\rket{x}$ for vectorizations of a $D\times D$ matrix $x$ in the $D^{2}$ dimensional ``double layer'' virtual space the transfer matrices act upon, the gauge conditions \eqref{eq:gauges} are equivalent to 
\begin{subequations}
\label{eq:gauges_TM}
\begin{align}
 \rbra{\unity}T_{L}&=\rbra{\unity}& T_{L}\rket{R}&=\rket{R} \label{eq:leftgauge_TM}\\
 T_{R}\rket{\unity}&=\rket{\unity}& \rbra{L}T_{R}&=\rbra{L}, \label{eq:rightgauge_TM}
\end{align}
\end{subequations}
i.e.\ $\unity$ and $R$ are the left and right dominant eigenvectors of $T_{L}$, while $L$ and $\unity$ are the left and right dominant eigenvectors of $T_{R}$.

As is standard practice in DMRG, we mix both of these representations and cast the state into the \textit{mixed canonical representation} 
\begin{subequations}
\label{eq:psi_mixed}
\begin{align}
\ket{\Psi(A)}&=\sum_{\bm{s}}(\ldots  A_{L}^{s_{n-1}} A_{C}^{s_{n}} A_{R}^{s_{n+1}} \ldots)\ket{\bm{s}}\label{eq:psiAC}\\
 &=\sum_{\bm{s}}(\ldots  A_{L}^{s_{n-1}}A_{L}^{s_{n}} C A_{R}^{s_{n+1}} A_{R}^{s_{n+2}} \ldots)\ket{\bm{s}},\label{eq:psiC}
\end{align} 
\end{subequations}
or diagrammatically,
\begin{align*}
\ket{\Psi(A)} &= 
\dots \begin{tikzpicture}[baseline = (X.base),every node/.style={scale=0.75},scale=.5]
\draw (0.5,0) -- (1,0);
\draw[rounded corners] (1,0.5) rectangle (2,-0.5);
\draw (1.5,0) node (X) {$A_L$};
\draw (2,0) -- (3,0); 
\draw[rounded corners] (3,0.5) rectangle (4,-0.5);
\draw (3.5,0) node {$A_L$};
\draw (4,0) -- (5,0);
\draw[rounded corners] (5,0.5) rectangle (6,-0.5);
\draw (5.5,0) node {$A_C$};
\draw (6,0) -- (9,0);
\draw[rounded corners] (9,0.5) rectangle (10,-0.5);
\draw (9.5,0) node {$A_R$};
\draw (10,0) -- (11,0);
\draw[rounded corners] (11,0.5) rectangle (12,-0.5);
\draw (11.5,0) node {$A_R$};
\draw (12,0) -- (12.5,0);
\draw (1.5,-.5) -- (1.5,-1); \draw (3.5,-.5) -- (3.5,-1); \draw (5.5,-.5) -- (5.5,-1); \draw (11.5,-.5) -- (11.5,-1);
\draw (9.5,-.5) -- (9.5,-1); 
\draw (5,1) node {$\phantom{X}$};
\end{tikzpicture}  \dots  \\
&= \dots \begin{tikzpicture}[baseline = (X.base),every node/.style={scale=0.75},scale=.5]
\draw (0.5,0) -- (1,0);
\draw[rounded corners] (1,0.5) rectangle (2,-0.5);
\draw (1.5,0) node (X) {$A_L$};
\draw (2,0) -- (3,0); 
\draw[rounded corners] (3,0.5) rectangle (4,-0.5);
\draw (3.5,0) node {$A_L$};
\draw (4,0) -- (5,0);
\draw[rounded corners] (5,0.5) rectangle (6,-0.5);
\draw (5.5,0) node {$A_L$};
\draw (6,0) -- (7,0);
\draw(7.5,0) circle (0.5);
\draw (8,0) -- (9,0);
\draw (7.5,0) node {$C$};
\draw[rounded corners] (9,0.5) rectangle (10,-0.5);
\draw (9.5,0) node {$A_R$};
\draw (10,0) -- (11,0);
\draw[rounded corners] (11,0.5) rectangle (12,-0.5);
\draw (11.5,0) node {$A_R$};
\draw (12,0) -- (12.5,0);
\draw (1.5,-.5) -- (1.5,-1); \draw (3.5,-.5) -- (3.5,-1);  \draw (5.5,-.5) -- (5.5,-1); \draw (11.5,-.5) -- (11.5,-1);
\draw (9.5,-.5) -- (9.5,-1); 
\draw (5,1) node {$\phantom{X}$};
\end{tikzpicture}  \dots  
\end{align*}
Here we have defined the \textit{center site tensor} $A^{s}_{C}$ (known as the single-site wave function $\Psi^{s}$ in DMRG)
\begin{equation}
\begin{split}
\AC &= \AL C = C \AR\\
\begin{tikzpicture}[baseline = (X.base),every node/.style={scale=0.750},scale=.55]
\draw (0,-0.5) -- (0.5,-0.5);
\draw[rounded corners] (0.5,0) rectangle (1.5,-1);  
\draw (1,-0.5) node (X) {$A_C$};
\draw (1,-1) -- (1,-1.5);
\draw (1.5,-0.5) -- (2,-0.5);
\end{tikzpicture} 
&= 
\begin{tikzpicture}[baseline = (X.base),every node/.style={scale=0.750},scale=.55]
\draw (2.5,0) -- (3,0);
\draw[rounded corners] (3,0.5) rectangle (4,-0.5);
\draw (3.5,0) node (X) {$A_L$};
\draw (3.5,-0.5) -- (3.5,-1);
\draw (4,0) -- (5,0);
\draw (5.5,0) circle (0.5);
\draw (5.5,0) node {$C$};
\draw (6,0) -- (6.5,0);
\end{tikzpicture} 
= 
\begin{tikzpicture}[baseline = (X.base),every node/.style={scale=0.750},scale=.55]
 \draw (0.5,0) -- (1,0);
\draw (1.5,0) circle (0.5);
\draw (1.5,0) node (X){$C$};
\draw (2,0) -- (3,0);
\draw[rounded corners] (3,0.5) rectangle (4,-0.5);
\draw (3.5,0) node {$A_R$};
\draw (3.5,-0.5) -- (3.5,-1);
\draw (4,0) -- (4.5,0);
\end{tikzpicture}
\end{split}
 \label{eq:AC}
\end{equation}
in terms of the bond matrix $C$, which constitutes the (invertible) gauge transformation relating $A_{L}$ and $A_{R}$ via $\AL = C\AR C^{-1}$. The singular values of $C$ then encode the entanglement spectrum of the state. Indeed, using $\AL C = C \AR$ we can verify that the left and right reduced density matrices in \eqref{eq:gauges} are given by $L=C^{\dagger}C$ and $R=CC^{\dagger}$. Furthermore, $\AL C = C \AR$  ensures that \eqref{eq:psiAC} and \eqref{eq:psiC} are translation invariant and that $A_{C}$ and $C$ can be shifted around arbitrarily. 
Normalization of the state, as well as of the reduced density matrices $L$ and $R$, corresponds to the single condition $\lVert C\rVert_2^2 = \Tr( CC^\dagger) = 1$. 

For ease of notation we further introduce the following partial states
\begin{subequations}
\label{eq:partial_LR}
\begin{align}
 \ket{\Psi_{L}^{\alpha}} &= \sum_{\bm{s}}(\ldots A_{L}^{s_{n-1}}A_{L}^{s_{n}})_{\alpha}\ket{\ldots s_{n-1}s_{n}} \label{eq:partial_psiL} \\
&= \dots 
\begin{tikzpicture}[baseline = (X.base),every node/.style={scale=0.750},scale=.55]
\draw (0.5,0) -- (1,0);
\draw[rounded corners] (1,0.5) rectangle (2,-0.5);
\draw (1.5,0) node (X) {$A_L$};
\draw (2,0) -- (3,0); 
\draw[rounded corners] (3,0.5) rectangle (4,-0.5);
\draw (3.5,0) node {$A_L$};
\draw (4,0) -- (5,0);
\draw[rounded corners] (5,0.5) rectangle (6,-0.5);
\draw (5.5,0) node {$A_L$};
\draw (7,0) node {$\alpha$};
\draw (6,0) -- (6.5,0);
\draw (1.5,-.5) -- (1.5,-1); \draw (3.5,-.5) -- (3.5,-1); \draw (5.5,-.5) -- (5.5,-1);
\end{tikzpicture}  \notag \\
\ket{\Psi_{R}^{\alpha}} &= \sum_{\bm{s}}(A_{R}^{s_{n}}A_{R}^{s_{n+1}}\ldots )_{\alpha}\ket{s_{n}s_{n+1}\ldots} \label{eq:partial_psiR} \\
  &= \begin{tikzpicture}[baseline = (X.base),every node/.style={scale=0.750},scale=.55]
\draw (4,0) node {$\alpha$};
\draw (4.5,0) -- (5,0);
\draw[rounded corners] (5,0.5) rectangle (6,-0.5);
\draw (5.5,0) node {$A_R$};
\draw (6,0) -- (7,0);
\draw[rounded corners] (7,0.5) rectangle (8,-0.5);
\draw (7.5,0) node {$A_R$};
\draw (8,0) -- (9,0);
\draw[rounded corners] (9,0.5) rectangle (10,-0.5);
\draw (9.5,0) node {$A_R$};
\draw (10,0) -- (10.5,0);
\draw (5.5,-.5) -- (5.5,-1); \draw (7.5,-.5) -- (7.5,-1); \draw (9.5,-.5) -- (9.5,-1);
\end{tikzpicture} \dots  \notag
\end{align}
\end{subequations}
with $n$ arbitrary, and use them to define the reduced basis states
\begin{subequations}
\label{eq:partial_center}
\begin{align}
  \ket{\Psi_{A_{C}}^{(\alpha,s,\beta)}} &= \ket{\Psi_{L}^{\alpha}}\ket{s}\ket{\Psi_{R}^{\beta}} \label{eq:partial_psiAC}\\
  \ket{\Psi_{C}^{(\alpha,\beta)}} &= \ket{\Psi_{L}^{\alpha}}\ket{\Psi_{R}^{\beta}}.\label{eq:partial_psiC}
\end{align}
\end{subequations}

\subsection{Effective Hamiltonian}
\label{sec:effH_NN}
The use of the mixed canonical representation \eqref{eq:psiAC} in DMRG is of significant importance for the stability, as it reduces the minimization of the (global) energy expectation value $\braket{\Psi|H|\Psi}/\braket{\Psi|\Psi}$ with respect to $A_{C}$ into a standard (hermitian) eigenvalue problem, instead of a generalized one. The effective Hamiltonian for this eigenvalue problem is the system Hamiltonian $H$ projected onto the degrees of freedom of $A_{C}$, and is known as the ``reduced'' or ``superblock'' Hamiltonian in DMRG.

We define the thermodynamic limit version of this reduced single-site Hamiltonian acting on $A_{C}$ as
\begin{align}
{H_{A_{C}}}_{(\alpha,s,\beta)}^{(\alpha',s',\beta')} &= \braket{\Psi_{A_{C}}^{(\alpha',s',\beta')}|H|\Psi_{A_{C}}^{(\alpha,s,\beta)}} \label{eq:HAC}\\
&= \cdots \begin{tikzpicture}[baseline = (X.base),every node/.style={scale=0.750},scale=.55]
\mpsT{-4.5}{1.5}{A_L} \mpsT{-2.5}{1.5}{A_L}  \mpsT{1.5}{1.5}{A_R} \mpsT{3.5}{1.5}{A_R} 
\mpsT{-4.5}{-1.5}{\bar{A}_L} \mpsT{-2.5}{-1.5}{\bar{A}_L}  \mpsT{1.5}{-1.5}{\bar{A}_R} \mpsT{3.5}{-1.5}{\bar{A}_R}
\lineH{-5.5}{-5}{+1.5} \lineH{-4}{-3}{+1.5}  \lineH{2}{3}{+1.5} \lineH{4}{4.5}{+1.5} 
\lineH{-5.5}{-5}{-1.5} \lineH{-4}{-3}{-1.5}  \lineH{2}{3}{-1.5} \lineH{4}{4.5}{-1.5} 
\lineV{-1}{-.5}{-4.5} \lineV{-1}{-.5}{-2.5} \lineV{.5}{1}{-4.5} \lineV{.5}{1}{-2.5} 
\lineV{.5}{1}{-0.5} 
\lineV{-1}{-0.5}{1.5} \lineV{0.5}{1}{1.5}  \lineV{-1}{-.5}{3.5} \lineV{.5}{1}{3.5}
\lineH{-5.5}{4.5}{0.5}
\draw (-0.5,0) node (X) {$H$};
\lineH{-5.5}{4.5}{-0.5}
\lineH{-2}{-1}{-1.5} \lineH{0}{1}{-1.5} \lineV{-1}{-.5}{-0.5}
\lineH{-2}{-1}{1.5} \lineH{0}{1}{+1.5} \lineV{0.5}{1}{-0.5}
\end{tikzpicture} \cdots \notag 
\end{align}

Additionally, we also define an effective Hamiltonian acting on the bond matrix $C$ as
\begin{align}
 {H_{C}}_{(\alpha,\beta)}^{(\alpha',\beta')} &= \braket{\Psi_{C}^{(\alpha',\beta')}|H|\Psi_{C}^{(\alpha,\beta)}} \label{eq:HC}\\
 &= \cdots \begin{tikzpicture}[baseline = (X.base),every node/.style={scale=0.750},scale=.55]
\mpsT{-4.5}{1.5}{A_L} \mpsT{-2.5}{1.5}{A_L} 
\mpsT{1.5}{1.5}{A_R} \mpsT{3.5}{1.5}{A_R} 
\mpsT{-4.5}{-1.5}{\bar{A}_L} \mpsT{-2.5}{-1.5}{\bar{A}_L}  \mpsT{1.5}{-1.5}{\bar{A}_R} \mpsT{3.5}{-1.5}{\bar{A}_R}
\lineH{-5.5}{-5}{+1.5} \lineH{-4}{-3}{+1.5} \lineH{-2}{-1}{1.5} \lineH{0}{1}{+1.5} \lineH{2}{3}{+1.5} \lineH{4}{4.5}{+1.5} 
\lineH{-5.5}{-5}{-1.5} \lineH{-4}{-3}{-1.5} \lineH{-2}{-1}{-1.5} \lineH{0}{1}{-1.5} \lineH{2}{3}{-1.5} \lineH{4}{4.5}{-1.5} 
\lineV{-1}{-.5}{-4.5} \lineV{-1}{-.5}{-2.5} \lineV{.5}{1}{-4.5} \lineV{.5}{1}{-2.5} 
\lineV{-1}{-0.5}{1.5} \lineV{0.5}{1}{1.5}  \lineV{-1}{-.5}{3.5} \lineV{.5}{1}{3.5}
\lineH{-5.5}{4.5}{0.5}
\draw (-0.5,0) node (X) {$H$};
\lineH{-5.5}{4.5}{-0.5}
\end{tikzpicture} \cdots \notag
\end{align}
which does not appear directly in the context of DMRG, but will be needed later for a consistent update of the state without breaking translation invariance. It can be interpreted as a ``zero site'' effective Hamiltonian, which would feature in an optimization of the global energy expectation value with respect to the Schmidt values. 

In an efficient implementation, these effective eigenvalue problems are typically solved using an iterative eigensolver, so that we only need to implement the action of $H_{A_{C}}$ and $H_{C}$ onto $A_{C}$ and $C$.

While the energy expectation value is extensive and thus divergent in the thermodynamic limit, the effective Hamiltonians $H_{A_C}$ and $H_C$ are well defined and finite in the thermodynamic limit if one properly subtracts the current energy expectation value from the Hamiltonian $H$. We demonstrate this procedure for the case of nearest neighbor interactions $H=\sum_{n}h_{n,n+1}$, where the two-site Hamiltonian $h_{n,n+1}$ acts on neighboring sites $n,n+1$ only. We refer to \App{sec:effH_construct} for the case of long range interactions and for general Hamiltonians given in terms of Matrix Product Operators (MPOs).

\begin{widetext}
In the case of nearest neighbor interactions, the action of $H_{A_{C}}$ onto $A_{C}$ splits up into four individual contributions, which follow from the decomposition $\ket{\Psi} = \sum_{\alpha,\beta,s}A^{s}_{C,(\alpha,\beta)} \ket{\Psi_L^\alpha} \ket{s} \ket{\Psi_R^\beta}$ (left block containing sites $n<0$, center site $n=0$, and right block containing sites $n>0$). The action of $H_{A_{C}}$ onto $A_{C}$ is given by
\begin{equation}
 \begin{split}
 {A}^{\prime s}_{C} &= 
 \sum_{tk\ell}h^{ts}_{k\ell}{A_{L}^{t}}^{\dagger}A_{L}^{k}A^{\ell}_{C} + h^{st}_{k\ell}A^{k}_{C}A_{R}^{\ell}{A_{R}^{t}}^{\dagger} + 
 H_{L} A^{s}_{C} + A^{s}_{C} H_{R}  \\
\begin{tikzpicture}[baseline = (X.base),every node/.style={scale=0.750},scale=.55]
\draw[rounded corners] (1,-.5) rectangle (2,.5);
\draw (0.5,0) -- (1,0); \draw (2,0) -- (2.5,0); \draw (1.5,-.5) -- (1.5,-1);
\draw (1.5,0) node {${A}^{\prime}_C$};
\end{tikzpicture} \;
&= 
\begin{tikzpicture}[baseline = (X.base),every node/.style={scale=0.750},scale=.55]
\draw (1,-1.5) edge[out=180,in=180] (1,1.5);
\draw[rounded corners] (1,2) rectangle (2,1);
\draw (2, 1.5) -- (3, 1.5);
\draw (1.5,-0.5) -- (1.5,-1);
\draw[rounded corners] (3,2) rectangle (4,1);
\draw (1.5, 1) -- (1.5, 0.5);
\draw (3.5, 1) -- (3.5, 0.5);
\draw (3.5, -0.5) -- (3.5, -3);
\draw[rounded corners] (1,-1) rectangle (2,-2);
\draw (1,0.5) rectangle (4,-.5);
\draw (2, -1.5) -- (3, -1.5);
\draw (4,+1.5) edge[out = 0, in =0] (4,-1.5);
\draw (1.5,1.5) node {$A_L$};
\draw (3.5,1.5) node {$A_C$};
\draw (3,-1.5) edge[out=0,in=90] (3.25,-2);
\draw (3.25,-2) edge[out=-90,in=0] (3,-2.5);
\draw (2.5,-2.5) -- (3,-2.5);
\draw (4,-1.5) edge[out=180,in=90] (3.75,-2);
\draw (3.75,-2) edge[out=-90,in=180] (4,-2.5);
\draw (4,-2.5) -- (4.5,-2.5);
\draw (1.5,-1.5) node {$\bar{A}_L$};
\draw (2.5,0) node {$h$}; 
\end{tikzpicture} 
+
\begin{tikzpicture}[baseline = (X.base),every node/.style={scale=0.750},scale=.55]
\draw (1,-1.5) edge[out=180,in=180] (1,1.5);
\draw[rounded corners] (1,2) rectangle (2,1);
\draw (2, 1.5) -- (3, 1.5);
\draw (3.5, -0.5) -- (3.5, -1);
\draw[rounded corners] (3,2) rectangle (4,1);
\draw (1.5, 1) -- (1.5, 0.5);
\draw (3.5, 1) -- (3.5, 0.5);
\draw (1.5, -0.5) -- (1.5, -3);
\draw[rounded corners] (3,-1) rectangle (4,-2);
\draw (1,0.5) rectangle (4,-.5);
\draw (2, -1.5) -- (3, -1.5);
\draw (4,+1.5) edge[out = 0, in =0] (4, -1.5);
\draw (1.5,1.5) node {$A_C$};
\draw (3.5,1.5) node {$A_R$};
\draw (1,-1.5) edge[out=0,in=90] (1.25,-2);
\draw (1.25,-2) edge[out=-90,in=0] (1,-2.5);
\draw (1,-2.5) -- (0,-2.5);
\draw (2,-1.5) edge[out=180,in=90] (1.75,-2);
\draw (1.75,-2) edge[out=-90,in=180] (2,-2.5);
\draw (2,-2.5) -- (3,-2.5);
\draw (3.5,-1.5) node {$\bar{A}_R$};
\draw (2.5,0) node {$h$};
\end{tikzpicture}
+
\begin{tikzpicture}[baseline = (X.base),every node/.style={scale=0.750},scale=.55]
\draw (0,0) circle (.5);
\draw (0,0) node (X) {$H_L$};
\draw (0,0.5) edge[out=90,in=180] (1,1.5);
\draw (0,-0.5) edge[out=270,in=180] (1,-1.5);
\draw[rounded corners] (1,2) rectangle (2,1);
\draw (1.5,1.5) node {$A_C$};
\draw (1,-1.5) edge[out=0,in=90] (1.25,-2);
\draw (1.25,-2) edge[out=-90,in=0] (1,-2.5);
\draw (1,-2.5) -- (0,-2.5);
\draw (2,-1.5) edge[out=180,in=90] (1.75,-2);
\draw (1.75,-2) edge[out=-90,in=180] (2,-2.5);
\draw (2,-2.5) -- (3,-2.5);
\draw (2,+1.5) edge[out = 0, in =0] (2, -1.5);
\draw (1.5,-3) -- (1.5,1);
\end{tikzpicture}
+
\begin{tikzpicture}[baseline = (X.base),every node/.style={scale=0.750},scale=.55]
\draw (1,-1.5) edge[out=180,in=180] (1,1.5);
\draw[rounded corners] (1,2) rectangle (2,1);
\draw (1.5,1.5) node {$A_C$};
\draw (1,-1.5) edge[out=0,in=90] (1.25,-2);
\draw (1.25,-2) edge[out=-90,in=0] (1,-2.5);
\draw (1,-2.5) -- (0,-2.5);
\draw (2,-1.5) edge[out=180,in=90] (1.75,-2);
\draw (1.75,-2) edge[out=-90,in=180] (2,-2.5);
\draw (2,-2.5) -- (3,-2.5);
\draw (2,1.5) edge[out=0,in=90] (3,0.5);
\draw (3,0) circle (.5);
\draw (3,-.5) edge[out = -90, in = 0] (2, -1.5);
\draw (3,0) node {$H_R$};
\draw (1.5,-3) -- (1.5,1);
\end{tikzpicture} 
 \end{split}
 \label{eq:HontoAC_NN}
\end{equation}
where the first two terms correspond to the Hamiltonian terms $h_{-1,0}$ and $h_{0,1}$ coupling the center site to the left and right block, respectively, and 
$H_L$ and $H_R$ sum up the contributions of all the Hamiltonian terms $h_{n,n+1}$ acting strictly to the left and to the right of the center site.
\end{widetext}

The environments $H_{L}$ and $H_{R}$ are usually constructed iteratively while sweeping through the (finite) lattice in conventional DMRG, or grown successively in every iteration in IDMRG. In the thermodynamic limit, these terms consist of a diverging number of individual local interaction contributions $h_{n,n+1}$, and care needs to be taken in their construction.

Indeed, the $k^{\rm th}$ contribution to $\rbra{H_{L}}$ comes from the Hamiltonian term $h_{-k-1,-k}$ and is given by $\rbra{h_{L}}[T_{L}]^{k-1}$. Likewise, $[T_{R}]^{k-1}\rket{h_{R}}$ is the $k^{\rm th}$ contribution to $\rket{H_{R}}$ stemming from $h_{k,k+1}$.
Here, we have used the definitions
\begin{equation}
\begin{split}
 h_{L}&=\sum_{stk\ell}h^{st}_{k\ell}{A_{L}^{t}}^{\dagger}{A_{L}^{s}}^{\dagger}{A_{L}^{k}}{A_{L}^{\ell}}\\
 h_{R}&=\sum_{stk\ell}h^{st}_{k\ell}{A_{R}^{k}}{A_{R}^{\ell}}{A_{R}^{t}}^{\dagger}{A_{R}^{s}}^{\dagger},
 \label{eq:hlr_NN}
 \end{split}
\end{equation}
or diagrammatically
\begin{align*}
 \begin{tikzpicture}[baseline = (X.base),every node/.style={scale=0.750},scale=.55]
\draw (0,0) circle (.5);
\draw (0,0) node (X) {$h_L$};
\draw (0,0.5) edge[out=90,in=180] (1,1.5);
\draw (0,-0.5) edge[out=270,in=180] (1,-1.5);
\end{tikzpicture}
&=
 \begin{tikzpicture}[baseline = (X.base),every node/.style={scale=0.750},scale=.55]
\draw (-1,-1.5) edge[out=180,in=180] (-1,1.5);
\draw[rounded corners] (-1,2) rectangle (0,1);
\draw[rounded corners] (-1,-1) rectangle (0,-2);
\draw (0,1.5) -- (1,1.5); \draw (0,-1.5) -- (1,-1.5);
\draw[rounded corners] (1,2) rectangle (2,1);
\draw[rounded corners] (1,-1) rectangle (2,-2);
\draw (-1,0.5) rectangle (2,-.5);
\draw (0.5,0) node {$h$};
\draw (-0.5,1.5) node {$A_L$}; \draw (1.5,1.5) node {$A_L$};
\draw (-0.5,-1.5) node {$\bar{A}_L$}; \draw (1.5,-1.5) node {$\bar{A}_L$};
\draw (-.5,1) -- (-.5,.5);  \draw (1.5,1) -- (1.5,.5);
\draw (-.5,-1) -- (-.5,-.5); \draw (1.5,-1) -- (1.5,-.5);
\draw (2,1.5) -- (2.5,1.5); \draw (2,-1.5) -- (2.5,-1.5);
\end{tikzpicture}
\\
\begin{tikzpicture}[baseline = (X.base),every node/.style={scale=0.750},scale=.55]
\draw (2,1.5) edge[out=0,in=90] (3,0.5);
\draw (3,0) circle (.5);
\draw (3,-.5) edge[out = -90, in = 0] (2, -1.5);
\draw (3,0) node (X) {$h_R$};
\end{tikzpicture} &= 
\begin{tikzpicture}[baseline = (X.base),every node/.style={scale=0.750},scale=.55]
\draw (-1.5,1.5) -- (-1,1.5); \draw (-1.5,-1.5) -- (-1,-1.5);
\draw[rounded corners] (-1,2) rectangle (0,1);
\draw[rounded corners] (-1,-1) rectangle (0,-2);
\draw (0,1.5) -- (1,1.5); \draw (0,-1.5) -- (1,-1.5);
\draw[rounded corners] (1,2) rectangle (2,1);
\draw[rounded corners] (1,-1) rectangle (2,-2);
\draw (-1,0.5) rectangle (2,-.5);
\draw (0.5,0) node {$h$};
\draw (-0.5,1.5) node {$A_R$}; \draw (1.5,1.5) node {$A_R$};
\draw (-0.5,-1.5) node {$\bar{A}_R$}; \draw (1.5,-1.5) node {$\bar{A}_R$};
\draw (-.5,1) -- (-.5,.5);  \draw (1.5,1) -- (1.5,.5);
\draw (-.5,-1) -- (-.5,-.5); \draw (1.5,-1) -- (1.5,-.5);
\draw (2,1.5) edge[out = 0, in =0] (2, -1.5);
\end{tikzpicture}
\end{align*}
Summing up all such local contributions gives rise to infinite geometric sums of the transfer matrices $T_{L/R}$
\begin{align}
 \rbra{H_{L}}&=\rbra{h_{L}}\sum_{k=0}^{\infty}[{T_{L}}]^{k}&
  \rket{H_{R}}&=\sum_{k=0}^{\infty}[{T_{R}}]^{k}\rket{h_{R}},
  \label{eq:InfGS_NN}
\end{align}
where $\rbra{H_{L}}$ can be presented diagrammatically as
\begin{align*}
\begin{tikzpicture}[baseline = (X.base),every node/.style={scale=0.750},scale=.55]
\draw (0,0) circle (.5);
\draw (0,0) node (X) {$H_L$};
\draw (0,0.5) edge[out=90,in=180] (1,1.5);
\draw (0,-0.5) edge[out=270,in=180] (1,-1.5);
\end{tikzpicture}
=
\begin{tikzpicture}[baseline = (X.base),every node/.style={scale=0.750},scale=.55]
\draw (0,0) circle (.5);
\draw (0,0) node (X) {$h_L$};
\draw (0,0.5) edge[out=90,in=180] (1,1.5);
\draw (0,-0.5) edge[out=270,in=180] (1,-1.5);
\end{tikzpicture} 
\left[
\unity + 
\begin{tikzpicture}[baseline = (X.base),every node/.style={scale=0.750},scale=.55]
\draw (.5,1.5) -- (1,1.5); \draw (.5,-1.5) -- (1,-1.5);
\draw[rounded corners] (1,2) rectangle (2,1);
\draw[rounded corners] (1,-1) rectangle (2,-2);
\draw (1.5,1.5) node {$A_L$}; \draw (1.5,-1.5) node {$\bar{A}_L$};
\draw (1.5,1) -- (1.5,-1); \draw (2,1.5) -- (2.5,1.5); \draw (2,-1.5) -- (2.5,-1.5);
\end{tikzpicture} +
\begin{tikzpicture}[baseline = (X.base),every node/.style={scale=0.750},scale=.55]
\draw (.5,1.5) -- (1,1.5); \draw (.5,-1.5) -- (1,-1.5);
\draw[rounded corners] (1,2) rectangle (2,1);
\draw[rounded corners] (1,-1) rectangle (2,-2);
\draw (1.5,1.5) node {$A_L$}; \draw (1.5,-1.5) node {$\bar{A}_L$};
\draw (1.5,1) -- (1.5,-1); \draw (2,1.5) -- (3,1.5); \draw (2,-1.5) -- (3,-1.5);
\draw[rounded corners] (3,2) rectangle (4,1);
\draw[rounded corners] (3,-1) rectangle (4,-2);
\draw (3.5,1.5) node {$A_L$}; \draw (3.5,-1.5) node {$\bar{A}_L$};
\draw (3.5,1) -- (3.5,-1); \draw (4,1.5) -- (4.5,1.5); \draw (4,-1.5) -- (4.5,-1.5);
\end{tikzpicture}
+ \;\ldots
\right]
\end{align*}
and likewise for $\rket{H_{R}}$. 

The transfer matrix $T_L$ has a dominant eigenvalue of magnitude one, with corresponding left and right eigenvectors $\rbra{\openone}$ and $\rket{R}$. The projection $\rbra{h_{L}}[T_{L}]^{k}\rket{R} = (h_{L}|R)$ is the energy density expectation value $e = \braket{\Psi|h_{-k-1,-k}|\Psi}$ and is independent of $k$. 
Subtracting the energy $e$ from the Hamiltonian as $\tilde{h} =  h - e \openone$, we can write $\rbra{h_L} = \rbra{\tilde{h}_L} + e \rbra{\openone}$. The second term is exactly proportional to the left eigenvector of eigenvalue $1$ and therefore gives rise to a diverging contribution in the geometric sum, corresponding to the total energy of the left half infinite block.
Since this contribution acts as the identity in the effective Hamiltonian $H_{A_C}$ [Eq.~\eqref{eq:HontoAC_NN}], we can however safely discard this diverging contribution without changing the eigenvectors of $H_{A_C}$. This corresponds to an overall energy shift of the left half infinite block such that $(H_{L}|R)=0$. For the remaining part $\rbra{\tilde{h}_L}$ the geometric sum converges. With $\rket{\tilde{h}_{R}}=\rket{h_{R}}-e\rket{\unity}$ the same comments apply to the construction of $\rket{H_R}$.

\begin{table*}[htb]
\begin{minipage}{\linewidth}
\begin{algorithm}[H]
\caption{Explicit terms of effective Hamiltonians with nearest neighbor interactions and their application}
\label{alg:Heff_NN}
\begin{algorithmic}[1]
  \Require two-site Hamiltonian $h$, current uMPS tensors $A_{L}$, $A_{R}$ in left and right gauge, left dominant eigenvector $\rbra{L}$ of $T_{R}$, right dominant eigenvector $\rket{R}$ of $T_{L}$, desired precision $\epsilon_{\rm S}$ for terms involving infinite geometric sums
  \Ensure Explicit terms of effective Hamiltonians $H_{A_{C}}$ and $H_{C}$, updated ${A}^{\prime}_{C}$ and $C^{\prime}$
  
  \Function{HeffTerms}{$H=h$,$A_{L}$,$A_{R}$,$L$,$R$,$\epsilon_{\rm S}$} \Comment{Calculates explicit terms of effective Hamiltonians}
    \State Calculate $h_{L}$ and $h_{R}$ from \eqref{eq:hlr_NN}
    \State Calculate $H_{L}$ and $H_{R}$ by iteratively solving \eqref{eq:HLR_NN_iterative} or (preferably) \eqref{eq:HLR_NN}, to precision $\epsilon_{\rm S}$
    \State $H_{A_{C}}\gets\{h,A_{L},A_{R},H_{L},H_{R}\}$
    \State $H_{C}\gets\{h,A_{L},A_{R},H_{L},H_{R}\}$
    \State \Return $H_{A_{C}},H_{C}$
  \EndFunction

  \Function{ApplyHAC}{$A_{C}$,$H_{A_{C}}$} 
    \Comment{Terms of $H_{A_{C}}$ from \Call{HeffTerms}{$H$,$A_{L}$,$A_{R}$,$L$,$R$,$\epsilon_{\rm S}$}}
    \State Calculate updated ${A}^{\prime}_{C}$ from \eqref{eq:HontoAC_NN}
    \State \Return ${A}^{\prime}_{C}$
  \EndFunction
   
  \Function{ApplyHC}{$C$,$H_{C}$} 
    \Comment{Terms of $H_{C}$ from \Call{HeffTerms}{$H$,$A_{L}$,$A_{R}$,$L$,$R$,$\epsilon_{\rm S}$}}
    \State Calculate updated ${C^{\prime}}$ from \eqref{eq:HontoC_NN}
    \State \Return ${C^{\prime}}$
  \EndFunction
\end{algorithmic}
\end{algorithm}
\end{minipage}
\caption{Pseudocode for obtaining the explicit terms of the effective Hamiltonians $H_{A_{C}}$ and $H_{C}$ for systems with nearest neighbor interactions and their applications onto a state.}
\label{tab:Heff_NN}
\end{table*}

We can evaluate $H_L$ and $H_R$ recursively as
\begin{equation}
\begin{split}
 \rbra{H_{L}^{[n+1]}} &= \rbra{H_{L}^{[n]}}T_{L} + \rbra{\tilde{h}_{L}}\\
 \rket{H_{R}^{[n+1]}} &= T_{R}\rket{H_{R}^{[n]}} + \rket{\tilde{h}_{R}}
\end{split}
 \label{eq:HLR_NN_iterative}
\end{equation} 
with initialization $\rbra{H_{L}^{[0]}}=\rbra{\tilde{h}_{L}}$ and $\rket{H_{R}^{[0]}}=\rket{\tilde{h}_{R}}$. We can repeat these recursions until e.g.\ $\lVert H_{L/R}^{[n+1]} - H_{L/R}^{[n]}\rVert$ drops below some desired accuracy $\epsilon_{\rm S}$. This strategy is conceptually simple and closely resembles the successive construction of the environments in the context of (I)DMRG, but is not very efficient, as its performance is comparable to that of a power method. 

A more efficient approach is to formally perform the geometric sums in \eqref{eq:InfGS_NN} explicitly, and to iteratively solve the resulting two systems of equations
\begin{equation}
\begin{split}
\rbra{H_{L}}[\unity - T_{L} + \rket{R}\rbra{\unity}] &= \rbra{h_{L}} -\rbraket{h_{L}}{R}\rbra{\unity}\\
[\unity - T_{R} + \rket{\unity}\rbra{L}]\rket{H_{R}} &= \rket{h_{R}}- \rket{\unity}\rbraket{L}{h_{R}}
\label{eq:HLR_NN}
\end{split}
\end{equation}
for $\rbra{H_{L}}$ and $\rket{H_{R}}$ to precision $\epsilon_{\rm S}$, as explained in \App{sec:invertE}.

So far, we have discussed the action of $H_{A_C}$. The action of $H_{C}$ onto $C$ follows simply from \eqref{eq:HontoAC_NN} by projecting onto $A_{L}$ or $A_{R}$. 
Using the defining property of $H_L$ or $H_R$, the result simplifies to
\begin{widetext}
\begin{equation}
 \begin{split}
 {C^{\prime}} & = \sum_{stk\ell} h^{st}_{k\ell}{A_{L}^{s}}^{\dagger}A_{L}^{k}C A_{R}^{\ell}{A_{R}^{t}}^{\dagger} + H_{L}C + C H_{R}.\\
\begin{tikzpicture}[baseline = (X.base),every node/.style={scale=0.750},scale=.55]
\draw (1.5,0) circle (.5);
\draw (0.5,0) -- (1,0); \draw (2,0) -- (2.5,0);
\draw (1.5,0) node (X) {${C^{\prime}}$};
\end{tikzpicture} \; &= 
\begin{tikzpicture}[baseline = (X.base),every node/.style={scale=0.750},scale=.55]
\draw (1,-1.5) edge[out=180,in=180] (1,1.5);
\draw[rounded corners] (1,2) rectangle (2,1);
\draw[rounded corners] (1,-1) rectangle (2,-2);
\draw (3.5,1.5) circle (.5);
\draw (2, 1.5) -- (3, 1.5); \draw (2, -1.5) -- (3, -1.5); 
\draw (4, 1.5) -- (5, 1.5); \draw (4, -1.5) -- (5, -1.5); 
\draw (1.5,-0.5) -- (1.5,-1); \draw (1.5, 1) -- (1.5, 0.5);
\draw (5.5, 1) -- (5.5, 0.5); \draw (5.5,-1) -- (5.5,-0.5);
\draw[rounded corners] (5,2) rectangle (6,1);
\draw[rounded corners] (5,-1) rectangle (6,-2);
\draw (6,+1.5) edge[out = 0, in =0] (6,-1.5);
\draw (1.5,1.5) node {$A_L$};
\draw (3.5,1.5) node {$C$};
\draw (5.5,1.5) node {$A_R$};
\draw (3,-1.5) edge[out=0,in=90] (3.25,-2);
\draw (3.25,-2) edge[out=-90,in=0] (3,-2.5);
\draw (2.5,-2.5) -- (3,-2.5);
\draw (4,-1.5) edge[out=180,in=90] (3.75,-2);
\draw (3.75,-2) edge[out=-90,in=180] (4,-2.5);
\draw (4,-2.5) -- (4.5,-2.5);
\draw (1.5,-1.5) node {$\bar{A}_L$};
\draw (5.5,-1.5) node {$\bar{A}_R$};
\draw (1,.5) rectangle (6,-.5);
\draw (3.5,0) node (X) {$h$}; 
\end{tikzpicture} 
+
\begin{tikzpicture}[baseline = (X.base),every node/.style={scale=0.750},scale=.55]
\draw (0,0) circle (.5);
\draw (0,0) node (X) {$H_L$};
\draw (0,0.5) edge[out=90,in=180] (1,1.5);
\draw (0,-0.5) edge[out=270,in=180] (1,-1.5);
\draw (1.5,1.5) circle (.5);
\draw (1.5,1.5) node {$C$};
\draw (1,-1.5) edge[out=0,in=90] (1.25,-2);
\draw (1.25,-2) edge[out=-90,in=0] (1,-2.5);
\draw (1,-2.5) -- (0,-2.5);
\draw (2,-1.5) edge[out=180,in=90] (1.75,-2);
\draw (1.75,-2) edge[out=-90,in=180] (2,-2.5);
\draw (2,-2.5) -- (3,-2.5);
\draw (2,+1.5) edge[out = 0, in =0] (2, -1.5);
\end{tikzpicture}
+
\begin{tikzpicture}[baseline = (X.base),every node/.style={scale=0.750},scale=.55]
\draw (1,-1.5) edge[out=180,in=180] (1,1.5);
\draw (1.5,1.5) circle (.5);
\draw (1.5,1.5) node {$C$};
\draw (1,-1.5) edge[out=0,in=90] (1.25,-2);
\draw (1.25,-2) edge[out=-90,in=0] (1,-2.5);
\draw (1,-2.5) -- (0,-2.5);
\draw (2,-1.5) edge[out=180,in=90] (1.75,-2);
\draw (1.75,-2) edge[out=-90,in=180] (2,-2.5);
\draw (2,-2.5) -- (3,-2.5);
\draw (2,1.5) edge[out=0,in=90] (3,0.5);
\draw (3,0) circle (.5);
\draw (3,-.5) edge[out = -90, in = 0] (2, -1.5);
\draw (3,0) node {$H_R$};
\end{tikzpicture}
 \end{split}
 \label{eq:HontoC_NN} 
\end{equation} 
\end{widetext}

The first two terms of \eqref{eq:HontoAC_NN} can be applied in $\O(d^{4}D^{3})$ operations\footnote{In many cases $h_{j,j+1}$ is sparse and the number $d_{h}$ of non-zero elements is usually of the order $\O(d^{2})$. The first two terms can then be applied in $\O(d_{h}D^{3})$ operations.}, and the last two terms in $\O(dD^{3})$ operations. For \eqref{eq:HontoC_NN} the first term can be applied in $\O(d^{4}D^{3})$ operations, and the last two terms in $\O(D^{3})$ operations. To generate the necessary terms for \eqref{eq:HontoAC_NN} and \eqref{eq:HontoC_NN} we have to iteratively evaluate two infinite geometric sums, involving $\O(D^{3})$ operations (when iteratively solving \eqref{eq:HLR_NN} the solutions from the previous iteration can be used as starting vectors to speed up convergence). A pseudocode summary for obtaining the necessary explicit terms of $H_{A_{C}}$ and $H_{C}$ and their applications onto a state is presented in \Tab{tab:Heff_NN}.

\subsection{Updating the state}
\label{sec:state_update}
In DMRG, we would update the state by replacing $A_C$  with the lowest eigenvector $\tilde{A}_{C}$ of $H_{A_{C}}$ and then shift the center site to the right by computing an orthogonal factorization $\tilde{A}^{s}_{C}=\tilde{A}^{s}_{L}\tilde{C}_R$, or to the left by computing $\tilde{A}^{s}_{C}=\tilde{C}_L\tilde{A}^{s}_{R}$. As such, the state gets updated by only replacing the current site with $\tilde{A}^{s}_{L}$ or $\tilde{A}^{s}_{R}$, while leaving all other sites untouched. However, applying this scheme in our setting would immediately destroy translation invariance after a single step.

We  want to construct an alternative scheme that applies global updates in order to preserve translation invariance at any time. Global updates can most easily be applied with an explicit uniform parameterization in terms of a single tensor $A$. On the other hand, DMRG experience teaches us that the stability is greatly enhanced when applying updates at the level of $A_C$ and $C$, which are isometrically related to the full state. 

We therefore calculate the lowest eigenvector $\tilde{A}_{C}$ of $H_{A_{C}}$ like in DMRG, but additionally also the lowest eigenvector $\tilde{C}$ of $H_{C}$. We then globally update the state by finding new $\tilde{A}_L$ and $\tilde{A}_R$ as the left and right isometric tensors that minimize 
$\sum_{s} \lVert \tilde{A}_{L}^{s} \tilde{C} - \tilde{A}_{C}^{s} \rVert^2$ and $\sum_{s} \lVert \tilde{C}\tilde{A}_{R}^{s} - \tilde{A}_{C}^{s} \rVert^2$ respectively. 
These minimization problems can be solved directly (not iteratively) and without inverting $\tilde{C}$ (see below).
As shown in \App{sec:theory}, at the variational optimum the values of these objective functions go to zero, and current $A_{C}$ and $C$ are the lowest eigenvectors of $H_{A_{C}}$ and $H_{C}$ respectively.

For the remainder of this section we omit tildes and use the following matricizations of the 3-index tensors
\begin{equation}
\begin{split}
\mathcal{A}_{L,(s\alpha,\beta)}&=A_{L,(\alpha,\beta)}^{s}\\
\mathcal{A}_{R,(\alpha,s\beta)}&=A_{R,(\alpha,\beta)}^{s}\\
\mathcal{A}^{[\ell]}_{C,(s\alpha,\beta)}=\mathcal{A}^{[r]}_{C,(\alpha,s\beta)}&=A_{C,(\alpha,\beta)}^{s}.
\end{split}
\label{eq:vectorizations}
\end{equation} 
We thus want to extract updated $A_L$ and $A_R$ from updated $A_C$ and $C$ by solving
\begin{subequations}
\label{eq:err}
\begin{align}
\epsilon_{L} &= \min_{\mathcal{A}_{L}^{\dagger}\mathcal{A}_{L}=\unity} \lVert \mathcal{A}_{C}^{[\ell]} - \mathcal{A}_{L}C\rVert_{2}\label{eq:err_l}\\
\epsilon_{R} &= \min_{\mathcal{A}_{R}\mathcal{A}_{R}^{\dagger}=\unity} \lVert \mathcal{A}_{C}^{[r]} - C\mathcal{A}_{R}\rVert_{2}.\label{eq:err_r}
\end{align}
\end{subequations}
In exact arithmetic, the solution of these minimization problems is known, namely $\mathcal{A}_L$ will be the isometry in the polar decomposition of $\mathcal{A}_C^{[l]} C^\dagger$ (and similar for $\mathcal{A}_R$, see Thm.~IX.7.2 in Ref.~\onlinecite{Bhatia}). Computing the singular value decompositions (SVD)
\begin{align}
 \mathcal{A}_{C}^{[\ell]}C^{\dagger}&=U^{[\ell]}\Sigma^{[\ell]}{V^{[\ell]}}^{\dagger}&
 C^{\dagger}\mathcal{A}_{C}^{[r]}&=U^{[r]}\Sigma^{[r]}{V^{[r]}}^{\dagger}
\end{align}
we thus obtain
\begin{align}
 \mathcal{A}_{L} &= U^{[\ell]}{V^{[\ell]}}^{\dagger}&
 \mathcal{A}_{R} &= U^{[r]}{V^{[r]}}^{\dagger}.\label{eq:ALAR_SVD}
\end{align}
Notice that close to (or at) an exact solution $A_C^s = A_L^s C = C A_R^s$, the singular values contained in $\Sigma^{[\ell/r]}$ are the square of the singular values of $C$, and might well fall below machine precision. Consequently, in finite precision arithmetic, corresponding singular vectors will not be accurately computed. 

An alternative that has proven to be robust and still close to optimal is given by directly using the following left and right polar decompositions
\begin{subequations}
\label{eq:isometricdecomposition}
\begin{align}
 \mathcal{A}_{C}^{[\ell]}&= U^{[\ell]}_{A_{C}}P^{[\ell]}_{A_{C}} &  C&=U^{[\ell]}_{C}P^{[\ell]}_{C}\\
 \mathcal{A}_{C}^{[r]}&= P^{[r]}_{A_{C}}U^{[r]}_{A_{C}} &  C&= P^{[r]}_{C}U^{[r]}_{C}
\end{align}
\end{subequations}
to obtain
\begin{align}
 \mathcal{A}_{L}&=U^{[\ell]}_{A_{C}} {U^{[\ell]}_{C}}^{\dagger}&
  \mathcal{A}_{R}&= {U^{[r]}_{C}}^{\dagger}U^{[r]}_{A_{C}},\label{eq:ALAR_alt}
\end{align}
where matrices $P$ are hermitian and positive. Alternative isometric decompositions might be considered in Eq.~\eqref{eq:isometricdecomposition}, though it is important that they are unique (e.g.\ QR with positive diagonal in $R$) in order to have $P^{[\ell / r]}_{A_{C}} \approx P^{[\ell/r]}_{C}$ close to convergence.

\subsection{The Algorithm: \algorithmname}
\label{sec:algo}

\begin{table*}[t]
\begin{minipage}{\linewidth}
\begin{algorithm}[H]
  \caption{variational uMPS algorithm for single-site unit cells}
  \label{alg:SS_VUMPS}
   \begin{algorithmic}[1]
   \Require Hamiltonian $H$, initial uMPS $A_{L}$, $A_{R}$, $C$, convergence threshold $\epsilon$
  \Ensure uMPS approximation $A_{L}$, $A_{R}$, $C$ of ground state of $H$, fulfilling fixed point relations \eqref{eq:FP1}, \eqref{eq:FP2} and \eqref{eq:FP3} up to precision $\epsilon$
    \Procedure{\algorithmname}{$H$,$A_{L}$,$A_{R}$,$C$,$\epsilon$} 
   \State initialize current precision $\epsilon_{\rm prec}>\epsilon$
   \label{SS_VUMPS:init}
   \While{$\epsilon_{\rm prec}>\epsilon$}
   \State (optional) Dynamically adjust bond dimension following \App{sec:bonddim} \label{SS_VUMPS:bonddim_increase}
   \State 
   \begin{minipage}[t]{0.9\linewidth}
    \flushleft
   Calculate explicit terms of effective Hamiltonians $H_{A_{C}},H_{C}\gets$\Call{HeffTerms}{$H$,$A_{L}$,$A_{R}$,$L$,$R$,$\epsilon_{\rm S}\leq\epsilon_{\rm prec}$} from Algorithm \ref{alg:Heff_NN}, \ref{alg:Heff_LR} or \ref{alg:Heff_MPO} 
   \end{minipage}
   \label{SS_VUMPS:HeffTerms}
   \State 
   \begin{minipage}[t]{0.9\linewidth}
    \flushleft
  Calculate ground state $\tilde{A}_{C}$ of effective Hamiltonian $H_{A_{C}}$ to precision $\epsilon_{\rm H}<\epsilon_{\rm prec}$ using an iterative eigensolver, calling \Call{ApplyHAC}{$A_{C}$,$H_{A_{C}}$} from Algorithm \ref{alg:Heff_NN}, \ref{alg:Heff_LR} or \ref{alg:Heff_MPO} 
   \end{minipage}
   \label{SS_VUMPS:HACGS}
   \State 
   \begin{minipage}[t]{0.9\linewidth}
    \flushleft
   Calculate ground state $\tilde{C}$ of effective Hamiltonian $H_{C}$ to precision $\epsilon_{\rm H}<\epsilon_{\rm prec}$ using an iterative eigensolver, calling \Call{ApplyHC}{$C$,$H_{C}$} from Algorithm \ref{alg:Heff_NN}, \ref{alg:Heff_LR} or \ref{alg:Heff_MPO} 
   \end{minipage}
   \label{SS_VUMPS:HCGS}
   \State Calculate new $\tilde{A}_{L}$ and $\tilde{A}_{R}$ from $\tilde{A}_{C}$ and $\tilde{C}$ using \eqref{eq:ALAR_SVD} or \eqref{eq:ALAR_alt}, depending on singular values of $\tilde{C}$\label{SS_VUMPS:new_ALAR}
   \State Evaluate new $\epsilon_{L}$ and $\epsilon_{R}$ from \eqref{eq:err} \label{SS_VUMPS:epsilon}
   \State (optional) Calculate current expectation values \label{SS_VUMPS:measure}
   \State Set $\epsilon_{\rm prec}\gets\max(\epsilon_{L},\epsilon_{R})$ and replace $A_{L}\gets\tilde{A}_{L}$, $A_{R}\gets\tilde{A}_{R}$ and $C\gets\tilde{C}$ \label{SS_VUMPS:nextstep}
   \EndWhile
   \State \Return $A_{L}$, $A_{R}$, $C$
  \EndProcedure
   \end{algorithmic}
\end{algorithm}
\end{minipage}
\caption{Pseudocode of the \algorithmname\ algorithm described in \Sec{sec:algo}. Terms within step \ref{SS_VUMPS:HeffTerms} involving the evaluation of infinite geometric sums usually require the left dominant eigenvector $L$ of $T_{R}$ and the right dominant eigenvector $R$ of $T_{L}$, for which $L=C^{\dagger}C$ and $R=CC^{\dagger}$ with current $C$ are a good enough approximation to current precision $\epsilon_{\rm prec}$ (see main text). Notice that this algorithm is free of any possibly ill-conditioned inverses and therefore has no convergence issues in the presence of small Schmidt values. It also does not require expensive reorthogonalizations of the state at intermediate iterations.}
\label{tab:SS_VUMPS}
\end{table*}

We are now ready to formulate our \textit{variational uniform MPS} (\algorithmname) algorithm. As shown in \App{sec:theory}, a variational minimum (vanishing energy gradient) in the manifold of uMPS is characterized by tensors $A_L$, $C$ and $A_R$ satisfying the conditions
\begin{subequations}
\label{eq:FP}
\begin{align} 
 \bm{H}_{A_{C}}\,\bm{A}_{C}&=E_{A_{C}}\, \bm{A}_{C}\label{eq:FP1}\\
  \bm{H}_{C}\,\bm{C}&=E_{C}\, \bm{C}\label{eq:FP2}\\
A_{C}^{s}&=A_{L}^{s}C=CA_{R}^{s}\label{eq:FP3}.
\end{align}
\end{subequations}
Here, bold symbols denote vectorizations of the MPS tensors and matricizations of the effective Hamiltonians, and $E_{A_{C}}$ and $E_{C}$ are the lowest eigenvalues of the effective Hamiltonians.\footnote{These values can be different and depend on the subtraction scheme for the divergent energy expectation value. If $h \to \tilde{h}$ is performed everywhere, we have $E_{A_C}=E_{C}=0$. If, in the case of nearest neighbor interactions, we only substitute $h \to \tilde{h}$ in the construction of $H_L$ and $H_R$, but not in the local terms, we will have $E_{A_C}=2E_{C}=2e$.}

When iterating the steps outlined in the previous subsections, convergence is obtained when these conditions are satisfied. In particular, starting with a properly orthogonalized initial trial state $\ket{\Psi(A)}$ of some bond dimension $D$, we begin by solving the two eigenvalue problems for the effective Hamiltonians $H_{A_{C}}$ and $H_{C}$. Since we are still far from the fixed point, the resulting lowest energy states $\tilde{A}_{C}$ and $\tilde{C}$ will in general not satisfy the gauge condition \eqref{eq:FP3} together with current $A_{L/R}$. 

Following the procedure of the previous section we can however find optimal approximations $\tilde{A}^{s}_{L}$ and $\tilde{A}^{s}_{R}$ for \eqref{eq:FP3} to arrive at an updated uMPS. Conversely, $\tilde{A}_{C}$ and $\tilde{C}$  will not be the correct lowest energy eigenstates of the new effective Hamiltonians $H_{\tilde{A}_{C}}$ and $H_{\tilde{C}}$ generated from $\tilde{A}_{L/R}$. We then use the updated state and reiterate this process of alternately solving the effective eigenvalue problems, and finding optimal approximations for $A_{L}$ and $A_{R}$ to update the state. For a pseudocode summary of this algorithm, see \Tab{tab:SS_VUMPS}. 

We now elaborate on the various steps in the \algorithmname\ algorithm. Firstly, extracting new $\tilde{A}_{L/R}$ from updated $\tilde{A}_C$ and $\tilde{C}$ can be done using the theoretically optimal (but numerically often inaccurate) Eq.~\eqref{eq:ALAR_SVD} or the more robust Eq.~\eqref{eq:ALAR_alt}, depending on the magnitude of the smallest singular value in $\tilde{C}$. As a good uMPS approximation will always involve small singular values, Eq.~\eqref{eq:ALAR_alt} is preferable most of the time, except maybe during the first few iterations. 

The maximum of the error quantities \eqref{eq:err}
\begin{equation}
\epsilon_{\rm prec} =\max(\epsilon_{L},\epsilon_{R})
\label{eq:prec}
\end{equation} 
provides an error measure for the fixed point condition in Eq.~\eqref{eq:FP3} and is used as a global convergence criterion. It measures the precision of the current uMPS ground state approximation. Within every iteration, we use iterative methods (e.g.\ some variation of Lanczos) to find the eigenvectors $\tilde{A}_C$ and $\tilde{C}$ of the Hermitian operators $H_{A_C}$ and $H_{C}$. As the goal is to drive the state towards the fixed point relations in Eqs.~\eqref{eq:FP1} and \eqref{eq:FP2}, it is not necessary to solve these eigenvalue problems to full machine precision. Rather, it is sufficient to use a tolerance $\epsilon_{H}$ chosen relative to $\epsilon_{\rm prec}$.
\footnote{Further approximations to comparable accuracy can be made within the construction of the effective Hamiltonians, e.g.\ when determining $H_L$ and $H_R$ to precision $\epsilon_{\rm S}$. There, the approximations $\tilde{R}=CC^{\dagger}$ and $\tilde{L}=C^{\dagger}C$ for the true $L$ and $R$ needed for some of these operations are good enough, if $\epsilon_{\rm S}$ is roughly of the same order of magnitude as $\epsilon_{\rm prec}$.}
A value of $\epsilon_{H}$ of the order of $\epsilon_{\rm prec}/100$ has proven to work well in practice. It is also worthwhile to use tensors from the previous iteration as initial guess for the iterative solvers to speed up convergence.

As the main part of the algorithm works at fixed bond dimension (i.e.\ it is a single-site scheme in DMRG terminology), one might choose to increase the bond dimension $D$ before starting a new iteration. We have developed a subspace expansion technique that works directly in the thermodynamic limit and is explained in \App{sec:bonddim}. 

While the true comparison of this algorithm with IDMRG \cite{DMRG1,IDMRG} and ITEBD \cite{ITEBD} will take place in \Sec{sec:numerics} by gathering actual numerical simulation results, we can already compare the theoretical properties of these algorithms. Neither IDMRG or ITEBD is truly solving the variational problem in the sense of directly trying to satisfy the fixed point conditions Eqs.~\eqref{eq:FP}. IDMRG closely resembles regular DMRG on a successively growing lattice, as it inserts and optimizes over new tensors in the center of the lattice in each step. Tensors from previous steps are not updated, as this would render the cost prohibitive. When this approach converges, the resulting fixed point tensors in the center can be assumed to specify the unit cell of an infinite MPS. \algorithmname\ has the immediate advantage that i) it directly works in the thermodynamic limit at all iterations and ii) it completely replaces the entire state after every iteration, thus moving faster through the variational manifold. In contrast, IDMRG keeps memory of earlier iterations and cannot guarantee a monotonically decreasing energy that converges to an optimum associated with a translation invariant MPS in which the effects of the boundary have completely disappeared. The advantages of VUMPS come with a greater computational cost per iteration, as two eigenvalue problems (for $A_C$ and for $C$) and -- in the case of nearest neighbor interactions -- two linear systems (for $H_L$ and $H_R$) have to be solved. IDMRG only solves a single eigenvalue problem and builds $H_L$ and $H_R$ step by step in every iteration. The latter approach is analogous to a power method for eigenvalue problems and, while very cheap, is expected to require many iteration steps to converge, especially for systems with large correlation lengths (e.g.\ close to criticality). 

ITEBD \cite{ITEBD} is based on evolving an initial state in imaginary time by using a Trotter decomposition of the evolution operator. Like \algorithmname, ITEBD works in the thermodynamic limit at any intermediate step, typically with a unit cell that depends on how the Hamiltonian was split into local terms in order to apply the Trotter decomposition. Furthermore, as every application of the evolution operator increases the virtual dimension of the MPS, truncation steps are required to restore the original (or any suitable) value of the bond dimension. While \algorithmname\ can take big steps through the variational space,  time steps in ITEBD have to be chosen sufficiently small (especially in the final steps of the algorithm) to eliminate the Trotter error, which negatively affects the rate of convergence (Ref.~\onlinecite{ITEBD_Phien} however proposes a scheme to effectively obtain a larger time step). Furthermore, the Trotter splitting essentially limits the applicability of ITEBD to short-range interactions and dictates the size of the unit cell of the resulting MPS, e.g.\ in the most common case of nearest neighbor interactions a two-site unit cell is obtained. (The approach of Ref.~\onlinecite{MPO3} to obtain a translation invariant MPS is restricted to certain Hamiltonians, but see Ref.~\onlinecite{LRITEBD} for an alternative proposal that can in fact also deal with long range interactions.)

Finally, we can also compare \algorithmname\ to the more recent \textit{time dependent variational principle} (TDVP),\cite{TDVP} which was implemented as an alternative approach to simulate real and imaginary time evolution within the manifold of MPS by projecting the evolution direction onto the MPS tangent space. This approach can be applied to translation invariant MPS, independent of the type of Hamiltonian. When used to evolve in imaginary time, it can be identified as a covariant formulation of a gradient descent method, in that it evolves the state in the direction of the gradient of the energy functional, preconditioned with the metric of the manifold. As such, the energy decreases monotonically and at convergence, an exact (local) minimum is obtained, as characterized by the vanishing gradient. However, in its original formulation, TDVP was not formulated in a center site form and was therefore unstable and restricted to small time steps. For finite systems, a different formulation of the TDVP algorithm was provided in Ref.~\onlinecite{TDVP_Uni}, which allows for taking the limit of the imaginary time step to infinite, and then becomes provably equivalent to the single-site DMRG algorithm. \algorithmname\ can be motivated from these developments, as explained in \App{sec:theory}.

We conclude this section by elaborating on how to incorporate symmetries in the algorithm. The construction of uMPS that is explicitly invariant under onsite unitary symmetries is equivalent to (I)DMRG \cite{DMRG_MC,MPS6_S} and (I)TEBD \cite{ITEBD_Hastings,Sukhi_symmetries}, and it is immediately clear that the various steps in \algorithmname\ have a corresponding covariant formulation. The same comments apply to time reversal symmetry, in which case everything can be implemented in real arithmetic, or to reflection symmetry, in which case $C$ and $A_C^s$  will be symmetric matrices and $A_R^s = {A_L^s}^T$ (which implies that $H_L$ and $H_R$ are also related). In all of these cases, the computational cost is reduced. However, explicitly imposing the symmetry in the MPS requires caution, as the physical system might have spontaneous symmetry breaking, or -- more subtly -- might be in a symmetry protected topological phase where the symmetries cannot be represented trivially on the MPS tensor. 

In the case of spontaneous symmetry breaking, MPS algorithms tend to converge to maximally symmetry broken states for which the entanglement is minimal. This is also the case for \algorithmname. One can control which state the algorithm converges to by suitably biasing the initial state or by adding small  perturbation terms to the Hamiltonian which explicitly break the symmetry, and which are switched off after a few iterations. 

Explicit conservation of translation symmetry was the very first requirement in the construction of \algorithmname. In the case of  spontaneous breaking of translation symmetry down to $N$-site translation symmetry (as e.g.\ in the case of a state with antiferromagnetic order), enforcing one-site translation symmetry would result in a (non-injective) equal weight superposition of all symmetry broken uMPS ground state approximations. In order to reach an optimal accuracy with a given bond dimension, such a superposition of $N$ states is however undesirable, as the effective bond dimension is reduced to $D/N$. In the case where this situation cannot be amended by a simple unitary transformation that restores one-site translation symmetry (such as e.g.\ flipping every second spin in the case of an antiferromagnet), it is preferable to choose an MPS ansatz with a $N$-site unit cell, such that the state can spontaneously break translation symmetry. The generalization of the algorithm to multi-site unit cells is described in the next section.

\subsection{Multi Site Unit Cell Implementations}
\label{sec:MS_VUMPS}

We now generalize the \algorithmname\ algorithm of the previous section for one-site translation invariant uMPS to the setting of translation invariance over $N$ sites. Such a uMPS ansatz is then parameterized by $N$ independent tensors ${A{(k)}}^{s}\in\mathbb{C}^{D\times d\times D}$, $k=1,\ldots,N$, which define the unit cell tensor
\begin{equation}
 \mathbb{A}^{\mathbbm{s}_{n}}=A(1)^{s_{nN+1}}\ldots A(N)^{s_{nN+N}},
\end{equation} 
where $\mathbbm{s}=(s_{1},\ldots,s_{N})$ is a combined index. We can then write the variational state as
\begin{equation*}
\ket{\Psi(\mathbb{A})}=\sum_{\bm{s}} (\ldots\mathbb{A}^{\mathbbm{s}_{n-1}}\mathbb{A}^{\mathbbm{s}_{n}}\mathbb{A}^{\mathbbm{s}_{n+1}}\ldots )\ket{\bm{s}}
\end{equation*} 
and the left and right orthonormal forms are given by the relations
\begin{subequations}
\label{eq:gauges_MSUC}
\begin{equation}
\begin{split}
 \sum_{s}{A(k)_{L}^{s}}^{\dagger}A(k)_{L}^{s}&=\unity \\  
 \sum_{s}A(k)_{L}^{s}\,R(k)\, {A(k)_{L}^{s}}^{\dagger}&=R(k-1)
\end{split}
\label{eq:leftgauge_MSUC}
\end{equation} 
and
\begin{equation}
\begin{split}
 \sum_{s}A(k)_{R}^{s}\,{A(k)_{R}^{s}}^{\dagger}&=\unity \\ 
 \sum_{s}{A(k)_{R}^{s}}^{\dagger}\,L(k-1)\, A(k)_{R}^{s}&=L(k),
\end{split}
\label{eq:rightgauge_MSUC}
\end{equation} 
\end{subequations}
where it is understood that $N+1\equiv1$ and $0\equiv N$.

Defining the bond matrices $C(k)$ as the gauge transformation that relates left and right canonical form via $C(k-1)A(k)^{s}_{R}=A(k)_L^s C(k)$, we have $R(k)=C(k)C(k)^{\dagger}$ and $L(k)=C(k)^{\dagger}C(k)$. We can then also cast $\ket{\Psi(\mathbb{A})}$ in a mixed canonical form similar to \eqref{eq:psiAC} with the center site tensor given by $A(k)_{C}^{s}=A(k)^{s}_{L}C(k)=C(k-1)A(k)^{s}_{R}$.

The variational minimum within this set of states is characterized by the following $3N$ fixed point relations
\begin{subequations}
\label{eq:FP_MSUC}
\begin{align} 
 \bm{H}_{A(k)_{C}}\,\bm{A}(k)_{C}&=E_{A(k)_{C}}\, \bm{A}(k)_{C}\label{eq:FP1_MSUC}\\
  \bm{H}_{C(k)}\,\bm{C}(k)&=E_{C(k)}\, \bm{C}(k)\label{eq:FP2_MSUC}\\
A(k)_{C}^{s}&= A(k)_{L}^{s}C(k)=C(k-1)A(k)_{R}^{s} \label{eq:FP3_MSUC}.
\end{align}
\end{subequations}
Notice that due to \eqref{eq:FP3_MSUC}, the relations for different $k$ are connected. There are several possible strategies for constructing algorithms which obtain states satisfying these conditions. 

In the following we present two approaches which have shown good performance and stable convergence, which we shall term the ``sequential'' and ``parallel'' methods. But let us first elaborate on computing effective Hamiltonians for multi-site unit cells, which works similarly in both methods. We again restrict to the case of nearest neighbor interactions, such that the effective Hamiltonians are constructed similar as in \Sec{sec:effH_NN}. To construct e.g.\ the left block Hamiltonian $H_L$, we first collect all local contributions from a single unit cell in $h_L$, before performing the geometric series of the transfer matrix, which now mediates a translation over an entire unit cell.

\subsubsection{Sequential Algorithm}
The sequential algorithm is inspired by finite size DMRG, in that we sweep through the unit cell, successively optimizing one tensor at a time while keeping tensors on other sites fixed. Notice that at site $k$ we however need \textit{two} updated bond matrices $\tilde{C}(k)_{L}=\tilde{C}(k-1)$ and $\tilde{C}(k)_{R}=\tilde{C}(k)$, in order to calculate updated $\tilde{A}(k)^{s}_{L/R}$ from $\tilde{A}(k)^{s}_{C}\approx \tilde{A}(k)^{s}_{L}\,\tilde{C}(k)_{R}\approx \tilde{C}(k)_{L}\, \tilde{A}(k)^{s}_{R}$. We thus have to amend steps \ref{SS_VUMPS:HeffTerms}, \ref{SS_VUMPS:HACGS} and \ref{SS_VUMPS:HCGS} of the single-site algorithm in Table \ref{tab:SS_VUMPS} by constructing and solving for \textit{two} effective Hamiltonians $H_{C(k-1)}$ and $H_{C(k)}$ instead of a single one.  
The newly optimized tensors then get replaced in \textit{all} unit cells of the infinite lattice, and contributions to the effective Hamiltonians have to be recalculated accordingly, before moving on to the next site. For a pseudocode summary see Algorithm~\ref{alg:MS_VUMPS_seq} in \Tab{tab:MS_VUMPS}.

One could now try to argue, that e.g.\ in a left to right sweep it is enough at site $k$ to calculate updated $\tilde{A}(k)_{C}$ and $\tilde{C}(k)_{R}=\tilde{C}(k)$ only, and to use $\tilde{C}(k-1)_{R}$ from the previous step at site $k-1$ as $\tilde{C}(k)_{L}$ for calculating $\tilde{A}(k)_{R}$. This approach however fails, as the effective Hamiltonian used for calculating $\tilde{A}(k)_{C}$ already contains updated $\tilde{A}(k-1)_{L/R}$, while the effective Hamiltonian used for calculating $\tilde{C}(k-1)_{R}$ does not, and we cannot determine $\tilde{A}(k)_{R}$ from $\tilde{A}(k)_{C}$ and $\tilde{C}(k-1)_{R}$. Rather, $\tilde{C}(k)_{L}$ has to be recalculated using an \textit{updated} effective Hamiltonian, which exactly leads to the sequential Algorithm \ref{alg:MS_VUMPS_seq}.

There is an additional subtlety that needs to be considered, in order for all tensors to fulfill the gauge constraints \eqref{eq:FP3_MSUC} to current precision. Bond matrices $\tilde{C}(k)$ are calculated as lowest energy eigenvectors of effective Hamiltonians $H_{C(k)}$ and are therefore only determined up to a phase. Consider $C(k)$ defined between sites $k$ and $k+1$. At step $k$ it is updated as $\tilde{C}(k)_{R}$ and used to calculate $\tilde{A}(k)^{s}_{L}$. In the next step $k+1$ however it is recalculated as $\tilde{C}(k+1)_{L}$ (with an updated effective Hamiltonian) and used to determine $\tilde{A}(k+1)^{s}_{R}$. At the fixed point we should then have $\tilde{C}(k)_{R}=\tilde{C}(k+1)_{L}=C(k)$, but this is only true if there is no phase ambiguity, which would also consequently lead to a phase mismatch between $\tilde{A}(k)_{L}$ and $\tilde{C}(k)$ after step $k+1$.
This issue does not pose a problem for algorithm convergence (during calculations, matrices $C(k)$ always appear as products of the form $C(k)^{\dagger}C(k)$ or $C(k)C(k)^{\dagger}$ and mismatching phases thus cancel out), but can be easily circumvented by employing a phase convention when calculating updated $\tilde{C}(k)$.

\subsubsection{Parallel Algorithm}
In the parallel approach, we choose to update an entire unit cell at once, using effective Hamiltonians generated from the same current state. To that end, we first generate all terms necessary for all $H_{A(k)_{C}}$ and $H_{C(k)}$. For the case of nearest neighbor interactions, the contributions $H_{L}$ and $H_{R}$ to the left and right environment outside the unit cell can be shared, so that the corresponding geometric sum only needs to be computed once, and contributions inside the unit cell are obtained through successive applications of transfer matrices. 

Next, we simultaneously and independently solve for the ground states $\tilde{A}(k)_{C}$ and $\tilde{C}(k)$ of all $2N$ effective Hamiltonians at once. Once these are obtained we again simultaneously and independently determine all updated $\tilde{A}(k)_{L}$ and $\tilde{A}(k)_{R}$, concluding one iteration for updating the entire unit cell. For a pseudocode summary see Algorithm~\ref{alg:MS_VUMPS_par} in \Tab{tab:MS_VUMPS}.

\begin{table*}[p]
\def\MSseqboxwidth{0.85}
\def\MSparboxwidth{0.85}

\begin{minipage}{\linewidth}
\begin{algorithm}[H]
  \caption{\textit{sequential} variational uMPS algorithm for multi-site unit cells}
  \label{alg:MS_VUMPS_seq}
   \begin{algorithmic}[1]
   \Require Hamiltonian $H$, initial uMPS $\{A_{L}\},\{A_{R}\},\{C\}$ of an $N$-site unit cell, convergence threshold $\epsilon$
  \Ensure uMPS approximation $\{A_{L}\},\{A_{R}\},\{C\}$ of ground state of $H$, fulfilling fixed point relations \eqref{eq:FP1_MSUC}, \eqref{eq:FP2_MSUC} and \eqref{eq:FP3_MSUC} up to precision $\epsilon$.
    \Procedure{\algorithmname MultiSequential}{$H$,$\{A_{L}\},\{A_{R}\},\{C\}$,$\epsilon$} 
   \State initialize current precision $\epsilon_{\rm prec}>\epsilon$
   \label{MS_VUMPS_seq:init}
   \While{$\epsilon_{\rm prec}>\epsilon$}
   \For{$n=1,\ldots,N}$
   \State (optional) Dynamically adjust bond dimension following \App{sec:bonddim} \label{MS_VUMPS_seq:bonddim_increase}
   \State 
   \begin{minipage}[t]{\MSseqboxwidth\linewidth}
   \flushleft
   Calculate explicit terms of effective Hamiltonians from a multi-site version $H_{A(n)_{C}},H_{C(n-1)},H_{C(n)}\gets$\Call{HeffTermsMulti}{$H$,$\{A_{L}\}$,$\{A_{R}\}$,$\{L\}$,$\{R\}$,$\epsilon_{\rm S}\leq\epsilon_{\rm prec}$} of Algorithm \ref{alg:Heff_NN}, \ref{alg:Heff_LR} or \ref{alg:Heff_MPO}
   \end{minipage}
   \State
   \begin{minipage}[t]{\MSseqboxwidth\linewidth}
    \flushleft
   Calculate ground state $\tilde{A}_{C}$ of effective Hamiltonian $H_{A(n)_{C}}$ to precision $\epsilon_{\rm H}<\epsilon_{\rm prec}$ using an iterative eigensolver, calling \Call{ApplyHAC}{$C$,$H_{A(n)_{C}}$} from Algorithm \ref{alg:Heff_NN}, \ref{alg:Heff_LR} or \ref{alg:Heff_MPO} 
   \end{minipage}
   \label{MS_VUMPS_seq:HACGS}
   \State
   \begin{minipage}[t]{\MSseqboxwidth\linewidth}
    \flushleft
   Calculate ground state $\tilde{C}_{L}$ of effective Hamiltonian $H_{C(n-1)}$ to precision $\epsilon_{\rm H}<\epsilon_{\rm prec}$ using an iterative eigensolver, calling \Call{ApplyHC}{$C$,$H_{C(n-1)}$} from Algorithm \ref{alg:Heff_NN}, \ref{alg:Heff_LR} or \ref{alg:Heff_MPO} 
   \qquad \Comment{To ensure gauge consistency, employ a phase convention for $\tilde{C}_{L}$}
   \end{minipage}
   \label{MS_VUMPS_seq:HCLGS}
   \State
   \begin{minipage}[t]{\MSseqboxwidth\linewidth}
    \flushleft
   Calculate ground state $\tilde{C}_{R}$ of effective Hamiltonian $H_{C(n)}$ to precision $\epsilon_{\rm H}<\epsilon_{\rm prec}$ using an iterative eigensolver, calling \Call{ApplyHC}{$C$,$H_{C(n)}$} from Algorithm \ref{alg:Heff_NN}, \ref{alg:Heff_LR} or \ref{alg:Heff_MPO} 
   \qquad \Comment{To ensure gauge consistency, employ a phase convention for $\tilde{C}_{R}$}
   \end{minipage}
   \label{MS_VUMPS_seq:HCRGS}
   \State Calculate new $\tilde{A}_{L}$ from $\tilde{A}_{C}$ and $\tilde{C}_{R}$ using \eqref{eq:ALAR_SVD} or \eqref{eq:ALAR_alt}, depending on singular values of $\tilde{C}_{R}$ \label{MS_VUMPS_seq:new_AL}
   \State Calculate new $\tilde{A}_{R}$ from $\tilde{A}_{C}$ and $\tilde{C}_{L}$ using \eqref{eq:ALAR_SVD} or \eqref{eq:ALAR_alt}, depending on singular values of$\tilde{C}_{L}$ \label{MS_VUMPS_seq:new_AR}
   \State Evaluate new $\epsilon_{L}(n)$ and $\epsilon_{R}(n)$ from \eqref{eq:err_l} and \eqref{eq:err_r} \label{MS_VUMPS_seq:epsilon}
   \State Replace $A(n)_{L}\gets\tilde{A}_{L}$, $A(n)_{R}\gets\tilde{A}_{R}$, $C(n-1)\gets\tilde{C}_{L}$ and $C(n)\gets\tilde{C}_{R}$ \label{MS_VUMPS_seq:reset}
   \EndFor
   \State Set $\epsilon_{\rm prec}\gets\max(\{\epsilon_{L}\},\{\epsilon_{R}\})$ \label{MS_VUMPS_seq:set_prec}
   \State (optional) Calculate current expectation values \label{MS_VUMPS_seq:measure}
   \EndWhile
   \State \Return $\{A_{L}\},\{A_{R}\},\{C\}$
  \EndProcedure
   \end{algorithmic}
\end{algorithm}
\end{minipage}

\begin{minipage}{\linewidth}
\begin{algorithm}[H]
  \caption{\textit{parallel} variational uMPS algorithm for multi-site unit cells}
  \label{alg:MS_VUMPS_par}
   \begin{algorithmic}[1]
   \Require Hamiltonian $H$, initial uMPS $\{A_{L}\},\{A_{R}\},\{C\}$ of an $N$-site unit cell, convergence threshold $\epsilon$
  \Ensure uMPS approximation $\{A_{L}\},\{A_{R}\},\{C\}$ of ground state of $H$, fulfilling fixed point relations \eqref{eq:FP1_MSUC}, \eqref{eq:FP2_MSUC} and \eqref{eq:FP3_MSUC} up to precision $\epsilon$.
    \Procedure{\algorithmname MultiParallel}{$H$,$\{A_{L}\},\{A_{R}\},\{C\}$,$\epsilon$} 
   \State initialize current precision $\epsilon_{\rm prec}>\epsilon$
   \label{MS_VUMPS_par:init}
   \While{$\epsilon_{\rm prec}>\epsilon$}
   \State (optional) Dynamically adjust bond dimension following \App{sec:bonddim} \label{MS_VUMPS_par:bonddim_increase}
   \For{$n=1,\ldots,N$}
   \State 
   \begin{minipage}[t]{\MSparboxwidth\linewidth}
   \flushleft
   Calculate explicit terms of effective Hamiltonians from a multi-site version $H_{A(n)_{C}},H_{C(n)}\gets$\Call{HeffTermsMulti}{$H$,$\{A_{L}\}$,$\{A_{R}\}$,$\{L\}$,$\{R\}$,$\epsilon_{\rm S}\leq\epsilon_{\rm prec}$}  of Algorithm \ref{alg:Heff_NN}, \ref{alg:Heff_LR} or \ref{alg:Heff_MPO}
   \end{minipage}
   \label{MS_VUMPS_par:HeffTerms}
   \State
    \begin{minipage}[t]{\MSparboxwidth\linewidth}
    \flushleft
   Calculate ground state $\tilde{A}(n)_{C}$ of effective Hamiltonian $H_{A(n)_{C}}$ to precision $\epsilon_{\rm H}<\epsilon_{\rm prec}$ using an iterative eigensolver, calling \Call{ApplyHAC}{$C$,$H_{A(n)_{C}}$} from Algorithm \ref{alg:Heff_NN}, \ref{alg:Heff_LR} or \ref{alg:Heff_MPO} 
   \end{minipage}
   \label{MS_VUMPS_par:HACGS}
   \State
   \begin{minipage}[t]{\MSparboxwidth\linewidth}
    \flushleft
   Calculate ground state $\tilde{C}(n)$ of effective Hamiltonian $H_{C(n)}$ to precision $\epsilon_{\rm H}<\epsilon_{\rm prec}$ using an iterative eigensolver, calling \Call{ApplyHC}{$C$,$H_{C(n-1)}$} from Algorithm \ref{alg:Heff_NN}, \ref{alg:Heff_LR} or \ref{alg:Heff_MPO} 
   \end{minipage}
   \label{MS_VUMPS_par:HCGS}
   \EndFor
   \For{$n=1,\ldots,N$}
   \State Calculate new $\tilde{A}(n)_{L}$ from $\tilde{A}(n)_{C}$ and $\tilde{C}(n)$ using \eqref{eq:ALAR_SVD} or \eqref{eq:ALAR_alt}, depending on singular values of $\tilde{C}(n)$ \label{MS_VUMPS_par:new_AL}
   \State Calculate new $\tilde{A}(n)_{R}$ from $\tilde{A}(n)_{C}$ and $\tilde{C}(n-1)$ using \eqref{eq:ALAR_SVD} or \eqref{eq:ALAR_alt}, depending on singular values of $\tilde{C}(n-1)$ \label{MS_VUMPS_par:new_AR}
   \State Evaluate new $\epsilon_{L}(n)$ and $\epsilon_{R}(n)$ from \eqref{eq:err_l} and \eqref{eq:err_r} \label{MS_VUMPS_par:epsilon}
   \EndFor
   \State Replace $\{A_{L}\}\gets\{\tilde{A}_{L}\}$, $\{A_{R}\}\gets\{\tilde{A}_{R}\}$ and $\{C\}\gets\{\tilde{C}\}$ \label{MS_VUMPS_par:reset}
   \State (optional) Calculate current expectation values \label{MS_VUMPS_par:measure}
   \State Set $\epsilon_{\rm prec}\gets\max(\{\epsilon_{L}\},\{\epsilon_{R}\})$ \label{MS_VUMPS_par:set_prec}
   \EndWhile
   \State \Return $\{A_{L}\},\{A_{R}\},\{C\}$
  \EndProcedure
   \end{algorithmic}
\end{algorithm}
\end{minipage}
\caption{Pseudocode for the two approaches for a multi-site unit cell implementation described in \Sec{sec:MS_VUMPS}.
Algorithm \ref{alg:MS_VUMPS_seq} sweeps through the unit cell and sequentially updates tensors site by site, replacing updated tensors in all unit cells before moving on to the next site. Algorithm \ref{alg:MS_VUMPS_par} updates the entire unit cell at once by independently updating tensors on each site.}
\label{tab:MS_VUMPS}
\end{table*}

\subsubsection{Juxtaposition of Both Approaches}
Several comments on the two presented algorithms are in order. First, the parallel algorithm requires substantially less computational effort, since the construction of the different effective Hamiltonians $H_{A(k)_C}$ can recycle the calculation of the infinite geometric sum. Therefore, updating an entire unit cell only requires to evaluate two infinite geometric sums and $2N$ effective eigenvalue problems. In the sequential algorithm, updating the environment after every tensor update requires to reevaluate the geometric sum, thus leading to $2N$ infinite geometric sums and $3N$ effective eigenvalue problems for updating the complete unit cell. Additionally, the parallel approach offers the possibility of parallelizing the solution of all $2N$ eigenvalue problems in one iteration, while in the sequential approach only $3$ eigenvalue problems can be solved in parallel for each site. However, while sweeping through the unit cell in the sequential approach, initial guesses for solving the infinite geometric sums can be generated easily from the previous iterations, and are usually much better than the initial guesses in the parallel algorithm. Equivalently, updated $\tilde{C}(k)$ obtained at site $k$ is a very good initial guess for its recalculation with updated environment on site $k+1$. Overall, the computational cost for the parallel update is still much cheaper, albeit less than expected. 

On the other hand, state convergence in terms of iterations is generally substantially faster in the sequential approach. This seems reasonable, as the optimization on a current site takes into account all previous optimization steps, whereas in the parallel approach, the optimizations on different sites within one iteration are independent of each other. This effect gets amplified with increasing unit cell size $N$, and the performance of the parallel approach decreases, while the performance of the sequential approach seems more stable against increasing $N$. 

In conclusion, while updating the entire unit cell is computationally cheaper in the parallel approach, the sequential algorithm usually requires a substantially smaller number of iterations due to faster convergence. While there are instances where one approach clearly outperforms the other by far, such cases are rare and strongly depend on initial conditions, and generally both approaches show comparable performance. For comparison benchmark results see \Sec{sec:bench_multi}.

\section{Test Cases and Comparison}
\label{sec:numerics}
In this section we test the performance of the new algorithm on several paradigmatic strongly correlated lattice models in the thermodynamic limit, with nearest neighbor as well as long range interactions. In \Sec{sec:models} we introduce and discuss the models under considerations. In \Sec{sec:bench1} we first test the convergence and stability of the single and multi-site implementations of the new algorithm. Lastly, we compare its performance against established conventional MPS methods for ground state search in \Sec{sec:bench2}.

\subsection{Models}
\label{sec:models}
As examples for spin chain models with nearest neighbor interactions we study the spin $S=1/2$ \textit{transverse field Ising} (TFI) model
\begin{equation}
 H_{\rm TFI}=-\sum_{j}X_{j}X_{j+1}-h\sum_{j}Z_{j}
 \label{eq:TFI_Ham}
\end{equation} 
and the \textit{XXZ} model for general spin $S$ 
\begin{equation}
 H_{\rm XXZ}=\sum_{j}X_{j}X_{j+1} + Y_{j}Y_{j+1} + \Delta Z_{j}Z_{j+1}.
 \label{eq:XXZ_Ham}
\end{equation} 
Here $X$, $Y$ and $Z$ are spin $S$ representations of the generators of $SU(2)$. The ground state energies are known exactly for the TFI model,\cite{Pfeuty} and for $S=1/2$ also for the XXZ model.\cite{Takahashi} For the $S=1$ XXZ model we focus on the isotropic antiferromagnetic case $\Delta=1$ and take the result of Ref.~\onlinecite{TDVP} for the ground state energy for $D=1024$ as quasi-exact result.

As a further example for a system with nearest neighbor interactions we also study the \textit{Fermi Hubbard} model
\begin{equation}
\begin{split}
 H_{\rm HUB}=&-t\sum_{\sigma,j}c_{\sigma,j}c^{\dagger}_{\sigma,j+1} - c^{\dagger}_{\sigma,j}c_{\sigma,j+1}\\
 &+ U\sum_{j}\left( n_{\uparrow,j}-\frac{1}{2} \right)\left( n_{\downarrow,j}-\frac{1}{2} \right),
 \end{split}
 \label{eq:HUB_Ham}
\end{equation} 
where $c_{\sigma,j}$, $c^{\dagger}_{\sigma,j}$ are creation and annihilation operators of electrons of spin $\sigma$ on site $j$, $n_{\sigma,j}=c^{\dagger}_{\sigma,j}c_{\sigma,j}$ and $n_{j}=n_{\uparrow,j} + n_{\downarrow,j}$ are the particle number operators. Again, the exact ground state energy is known.\cite{Lieb_Wu68,Essler}

As an example for an exactly solvable model with (algebraically decaying) long range interactions we consider the Haldane-Shastry model\cite{Haldane88,Shastry88}
\begin{equation}
 H_{\rm HS}=\sum_{j}\sum_{n>0}n^{-2}[X_{j}X_{j+n} + Y_{j}Y_{j+n} + Z_{j}Z_{j+n}],
 \label{eq:HS_Ham}
\end{equation}
where $X$, $Y$ and $Z$ are again spin $S=1/2$ representations of the generators of $SU(2)$. In order to efficiently compute the terms of the effective Hamiltonian (see \App{sec:effH_LR}), we expand the distance function $f(n)=n^{-2}$ in a sum of $K=20$ exponentials, with maximum residual less than $10^{-6}$ for a fit over $N=1000$ sites.

Finally, as a state of the art problem of current interest, we also consider the two-dimensional antiferromagnetic $S=1/2$ Heisenberg model on a cylinder of infinite length, but finite circumference $W$
\begin{equation}
\begin{split}
 H^{\rm cyl}_{\rm XXZ}=&\sum_{i,j} X_{[i,j]}(X_{[i,j+1]} + X_{[i+1,j]}) \\
  &+ Y_{[i,j]}(Y_{[i,j+1]} + Y_{[i+1,j]})\\
  &+ Z_{[i,j]}(Z_{[i,j+1]} + Z_{[i+1,j]}),
  \end{split}
 \label{eq:XXZ_Ham_cyl}
\end{equation} 
where $[i,j]$ denotes the location on the cylinder with $i\in\mathbb{Z}$ and $1\leq j\leq W$ (assuming periodic boundary conditions, i.e. $W+1\equiv 1$). Following popular procedure (as e.g. also done in Ref.~\onlinecite{Osorio17}), we map the model onto a one-dimensional chain following a sawtooth path along the cylinder, such that the longest range spin-spin interactions are over $W$ sites. We take the energies obtained from Loop QMC for several cylinder circumferences $W$ presented in Table I in Ref.~\onlinecite{Osorio17} as quasi-exact ground state energies.

\subsection{Performance Benchmarks}
\label{sec:bench1}

\begin{figure*}[t]
 \centering
 \includegraphics[width=0.48\linewidth,keepaspectratio=true]{\figpath/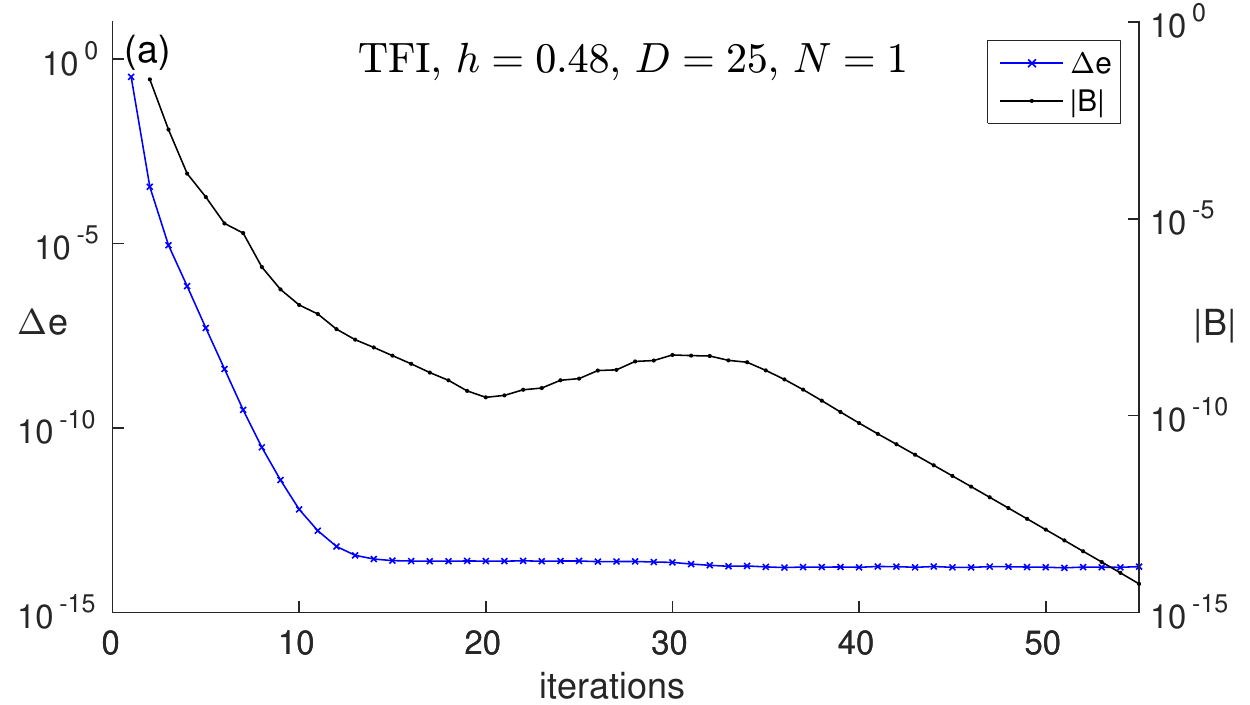}
 \includegraphics[width=0.48\linewidth,keepaspectratio=true]{\figpath/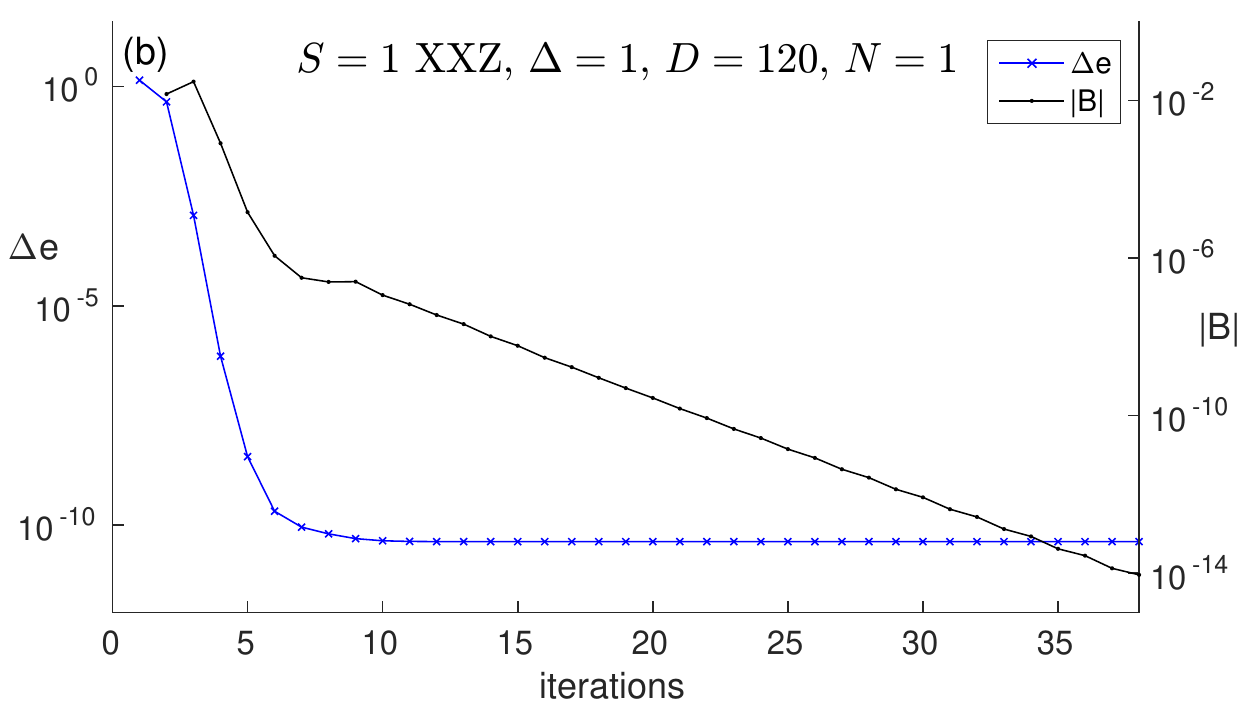}
 \includegraphics[width=0.48\linewidth,keepaspectratio=true]{\figpath/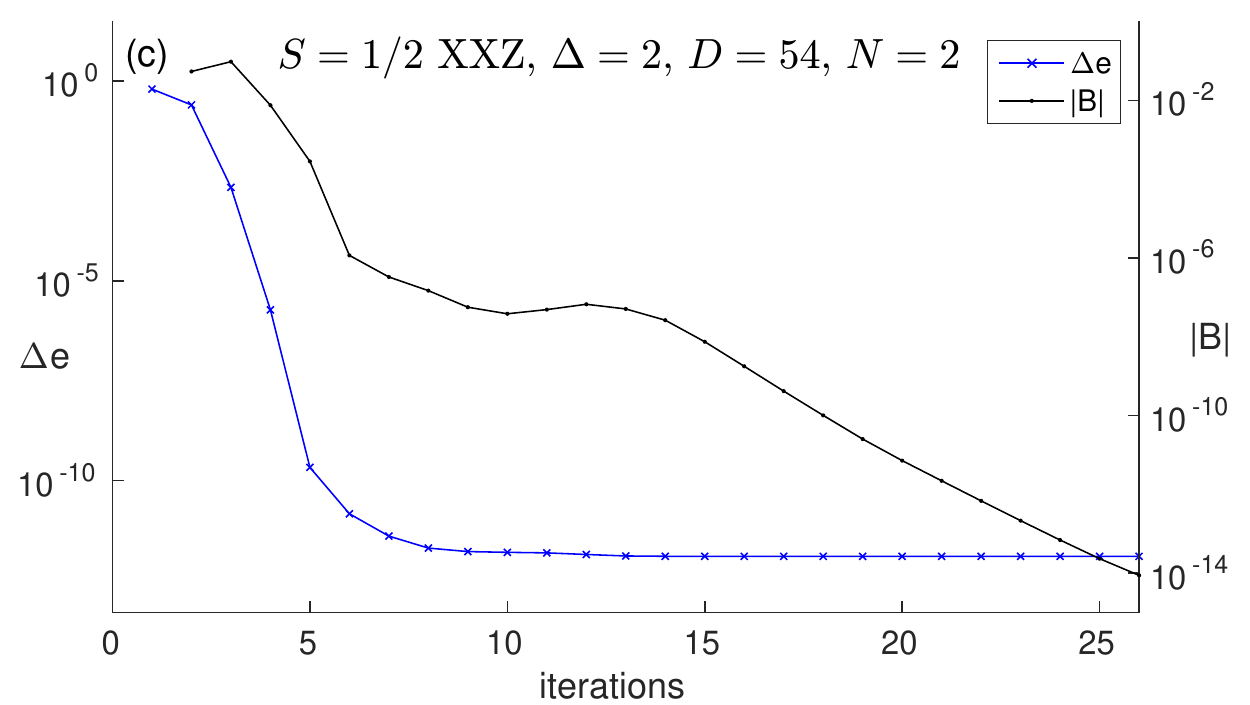}
 \includegraphics[width=0.48\linewidth,keepaspectratio=true]{\figpath/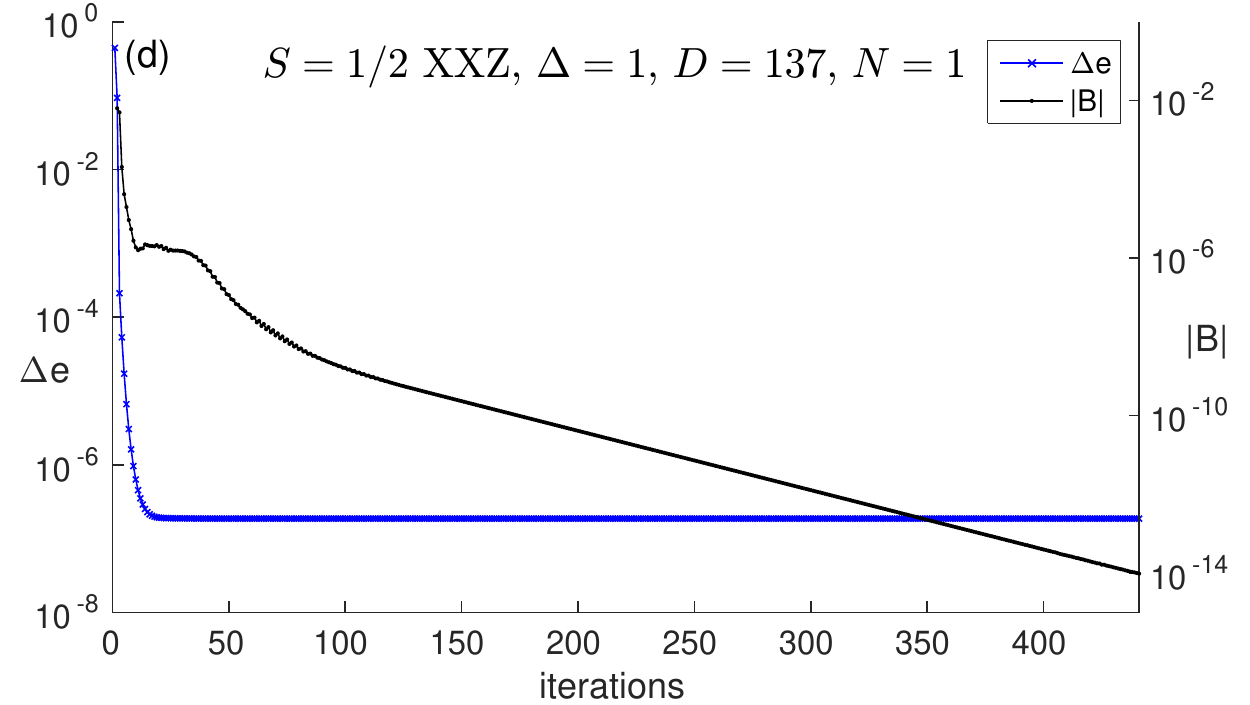}
 \includegraphics[width=0.48\linewidth,keepaspectratio=true]{\figpath/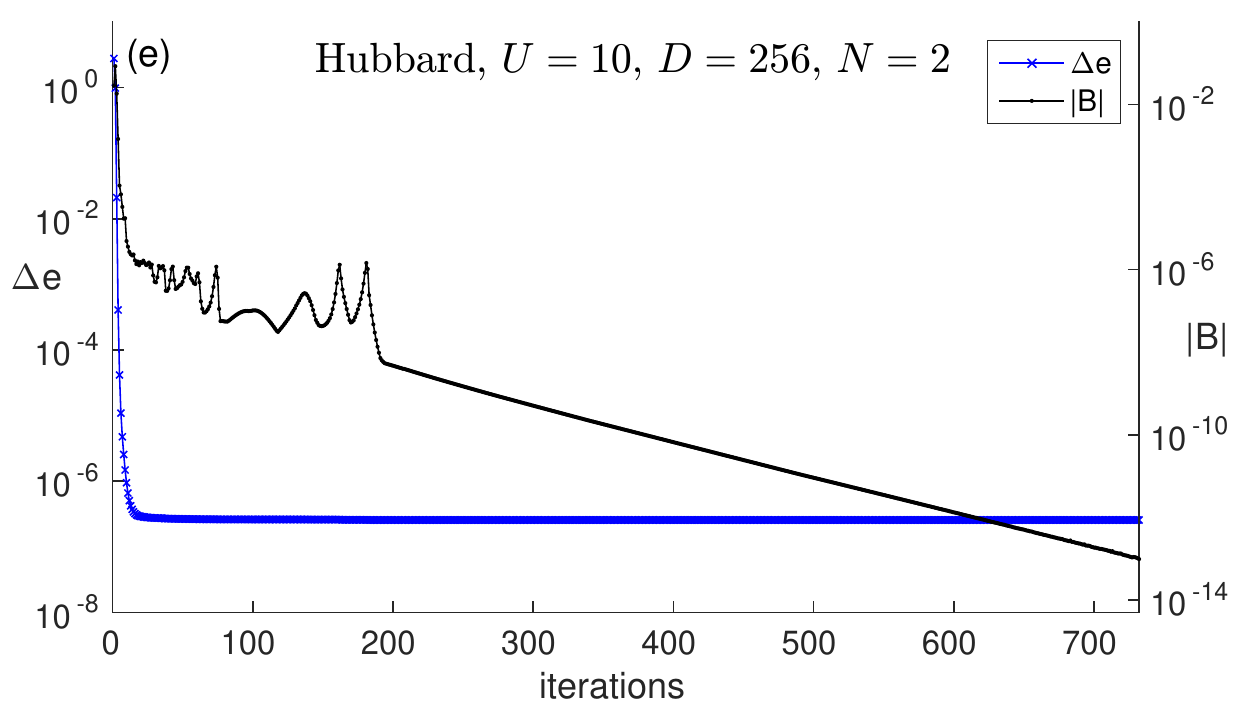}
 \includegraphics[width=0.48\linewidth,keepaspectratio=true]{\figpath/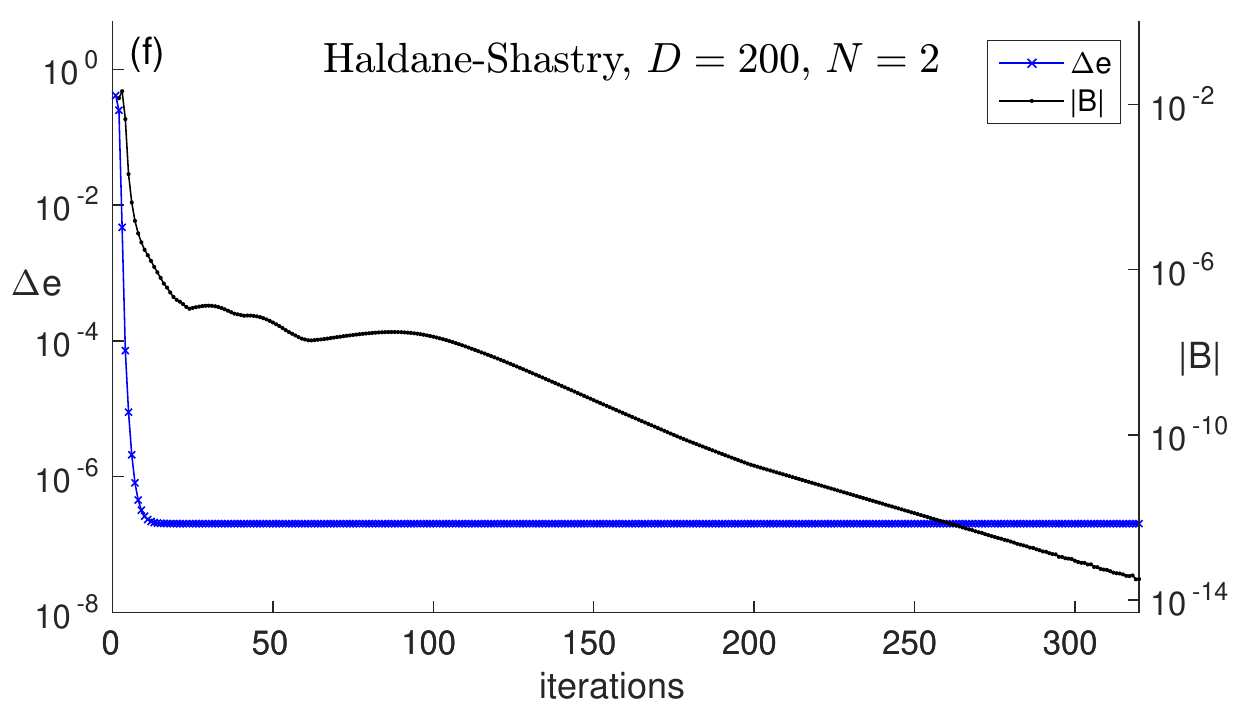}
 \includegraphics[width=0.48\linewidth,keepaspectratio=true]{\figpath/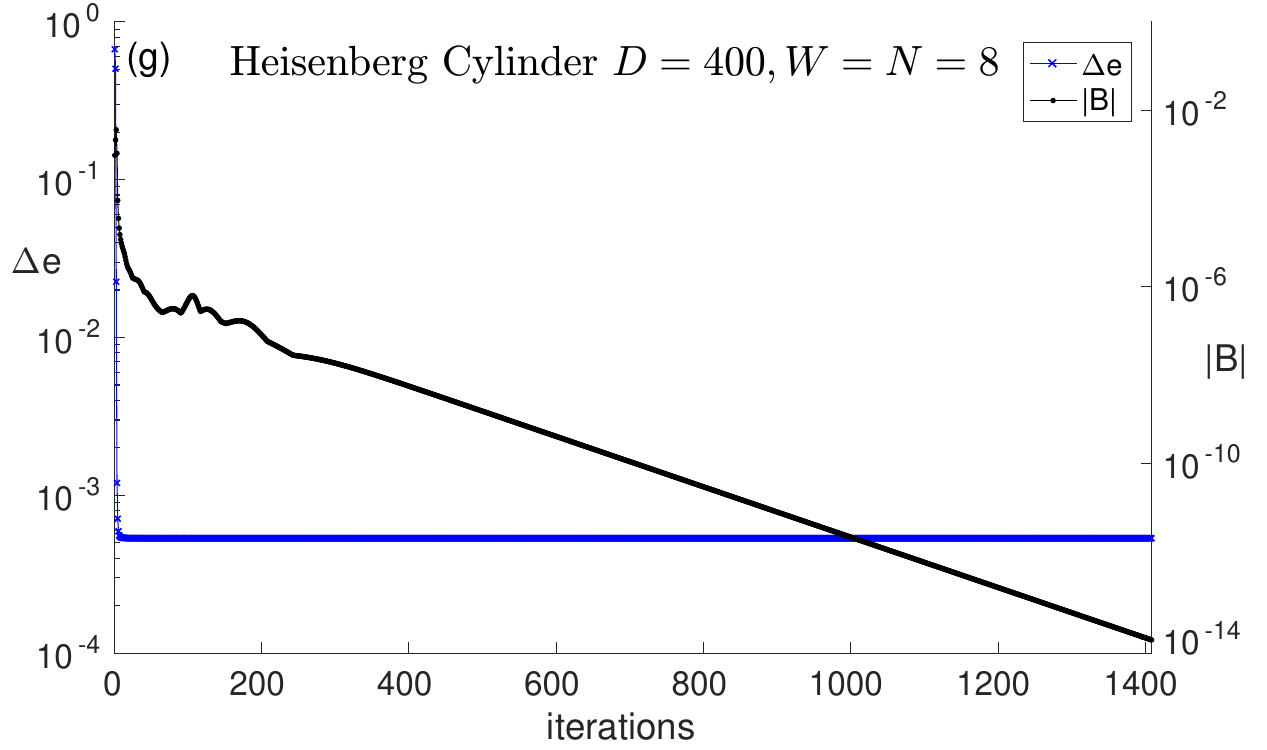}
 \includegraphics[width=0.48\linewidth,keepaspectratio=true]{\figpath/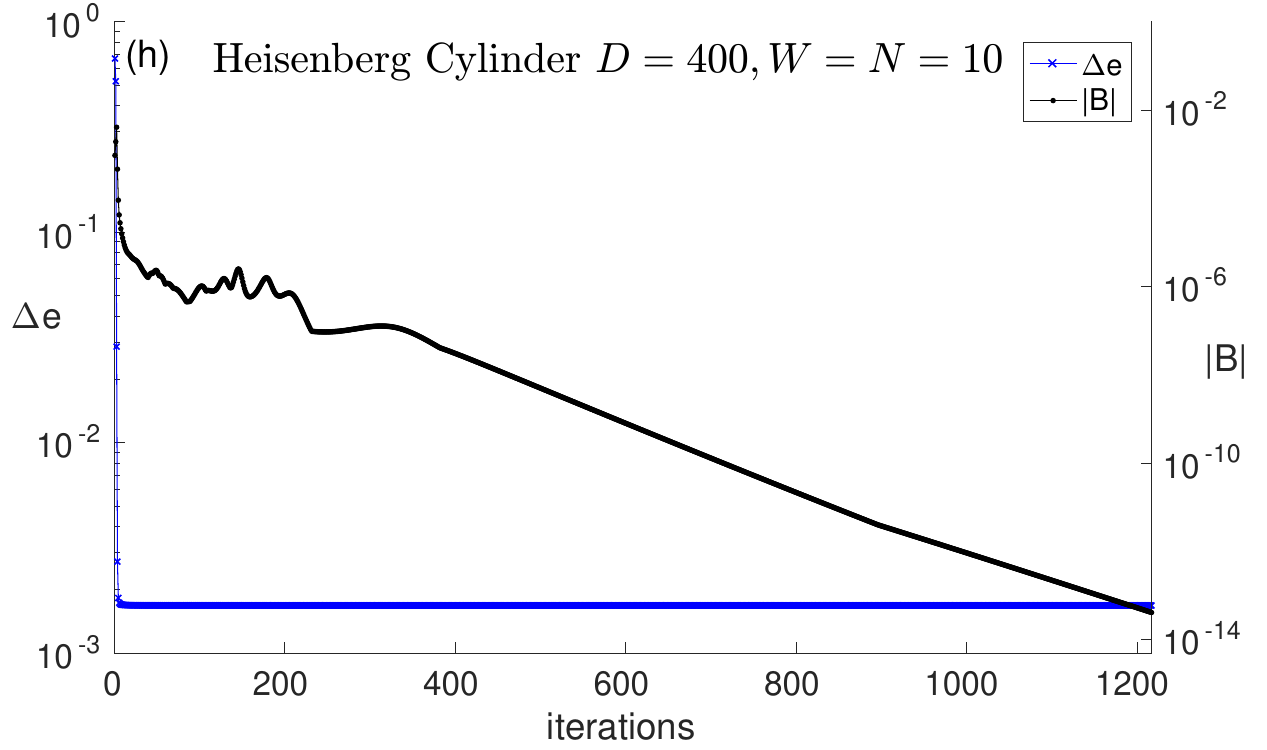}
 \caption{Plot of energy density error $\Delta e$ and gradient norm $\gradnorm$ for \algorithmname\ with a single-site or $N$-site unit cell: 
 (a) TFI model in the gapped symmetry broken phase at $h=0.48$ and $D=25$, 
 (b) gapped isotropic $S=1$ XXZ antiferromagnet and $D=120$, 
 (c) $S=1/2$ XXZ antiferromagnet in the gapped symmetry broken phase at $\Delta=2$ and $D=54$, 
 (d) critical isotropic $S=1/2$ XXZ antiferromagnet at $\Delta=1$ and $D=137$,  
 (e) critical Fermi Hubbard model at $U=10$ and $D=256$, 
 (f) critical $S=1/2$ Haldane-Shastry model at $D=200$, and 
 the isotropic $S=1/2$ Heisenberg antiferromagnet on a Cylinder of circumference $W=8$ (g) and $W=10$ (h).
 The uMPS ground state approximations of (c), (e), (f), (g) and (h) are not translation invariant and have been obtained from Algorithm~\ref{alg:MS_VUMPS_seq} with a multi-site unit cell. Regardless of the criticality of the model, \algorithmname\ converges exponentially fast in gradient norm $\gradnorm$. Notice that at the point where the energy has already converged to machine precision, the gradient is still quite far from zero, and the state thus still some distance from the variational optimum.
 }
 \label{fig:VUMPS_EF}
\end{figure*}

We performed convergence benchmarks for several instances of the models introduced in the previous section, using simple implementations of \algorithmname\ for single or multi-site unit cells presented in Algorithm \ref{alg:SS_VUMPS}, \ref{alg:MS_VUMPS_seq} and \ref{alg:MS_VUMPS_par}, without explicitly exploiting any symmetries. 
Hereto, we consider firstly the error in the variational energy density
\begin{equation}
 \Delta e=e-e_{\rm exact}
 \label{eq:deltae}
\end{equation} 
as a function of the number of iterations. Here $e_{\rm exact}$ is the exact analytic (or quasi-exact numerical) ground state energy density of the model under consideration. \algorithmname\ as formulated in Table~\ref{tab:SS_VUMPS} has its internal convergence measure used to determine when to stop the iteration loop, as well as to set the tolerance in the iterative solvers used within every single outer iteration. However, as a more objective quantity that measures the distance to the variational minimum, we also compute the norm of the energy gradient; it is an absolute measure of convergence which is independent of any prior iterations, as opposed to relative changes in e.g.\ the energy or Schmidt spectrum between iterations.
We denote this quantity as $\gradnorm$, the two-norm of a $D\times d\times D$ tensor $B$, which can be worked out to be given by (see \App{sec:grad_effH})
\begin{align}
B^s &= {A}^{\prime s}_{C} - A^{s}_{L}C' &\text{or}&& B^s &= {A}^{\prime s}_{C} - C' A^{s}_{R}.
\end{align}
The efficient and accurate computation of the gradient norm is further discussed in \App{sec:converr}. To obtain the energy gradient $\gradnorm$ of an $N$-site unit cell, it is equivalent to determine the gradients $B(k)$ for each site independently and to calculate the norm of the concatenation of all $N$ gradients. 

A well known-property of the variational principle is that the energy expectation value itself converges quadratically faster than the state. When the state has converged to some accuracy $\gradnorm$, the energy density has already converged to precision $\O(\gradnorm^{2})$ which can therefore be well beyond machine precision. The convergence measure $\gradnorm$ does however dictate the convergence of other observables which are not diagonal in the energy eigenbasis. Note, however, that we are here referring to convergence towards the value at the variational optimum, not towards the exact value. The error between the variational optimum and the exact ground state can be quantified using e.g.\ the energy variance, or -- in the context of DMRG -- the truncation error. Both quantities are also discussed in \App{sec:converr}. We show results for the truncation error further down in \Sec{sec:extrapolation}, and when comparing \algorithmname\ to IDMRG and ITEBD in \Sec{sec:bench2}.

We show results for examples of 3 gapped and 5 critical systems in \Fig{fig:VUMPS_EF}. Specifically, as examples for gapped systems we considered (a) the TFI model \eqref{eq:TFI_Ham} in the symmetry broken ferromagnetic phase at $h=0.48$, (b) the isotropic $S=1$ Heisenberg antiferromagnet, i.e.\ the $S=1$ XXZ model \eqref{eq:XXZ_Ham} at $\Delta=1$ and (c) the $S=1/2$ XXZ model \eqref{eq:XXZ_Ham} in the symmetry broken antiferromagnetic phase at $\Delta=2$. As examples for gapless systems we considered (d) the isotropic $S=1/2$ Heisenberg antiferromagnet, i.e.\ the $S=1/2$ XXZ model \eqref{eq:XXZ_Ham} at $\Delta=1$, (e) the repulsive Fermi Hubbard model \eqref{eq:HUB_Ham} at $U=10$ and half filling and (f) the Haldane-Shastry model \eqref{eq:HS_Ham}. Finally, we also show results for the isotropic $S=1/2$ Heisenberg antiferromagnet on a cylinder \eqref{eq:XXZ_Ham_cyl} for different widths $W$ in (g) and (h).

Out of the gapped systems, only the antiferromagnetic ground state of (c) physically breaks translation invariance by spontaneously breaking the $\mathbb{Z}_{2}$ spin-flip symmetry; we therefore choose a two-site unit cell in this case. The critical systems physically show no spontaneous symmetry breaking. However, for uMPS ground state approximations in the gapless case it is often energetically beneficial to artificially break symmetries (which are restored in the limit of infinite bond dimension). In all three cases, the optimal uMPS ground state approximation artificially breaks a SU(2) symmetry and develops antiferromagnetic order, breaking translation invariance. We therefore choose a two-site unit cell in the case of the Hubbard model (e) and the Haldane-Shastry model (f). In the case of the Heisenberg antiferromagnet (d), translation invariance can be restored through a unitary transformation by rotating every second spin by $\pi$ around the $z$-axis, transforming $H_{\rm XXZ}(\Delta)\to -H_{\rm XXZ}(-\Delta)$, and the artificially symmetry broken ground state becomes ferromagnetically ordered along the $x$ and $y$ directions. We can therefore choose a single-site unit cell for (d), and the staggered magnetization along $z$ is thus zero. A similar approach could be chosen to restore translation invariance also for the gapped antiferromagnet (c) and the Hubbard model (e), but we do not choose to do so for demonstrative reasons. For (g) and (h), we choose a similar strategy as for (d) and rotate every second spin by $\pi$ around the $z$-axis, allowing the usage of a unit-cell size $N=W$ instead of $N=2W$, which would be necessary to accommodate an antiferromagnetically ordered ground state.

To summarize, we used the single-site Algorithm~\ref{alg:SS_VUMPS} for (a), (b) and (d), and the sequential Algorithm~\ref{alg:MS_VUMPS_seq} with a two-site unit cell for (c), (e) and (f) and a $W$-site unit cell for (g) and (h). For a comparison between the sequential and parallel approach see \Sec{sec:bench_multi}.

\begin{figure}[t]
 \centering
\includegraphics[width=\linewidth,keepaspectratio=true]{\figpath/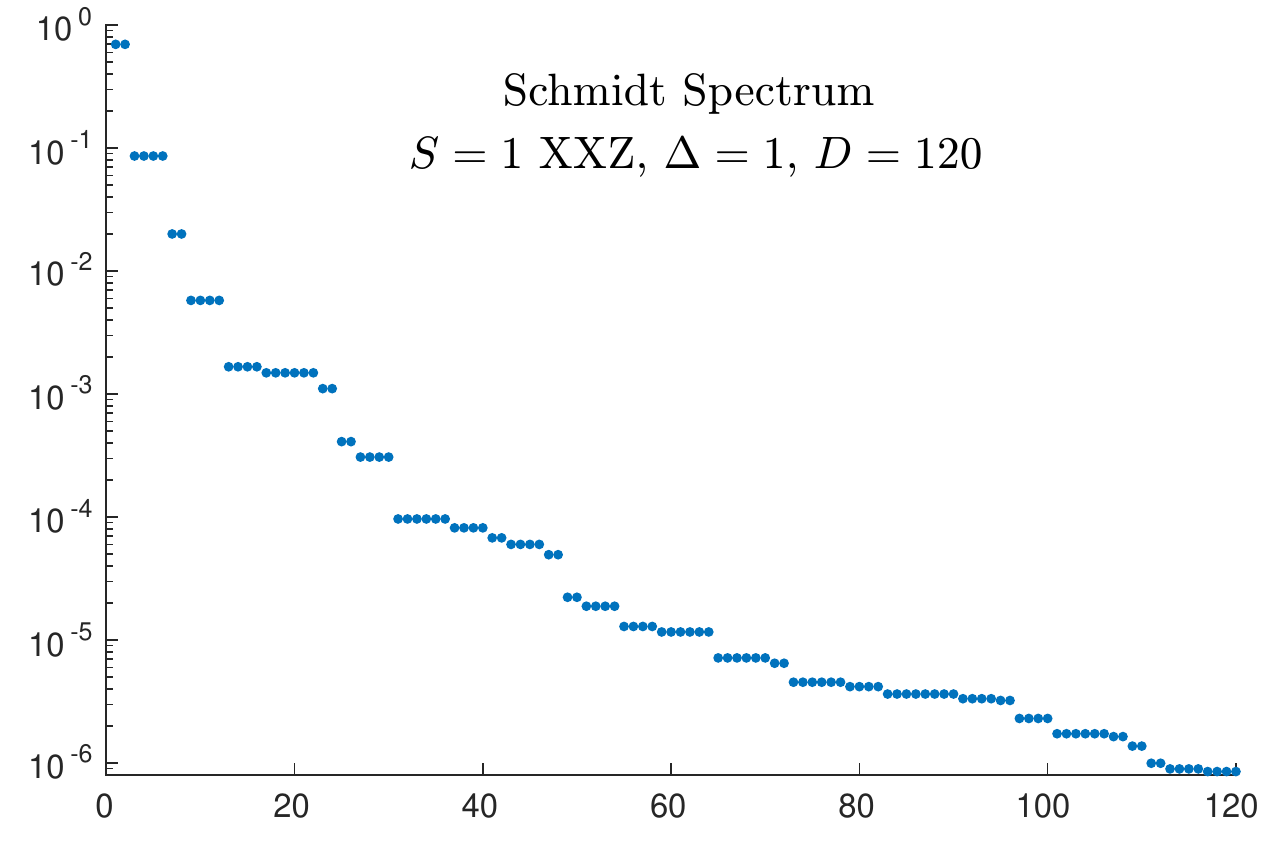}
 \begin{tabularx}{\linewidth}{XX}
 \hline\hline
   0.696198978154358 &  0.001665909341209 \\
   0.696198978154358 &  0.001487766860776 \\
   0.086098881485240 &  0.001487766860776 \\
   0.086098881485240 &  0.001487766860776 \\
   0.086098881485239 &  0.001487766860776 \\
   0.086098881485239 &  0.001487766860776 \\
   0.020013261627349 &  0.001487766860776 \\
   0.020013261627349 &  0.001106527294364 \\
   0.005770050481551 &  0.001106527294364 \\
   0.005770050481551 &  0.000410363691742 \\
   0.005770050481551 &  0.000410363691742 \\
   0.005770050481551 &  0.000307343959372 \\
   0.001665909341209 &  0.000307343959372 \\
   0.001665909341209 &  0.000307343959372 \\
   0.001665909341209 &  0.000307343959372 \\
    \hline\hline
 \end{tabularx}
 \caption{\textit{Top:} Schmidt spectrum of the $S=1$ Heisenberg antiferromagnet for $D=120$, converged to gradient norm $\gradnorm<10^{-15}$. \textit{Bottom:} The table shows the first 30 Schmidt values in descending order. The degeneracies are reproduced to 15 digits of precision, without exploiting any symmetries.}
 \label{fig:S1_schmidt}
\end{figure}

\begin{figure}[t]
 \centering
 \includegraphics[width=\linewidth,keepaspectratio=true]{\figpath/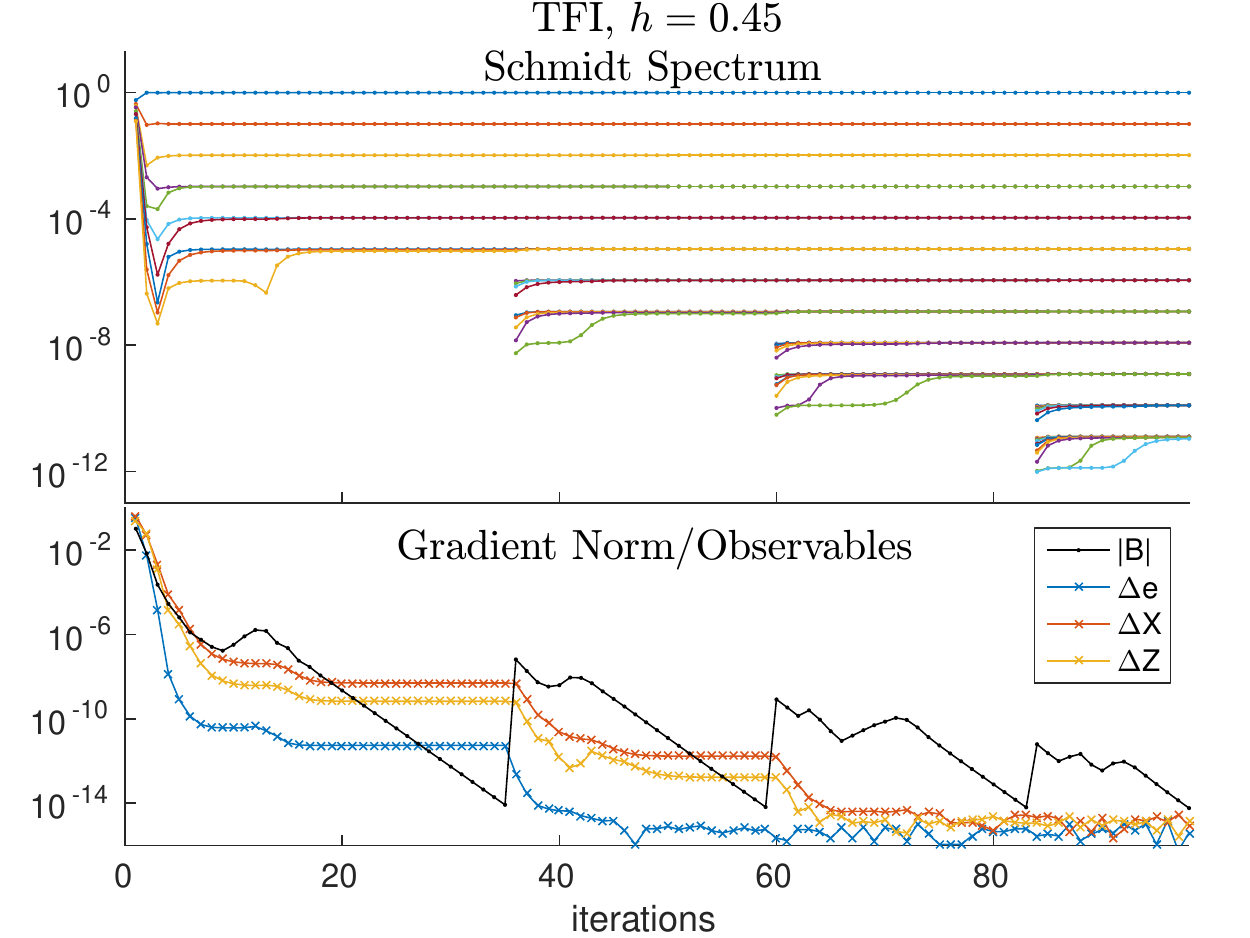}
 \caption{Evolution of the Schmidt spectrum (top) and the gradient norm $\gradnorm$ and various observables (bottom) with iteration number for the TFI model at $h=0.45$. Here we defined the deviation of an observable $O$ from its exact value as $\Delta O=\lvert\braket{O} - \braket{O}_{\rm exact}\rvert$, similar to \eqref{eq:deltae} for the energy.
 We used bond dimensions $D=[9,19,33,55]$, i.e we increased the bond dimension three times during the optimization process as soon as $\gradnorm$ dropped below $10^{-14}$ (at iterations 36, 60 and 84).  It is apparent that while high lying Schmidt values converge quite quickly, the better part of the final iterations goes into converging low lying Schmidt values. Moreover one can see that there is quite some rearrangement of exactly these low lying Schmidt values every time $\gradnorm$ reaches a local maximum (e.g.\ around iterations 12, 40, 70, 88 and 92) during a non-monotonous phase of gradient evolution (see also \Sec{sec:grad_conv}).
 }
 \label{fig:schmidt_evo}
\end{figure}

\subsubsection{General Convergence}
\label{sec:gen_conv}
Above all we observe that \algorithmname\ shows unprecedented fast convergence, both in the energy density $e$ and the norm of its gradient $\gradnorm$, and excellent accuracy of the final ground state approximation. Observe in \Fig{fig:VUMPS_EF} that in all cases the energy is already well converged to machine precision after $\O(10-50)$ iterations, while the state is still quite some distance from the variational optimum, according to the gradient norm $\gradnorm$. Further optimizing the state, this quantity can also be converged to essentially machine precision (even in the presence of small Schmidt values), while the energy virtually doesn't change anymore. The resulting final state then corresponds to the variationally optimal state for the given bond dimension. This is very useful in the case where the quantum state itself is required to be accurate to high precision, e.g.\ when used as a starting state for real time evolution or as a starting point to compute excited states and scattering thereof.\cite{haegeman2012variational,vanderstraeten2014s,vanderstraeten2015scattering} 

The Schmidt spectrum of the ground state of the $S=1$ Heisenberg antiferromagnet at $D=120$, converged to gradient norm $\gradnorm<10^{-15}$, is depicted in \Fig{fig:S1_schmidt}. It can be seen that the degeneracies are reproduced perfectly to the same precision, without explicitly exploiting any symmetries in the implementation of the algorithm.

In cases where the final desired bond dimension $D_{\rm final}$ is not known beforehand, one can successively enlarge a state of some small initial bond dimension every few iterations until the state fulfills the desired criteria, e.g.\ current bond dimension above some threshold, truncation error (see below) or smallest Schmidt value below some threshold, etc. This strategy is particularly useful when using an implementation exploiting physical symmetries of the system, such as e.g.\ conservation of magnetization or particle number, as the correct number and size of the required symmetry sectors in the MPS tensors is generally not known beforehand.\cite{QN_implementation} On the other hand, if $D_{\rm final}$ is known beforehand, it generally appears to be more efficient to immediately start from an initial state with $D=D_{\rm final}$. The gain in computational time due to the cheaper initial iterations with small bond dimension is usually outweighed by a considerable number of required additional iterations. On the other hand, for some hard problems (e.g.\ the Hubbard model) stability and convergence speed can profit from a strategy of sequentially increasing the bond dimension from some small initial value.

To conclude the discussion of general convergence, we plot the evolution of the Schmidt spectrum, as well as the gradient norm $\gradnorm$ and various observables vs. iteration number during a ground state optimization for the TFI model \eqref{eq:TFI_Ham} in the ferromagnetic phase at $h=0.45$ in \Fig{fig:schmidt_evo}. During the simulation we used a sequence of bond dimensions $D=[9,19,33,55]$, where we started with an initial random state with $D=9$ and increased the bond dimension to the next value as soon as $\gradnorm$ dropped below $10^{-14}$. We chose this set of bond dimension in order to not cut any degenerate multiplets of Schmidt values (see also \Sec{sec:degen_schmidt}). It can be seen that the high lying Schmidt values converge quite quickly, while most of the computational time goes into converging the low lying Schmidt values. Moreover, there is quite some rearrangement of the small Schmidt values every time the gradient norm $\gradnorm$ reaches a local maximum during a phase of non-monotonous evolution (see also next subsection). Lastly, from the evolution of the errors of the local observables $\braket{X}$ and $\braket{Z}$ it is apparent that they require a substantially higher bond dimension of $D=55$ to reach the same accuracy as the energy, which is already correct to machine precision at $D=19$.

\begin{figure}[t]
 \centering
 \includegraphics[width=\linewidth,keepaspectratio=true]{\figpath/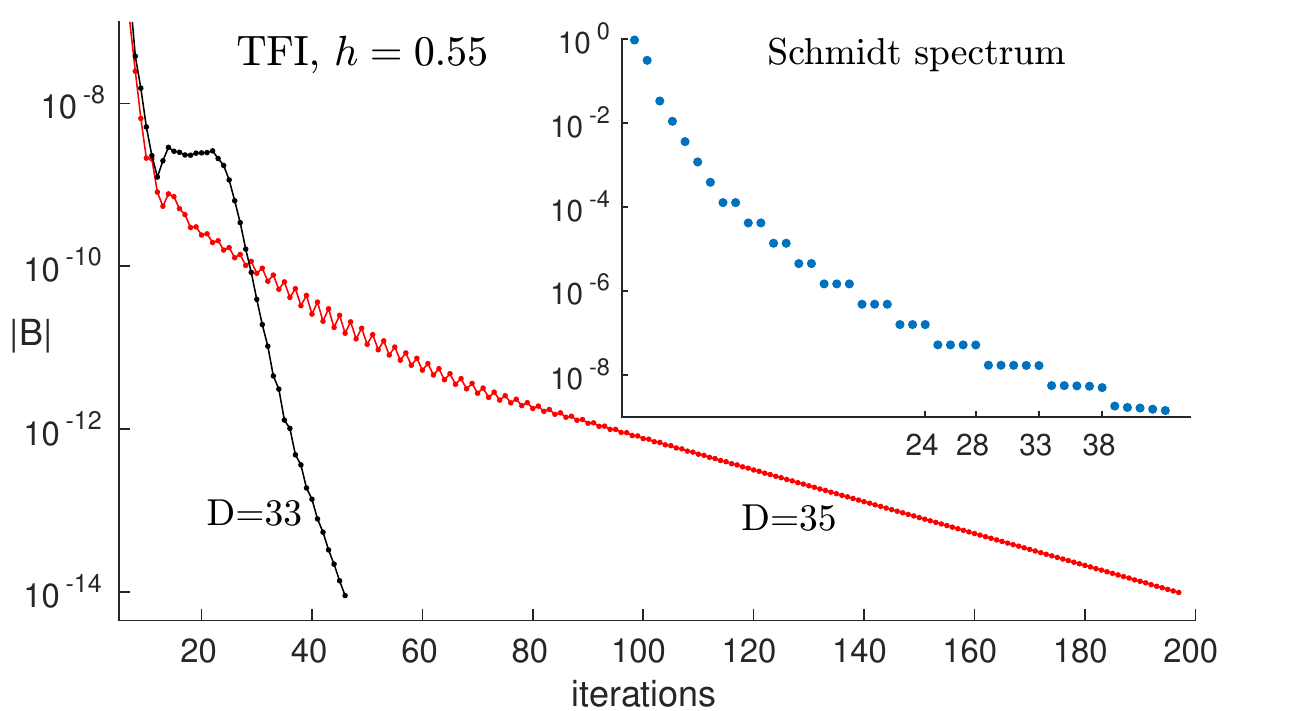}
 \caption{Comparison of the convergence rate of the gradient norm $\gradnorm$ for the TFI model at $h=0.55$, with $D=33$ and $D=35$. Convergence is roughly 4 times faster for $D=33$ as compared to $D=35$. The inset shows the Schmidt spectrum of the ground state (up to $D=43$). For $D=35$ the smallest Schmidt values form an incomplete degenerate multiplet, whereas for $D=33$ the multiplet is complete.}
 \label{fig:convrate_degeneracies}
\end{figure}

\subsubsection{Different Regimes of Gradient Norm Convergence}
\label{sec:grad_conv}
Depending on the complexity of the model, the gradient norm shows a period of irregular non-monotonous behavior before entering a regime of monotonous convergence. This can be understood as the (random) initial state having to adapt its initial structure (e.g.\ the Schmidt spectrum) to the requirements of the best variational ground state approximation. Monitoring the Schmidt values during this period nicely shows how groups of Schmidt values slightly rearrange to the correct structure (see e.g.\ \Fig{fig:schmidt_evo}) . This period is usually more dominant in critical systems -- as can be seen in \Fig{fig:VUMPS_EF} (d) - (f), where it takes $\O(50-100)$ iterations -- and of course strongly depends on the chosen initial state. One could argue, that the jumps in parameter space caused by the algorithm during this period are too big for the state to find the correct structure quickly, hindering a fast crossover to the regime of monotonous convergence. However, an approach of preconverging the state using smaller steps through parameter space -- e.g.\ by means of imaginary time evolution with moderate time steps -- has proven to be even slower in all cases tried. Thus, the best choice is still to use \algorithmname\ during the entire optimization process. We want to emphasize here, 
that we have never observed a stagnation of the algorithm during this initial regime; the algorithm always reached the monotonous regime eventually in all cases, and instances where the algorithm remains in the irregular regime for an unusually long time are rare and only occur in the case of particularly hard problems.

\begin{figure*}[t]
 \centering
 \includegraphics[width=\linewidth,keepaspectratio=true]{\figpath/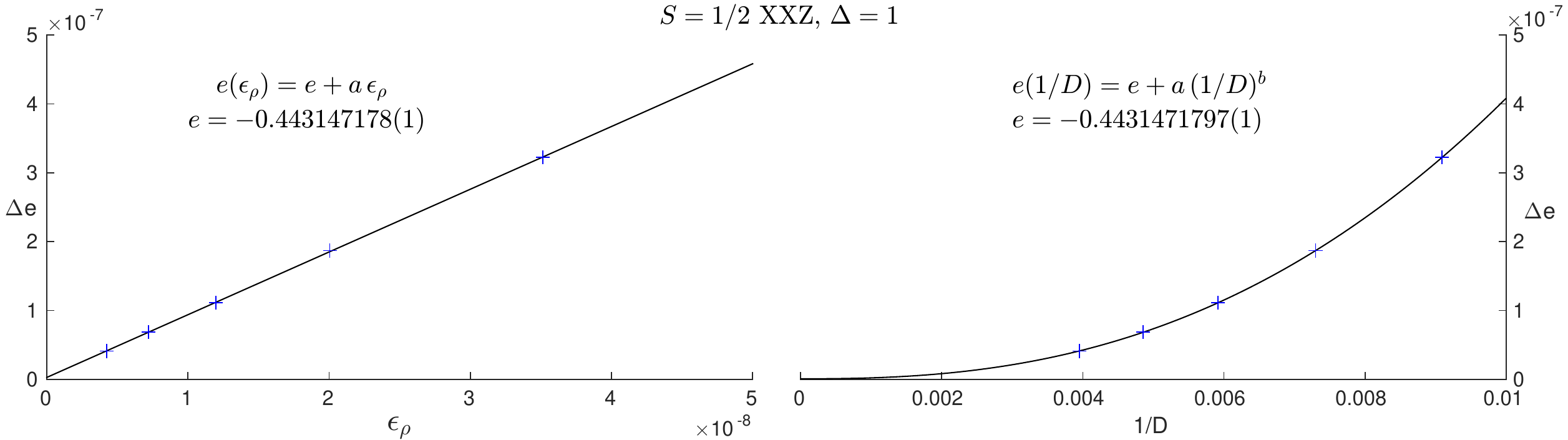}
 \caption{Scaling of the variational ground state energy $e$ with truncations error $\epsilon_{\rho}$ and bond dimension $D$ for the isotropic $S=1/2$ Heisenberg antiferromagnet. We plot the energy $e$ vs. truncation error $\epsilon_{\rho}$ on the left, and vs. inverse bond dimension $1/D$ on the right. The exact ground state energy is given by $e_{\rm exact}=-0.4431471805599453$ to 16 digits of precision. We obtain estimates from a linear fit $e(\epsilon_{\rho})=e + a\,\epsilon_{\rho}$ (left) and a power law fit $e(1/D)=e + a\,(1/D)^{b}$ (right), where the estimate on the right has one more digit of precision and is roughly 4 times more accurate than the estimate on the left.}
 \label{fig:energy_extrapolation}
\end{figure*}

As soon as the gradient norm reaches the monotonous regime, it always converges exponentially fast. Surprisingly, this is true even for critical systems, where one would in principle expect algebraic convergence. This can be qualitatively understood from the theory of finite entanglement scaling,\cite{FES_Luca,FES_Pollmann,FES_Vid} which states that the MPS approximation itself introduces a small perturbation away from criticality, and thus a finite gap. However, as \algorithmname\  improves convergence speed over existing methods (see also \Sec{sec:bench2}), it is ideally suited to study critical systems via the theory of finite entanglement scaling, which still requires that one finds the optimal MPS approximation in the first place.

\subsubsection{Degenerate Schmidt Values}
\label{sec:degen_schmidt}
In the presence of multiplets of degenerate Schmidt values, the convergence rate is severely affected if the smallest few Schmidt values are part of an incomplete multiplet, i.e.\ if the last multiplet is ``cut''. In that case the algorithm still shows stable convergence, albeit at a greatly reduced rate. For an example in the TFI model see \Fig{fig:convrate_degeneracies}. This issue can be easily circumvented by ensuring that the smallest few Schmidt values are part of a complete multiplet when dynamically increasing the bond dimension, or by choosing a viable (or reducing from some) fixed initial bond dimension.

\subsubsection{Energy Convergence with Bond Dimension}
\label{sec:extrapolation}
In a careful MPS study, variational energies obtained for different bond dimension $D$ are compared in order to extrapolate to the exact $D\to\infty$ limit. This can be done by plotting the energy $e(D)$ as a function of bond dimension against the inverse of the bond dimension $1/D$. The infinite $D$ limit is then obtained by fitting with a power law form and extrapolating to $1/D\to0$. In DMRG, another popular measure for the quality of an MPS approximation is given by the truncation error or discarded weight $\epsilon_{\rho}$, defined in Eq.~\eqref{eq:truncation_error}. The variational energy is found to scale linearly with $\epsilon_{\rho}$\cite{White_Huse93,Legeza_Fath96} and an extrapolation to $\epsilon_{\rho}\to 0$ is thus generally easier and more stable. For further details on assessing the quality of the ground state approximation we refer to \App{sec:converr}.

We show an example for both extrapolation schemes for the isotropic $S=1/2$ Heisenberg antiferromagnet in \Fig{fig:energy_extrapolation}. The exact ground state energy is given by $e_{\rm exact}=\frac{1}{4}-\log(2)$, or as numerical value by $e=-0.4431471805599453$ to 16 digits of precision. On the left we plot the energy vs. truncation error and obtain an estimate $e_{T}=-0.443147178(1)$ with 9 digits of precision  from a linear fit $e(\epsilon_{\rho})=e + a\,\epsilon_{\rho}$. On the right we plot the energy vs. inverse bond dimension and obtain an estimate $e_{D}=-0.4431471797(1)$ with 10 digits of precision from a power law fit $e(1/D)=e + a\,(1/D)^{b}$. Comparing to $e_{\rm exact}$ we observe that $e_{T}$ has an error of $\Delta e_{T}\approx 3\times 10^{-9}$, while $e_{D}$ has an error $\Delta e_{D}\approx 8\times 10^{-10}$.


\begin{figure*}[ht]
 \centering
 \includegraphics[width=0.48\linewidth,keepaspectratio=true]{\figpath/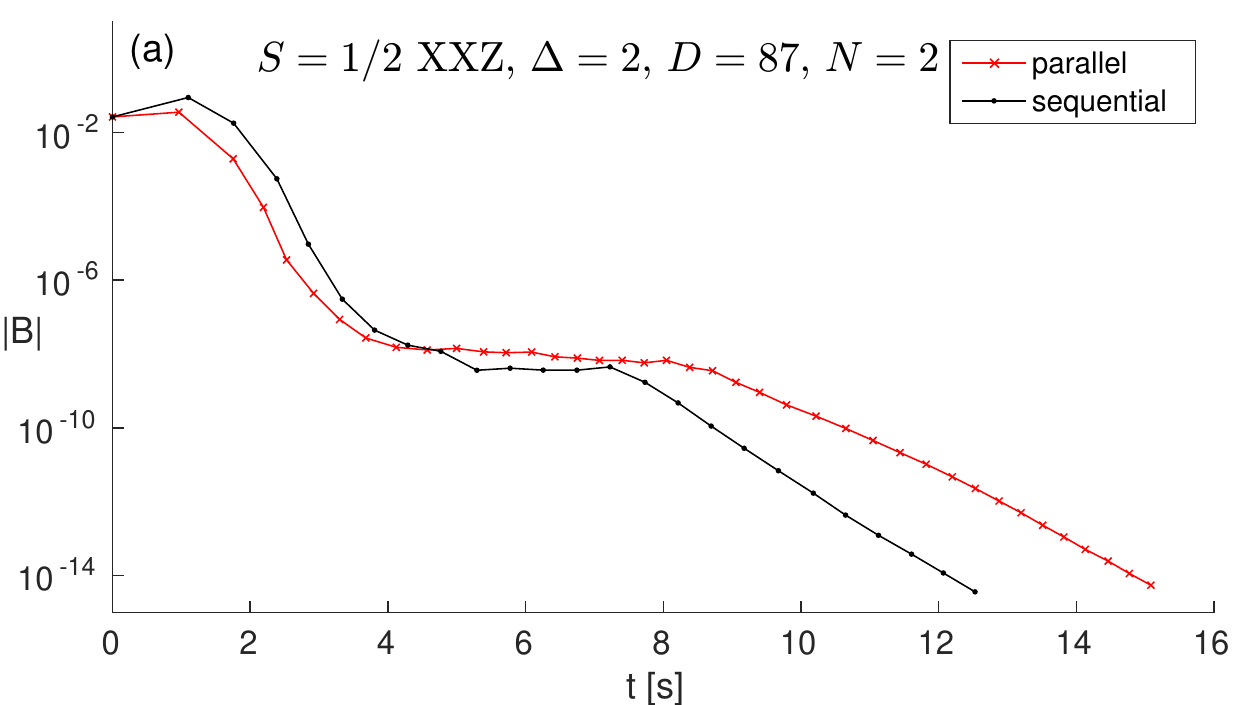}
 \includegraphics[width=0.48\linewidth,keepaspectratio=true]{\figpath/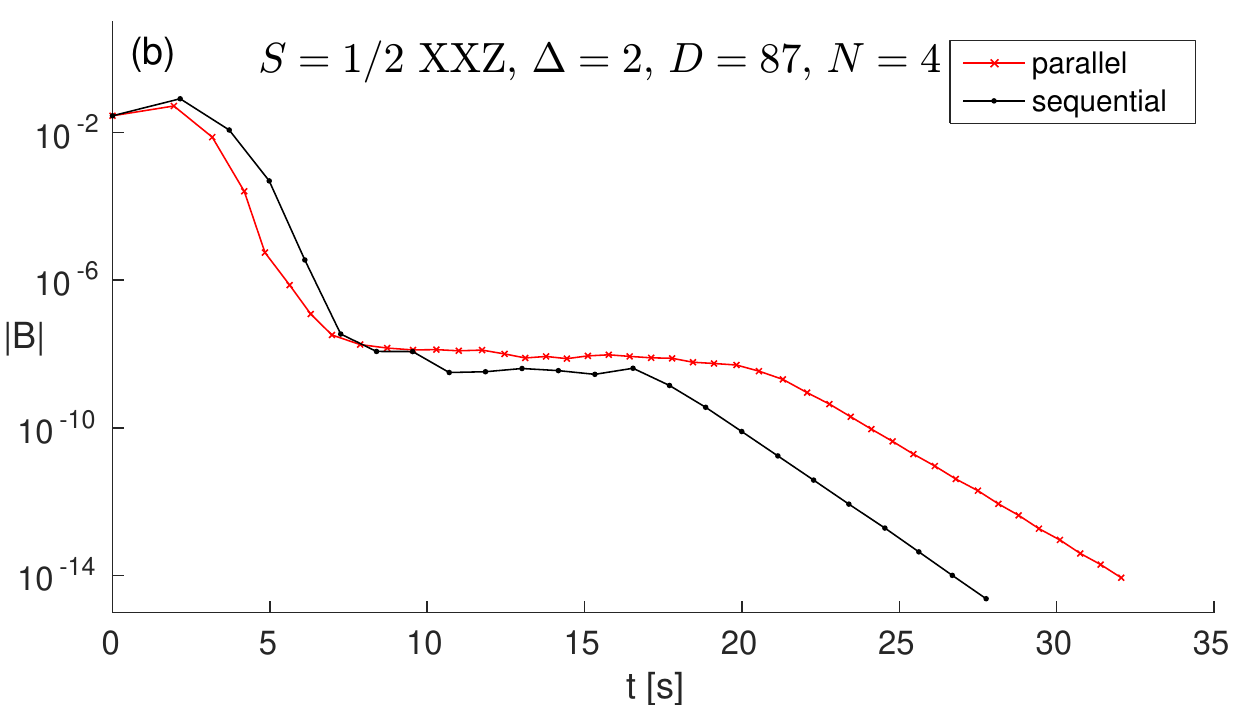}
 \includegraphics[width=0.48\linewidth,keepaspectratio=true]{\figpath/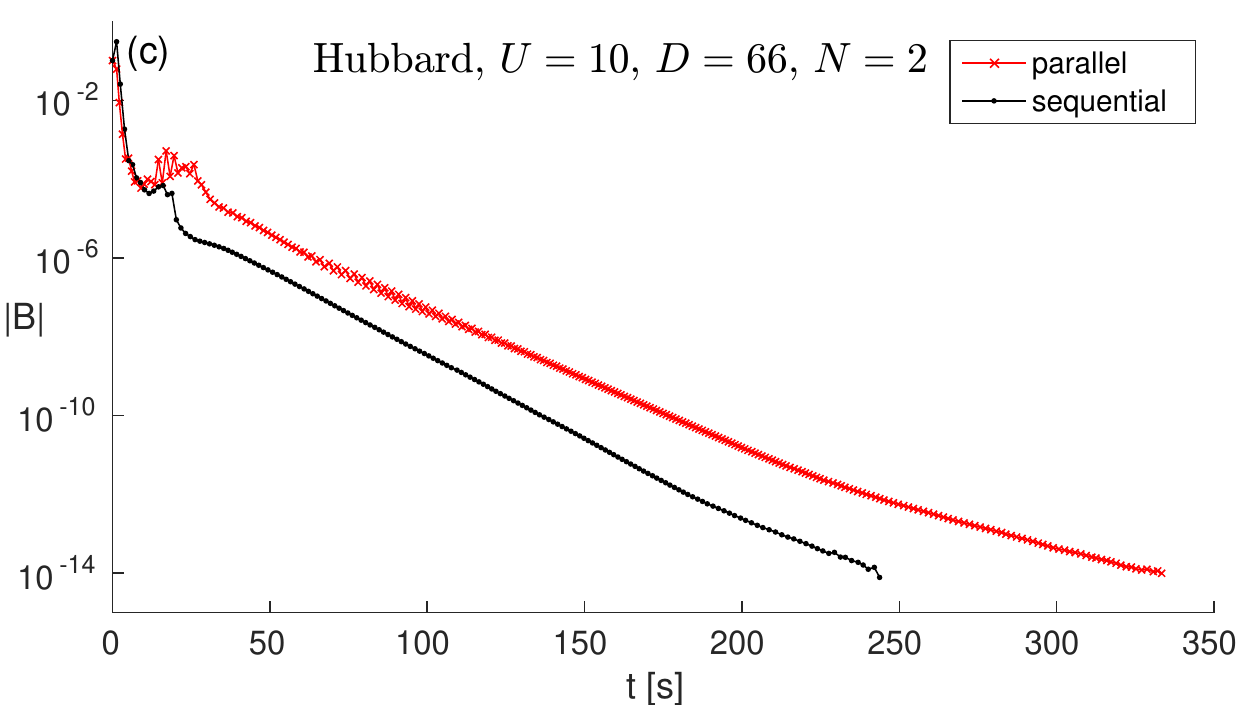}
 \includegraphics[width=0.48\linewidth,keepaspectratio=true]{\figpath/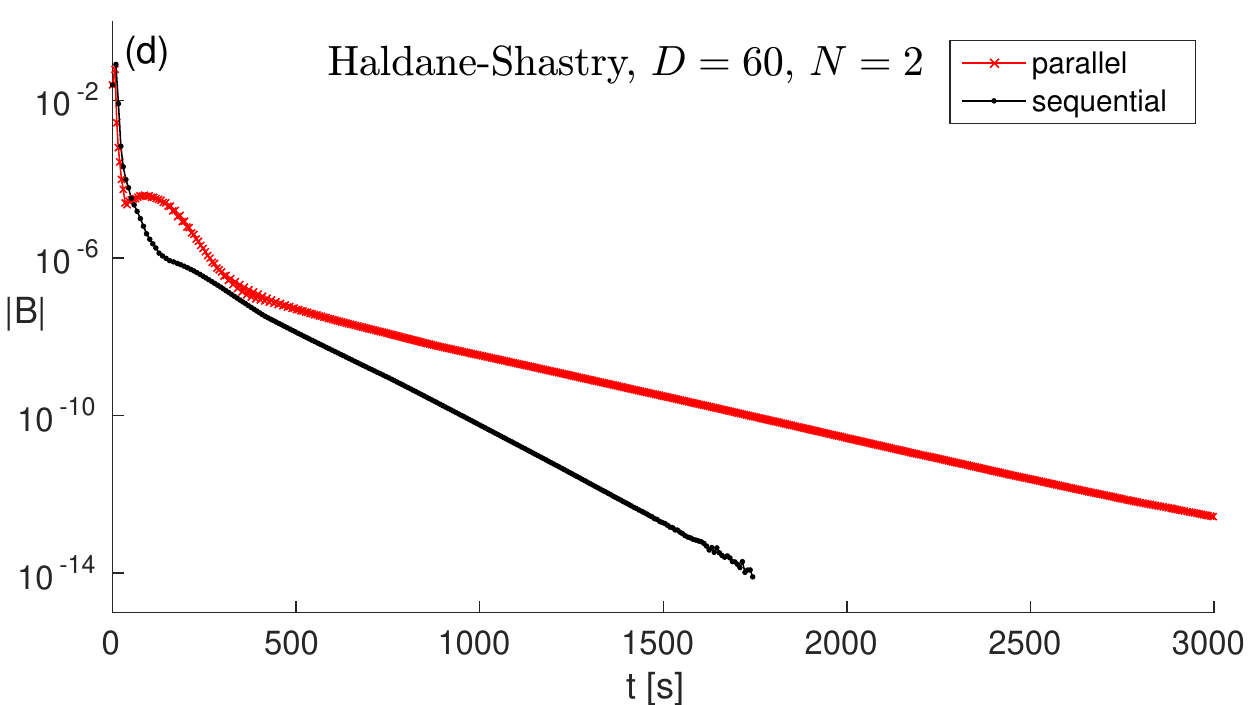}
 \caption{Example performance comparisons between the sequential and parallel algorithm presented in \Sec{sec:MS_VUMPS}. We show examples for 
 the gapped $S=1/2$ XXZ antiferromagnet at $\Delta=2$ and $D=87$ with (a) a two-site unit cell and (b) a four-site unit cell, as well as (c) the critical Hubbard model at $U=5$ and $D=66$ with a two-site unit cell and (d) the critical Haldane-Shastry model at $D=60$ with a two-site unit cell.
 In each example, both approaches were initialized in the same (random) initial state. It can be seen that while the parallel approach usually takes less time for one unit cell update iteration, the sequential approach shows faster convergence in number of iterations. Both approaches show similar performance in the considered cases, except for example (d).
}
 \label{fig:seq_vs_par}
\end{figure*}

\subsubsection{Multi Site Unit Cell Implementations}
\label{sec:bench_multi}
Lastly we discuss and compare the performance of the \textit{sequential} and \textit{parallel} algorithms for multi-site unit cells presented in \Sec{sec:MS_VUMPS}. As the two methods differ in their convergence with the number of iterations, as well as in the computational effort for each iteration of updating the entire unit cell, we compare the rate of convergence with absolute computing time $t$ in seconds. To that end we only time operations that are absolutely necessary for each algorithm, i.e.\ we do not time measurements, data storage etc. We further start from the same (random) initial state and keep the bond dimension fixed throughout the entire simulation for both methods to make the simulations as comparable as possible. All calculations are performed using a non-parallelized MATLAB implementation on a single core of a standard laptop CPU. 

We find that in general for gapped systems, the sequential approach outperforms the parallel approach, while in critical systems no definite statement about better performance can be made. There are instances where one algorithm takes substantially longer than the other to reach the regime of monotonous convergence, but such cases are rare and strongly depend on the model and initial state. 
Once both algorithms are in the monotonous regime, convergence speed in terms of absolute computing time is similar, with the sequential approach generally taking longer for each iteration, but the parallel approach generally requiring more iterations to reach convergence. Overall both approaches thus generally show comparable performance, with the sequential approach appearing to be slightly more stable and reliable in the cases we considered. \Fig{fig:seq_vs_par} shows examples for two gapped and two gapless systems.

\subsection{Comparison with IDMRG and ITEBD}
\label{sec:bench2}
We further benchmark the performance of \algorithmname\ against a standard two-site IDMRG implementation,\cite{DMRG1,DMRG2,IDMRG} and a standard two-site ITEBD implementation,\cite{ITEBD,ITEBD_Hastings} and compare the rate of convergence of the energy error $\Delta e$ and the norm of the energy gradient $\gradnorm$ between the three methods. For \algorithmname, we solve the effective eigenvalue problems in each iteration to precision $\epsilon_{H}=\epsilon_{\rm prec}/100$ with $\epsilon_{\rm prec}$ the current precision according to \eqref{eq:prec}. For IDMRG we solve the effective two-site eigenvalue problem in each iteration to precision $\epsilon_{H}=(1-F)/100$, with $F$ the current orthogonality fidelity $F$ (see Sec. III.A in Ref.~\onlinecite{IDMRG}). For ITEBD we employ a fourth order Suzuki Trotter decomposition and measure every 10 time steps. We use a sequence of time steps $\delta t\in[10^{-1},10^{-2},10^{-3},10^{-4},10^{-5},10^{-6}]$ where we decrease the time step as soon as the change in Schmidt values per unit of imaginary time drops below a certain threshold. Naturally, the strategy of time step reduction should be optimized carefully for each model under consideration, however we choose the same strategy for all example cases to maintain comparability.

We also explicitly calculate all necessary quantities for obtaining a truly variational energy (e.g.\ by reorthogonalizing the unit cell) and for measuring the energy gradient, even if these quantities are not  necessary for the respective algorithm itself.

As the three methods differ quite substantially in the number of iterations required for convergence, as well as the computational effort for each iteration, we again compare convergence against absolute computing time $t$ in seconds, where we only time operations that are absolutely necessary for each algorithm and we do not time measurements, data storage, reorthogonalizing, etc. We further start from the same (random) initial state and keep the bond dimension fixed throughout the entire simulation for all three methods to make the simulations as comparable as possible. Again, all calculations were performed using a non-parallelized MATLAB implementation on a single core of a standard laptop CPU.

\begin{figure*}[p]
 \centering
 \includegraphics[width=\linewidth,keepaspectratio=true]{\figpath/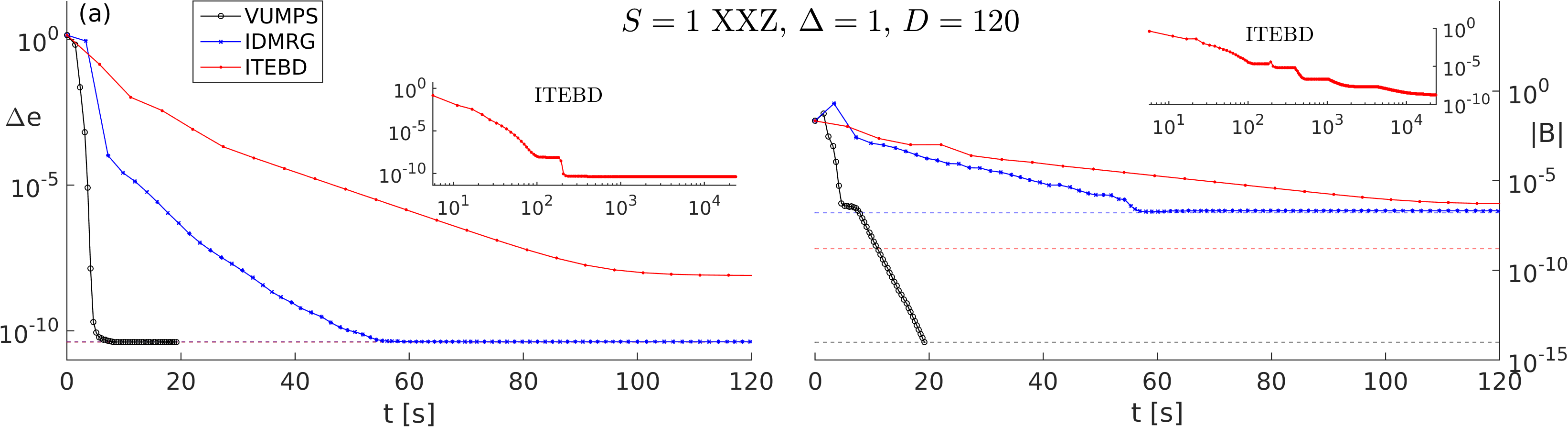}
 \includegraphics[width=\linewidth,keepaspectratio=true]{\figpath/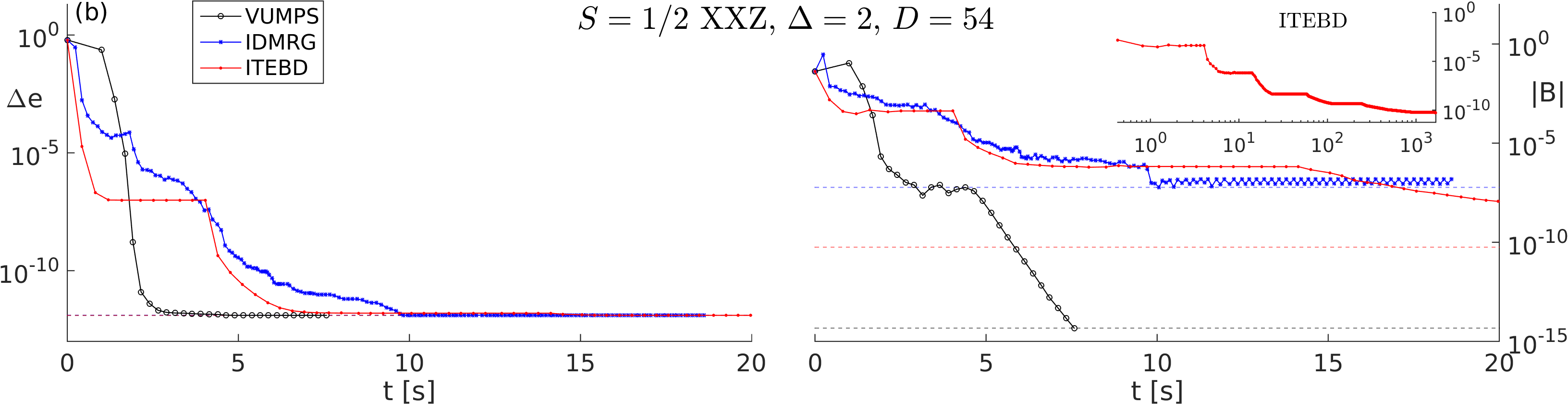}
 \includegraphics[width=\linewidth,keepaspectratio=true]{\figpath/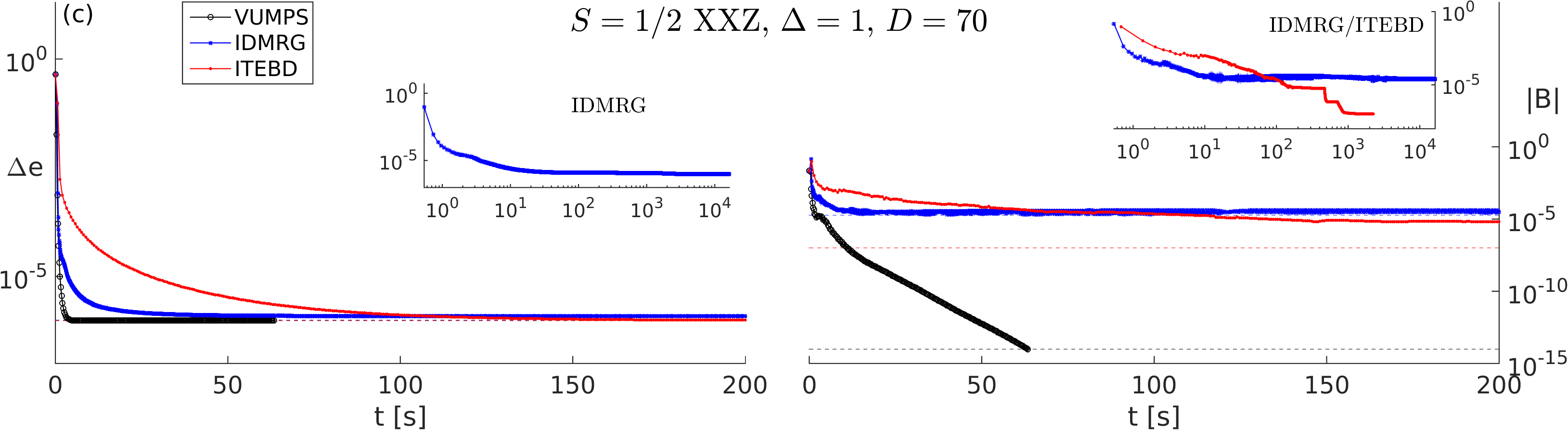}
 \includegraphics[width=\linewidth,keepaspectratio=true]{\figpath/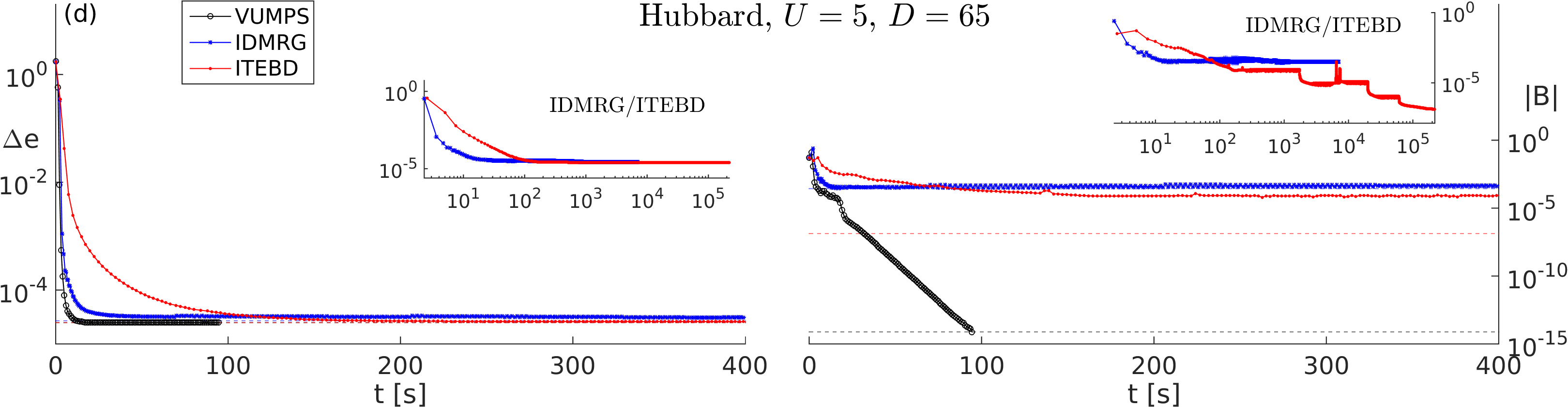}
 \caption{Comparative benchmark plots for \algorithmname, IDMRG and ITEBD. We plot the error $\Delta e$ on the left, and the gradient norm $\gradnorm$ on the right, vs. total computing time $t$ in seconds for 
 (a) the gapped isotropic $S=1$ XXZ antiferromagnet at $D=120$, 
 (b) the gapped $S=1/2$ XXZ antiferromagnet at $\Delta=2$ and $D=54$,
 (c) the critical isotropic $S=1/2$ XXZ antiferromagnet at $D=70$, and
 (d) the critical Hubbard model at $U=5$ and $D=65$.
The dashed lines are a guide to the eye and denote the minimum values of $\Delta e$ and $\gradnorm$ obtained by the respective algorithm. The insets show a plot of the entire ITEBD and/or IDMRG simulation with logarithmic time scale.
It is obvious that \algorithmname\ reaches convergence orders of magnitude faster than IDMRG or ITEBD, especially for critical systems. Notice also that, while $\Delta e$ differs only by a few percent between the different algorithms (up to $\approx 10\%$ for (d)), \algorithmname\ always manages to also converge $\gradnorm$ essentially to machine precision, whereas IDMRG and ITEBD stagnate at some substantially higher values, remaining quite far from the variational optimum.
 }
 \label{fig:Timings}
\end{figure*}

We show example comparisons for two gapped and two critical models in \Fig{fig:Timings}, similar to the cases studied in the previous section. Specifically, we show results for (a) the gapped isotropic $S=1$ Heisenberg antiferromagnet, i.e.\ the $S=1$ XXZ model \eqref{eq:XXZ_Ham} at $\Delta=1$, (b) the $S=1/2$ XXZ model \eqref{eq:XXZ_Ham} in the gapped symmetry broken antiferromagnetic phase at $\Delta=2$, (c) the critical isotropic $S=1/2$ Heisenberg antiferromagnet, i.e.\ the $S=1/2$ XXZ model \eqref{eq:XXZ_Ham} at $\Delta=1$, and (d) the critical Fermi Hubbard model \eqref{eq:HUB_Ham} at $U=5$ and half filling. We plot the energy error $\Delta e$ on the left and the gradient norm $\gradnorm$ on the right, vs. absolute computing time $t$ in seconds. For \algorithmname\ we used a single-site unit cell for (a) and (c), and a two-site unit cell for (b) and (d).

Above all we observe that \algorithmname\ clearly outperforms both IDMRG and ITEBD by far, both in convergence speed and accuracy of the final state, especially for critical systems. In all shown cases, the final energy error $\Delta e$ of all three algorithms only differ by a few percent; \algorithmname\ however always yields the best variational energy, often already after a few seconds, and thus converges in energy much faster than IDMRG or ITEBD -- in the case of critical systems even by orders of magnitude. Observe that especially for the two critical systems (c) and (d) a large part of the computational time of IDMRG and ITEBD goes in converging the last few digits of the energy (see also insets in \Fig{fig:Timings}). For (d) in particular, the final energy error obtained by IDMRG is still almost $10\%$ higher than the value obtained by VUMPS.

In terms of convergence of the energy gradient $\gradnorm$, we observe that IDMRG and ITEBD perform quite poorly. Surprisingly, IDMRG usually stagnates at some value $\gradnorm>10^{-7}$. ITEBD on the other hand would be in principle capable of converging $\gradnorm$ essentially also to machine precision, albeit at prohibitively long simulation times, as the limiting factor appears to be the Trotter error, requiring very small time steps; we therefore also only reach values of $\gradnorm\gtrsim10^{-10}$ with ITEBD within reasonable simulation times.
\footnote{The fact that ITEBD usually yields better variational states than IDMRG for the same bond dimension has already been observed in Ref.~\onlinecite{IDMRG}.}
\algorithmname\ on the other hand is always capable to converge $\gradnorm$ essentially to machine precision, and does so -- contrary to other methods -- exponentially fast and with unprecedented speed. 
For instance, in the case of the Hubbard model in example (d), ITEBD only reached a gradient norm of $\gradnorm=1.3\times10^{-7}$ after $\approx60$ hours of absolute computing time, while \algorithmname\ already reached this value after only $\approx 30$ seconds, and converged further to $\gradnorm<10^{-14}$ in $\approx 90$ seconds. IDMRG on the other hand stagnated at a quite high value of $\gradnorm\approx 2\times10^{-4}$.

Finally, we also compare variational ground state energies for the $S=1/2$ Heisenberg antiferromagnet on a cylinder with results from Ref.~\onlinecite{Osorio17,OsorioPriv}. We have already seen in \Fig{fig:VUMPS_EF} that \algorithmname\ shows excellent convergence speed also for this model. We show a comparison of obtained energies for cylinder circumferences $W=6,8$ and bond dimensions $D=256,512$ in Table~\ref{tab:IDMRG_vs_VUMPS_cyl}. There it is shown that for all cases, \algorithmname\ yields lower energies with differences roughly around the 5th-7th significant digit.\footnote{IDMRG simulations have been performed without a rotation by $\pi$ of every second spin and with a $N=2W$ unit cell to accommodate an antiferromagnetically ordered ground state.\cite{OsorioPriv}} We conclude that \algorithmname\ is thus perfectly suited for a fast and accurate study of two dimensional models of current interest on infinite cylinders and improves over current state of the art methods.

\begin{table}
 
 \begin{tabular}{r|c|c|}
  W=6			&D=256	&D=512\\
  \hline
   \algorithmname	&-0.672544677277	&-0.672724840927\\
  IDMRG			&-0.672518318233	&-0.672724792693
 \end{tabular} 
 
 \begin{tabular}{r|c|c|}
  W=8			&D=256	&D=512\\
  \hline
   \algorithmname	&-0.669761862738	&-0.670380353990\\
  IDMRG			&-0.669761333217	&-0.670379398859
 \end{tabular} 
 
 
\caption{Comparison of obtained variational ground state energies for the $S=1/2$ Heisenberg antiferromagnet on a cylinder \eqref{eq:XXZ_Ham_cyl} for circumference $W=6,8$ and bond dimensions $D=256,512$. \algorithmname\ improves on the values obtained by IDMRG around the 5th-7th significant digit.}
\label{tab:IDMRG_vs_VUMPS_cyl}
\end{table} 

\subsubsection{Observables}
We also measure and compare the regular and staggered (averaged) magnetizations $m_{\rm r}$ and $m_{\rm s}$ of the final state after convergence for the Hubbard model in example (d), as in this case all three methods use a two-site unit cell.
The exact ground state is SU(2) symmetric and thus has zero magnetization; a finite $D$ ground state approximation however artificially breaks this symmetry. The final values for the regular magnetization $m_{\rm r}$ are zero to machine precision for both \algorithmname\ and IDMRG, but $m_{\rm r}=8\times10^{-12}$ for ITEBD. The staggered magnetization is $m_{\rm s}=0.011162$ for \algorithmname, $m_{\rm s}=0.080768$ for IDMRG, and $m_{\rm s}=0.034797$ for ITEBD. Both the regular and staggered magnetizations are thus smallest for the final state obtained from \algorithmname. For IDMRG the staggered magnetization is highest, but the regular magnetization is zero, which in turn is finite for ITEBD. This result is not surprising, as \algorithmname\ yields the best variational state out of the three methods.

\subsubsection{Truncation Error}
As a last figure of merit, popular in DMRG studies as a measure of the quality of the MPS ground state approximation and used for extrapolations to the exact infinite $D$ limit, we also calculate the truncation error or discarded weight $\epsilon_{\rho}$ of the final state, defined in Eq.~\eqref{eq:truncation_error}. 
In the case of the Hubbard model in example (d), we obtain a truncation error 
$\epsilon_{\rho}=2.54438\times 10^{-6}$ 
from IDMRG, and a slightly lower 
$\epsilon_{\rho}=2.45138\times 10^{-6}$ 
from \algorithmname.

\section{Conclusion and Outlook}
\label{sec:conclusion}
We have introduced a novel algorithm for calculating MPS ground state approximations of strongly correlated one dimensional quantum lattices models with nearest neighbor or long range interactions, in the thermodynamic limit. It combines ideas from conventional DMRG and tangent space methods by variationally optimizing a uniform MPS by successive solutions of effective eigenvalue problems. The algorithm can easily be implemented by extending an existing single-site (I)DMRG implementation with routines for i) calculating effective Hamiltonian contributions from infinite environments and ii) solving an effective ``zero site'' eigenvalue problem in addition to the usual single-site problem. The new algorithm is free of any ill-conditioned inverses and therefore does not suffer from small Schmidt values, contrary to other tangent space methods such as TDVP. Additionally, as it does not rely on imaginary time evolution, it is especially fit for studying systems with long range interactions.

We described and benchmarked implementations for uniform MPS with both single-site and multi-site unit cells. We observed that the new algorithm clearly outperforms existing methods such as IDMRG and ITEBD, both in convergence speed and accuracy of the final state at convergence. This is especially also the case for systems with long range interactions or simulations of two dimensional models on infinite cylinders. The energy converges with unprecedented speed  after $\O(10-50)$ iterations, even in critical systems (where this is orders of magnitude faster than conventional methods). The algorithm further proceeds to converge the state to the variational optimum by minimizing the energy gradient essentially to machine precision; it does so exponentially fast, even for critical systems, contrary to other methods. The new algorithm is thus the perfect choice for studying critical systems. 
Additionally, a state converged to the variational optimum is particularly useful in cases where the quantum state itself is required to be accurate to high precision, e.g. when used as a starting state for real time evolution or for a variational calculation of elementary excitations.\cite{haegeman2012variational,vanderstraeten2014s,vanderstraeten2015scattering}

It is straightforward to include physical symmetries that come with good quantum numbers (such as e.g.\ conserved magnetization or particle number) after a proper definition of a symmetric uniform MPS unit cell, where absolute (diverging) values of these quantum numbers are replaced by densities. All steps of the algorithm then immediately also apply to MPS tensors with good quantum numbers. Symmetric ground states obtained this way are an excellent starting point for obtaining elementary excitations with well defined quantum numbers following Ref.~\onlinecite{haegeman2012variational}, which for instance enables to target elementary excitation that lie within a multi-particle continuum.\cite{QN_implementation}

Within the same framework it is also very natural to recover real or imaginary time evolution by replacing the effective Hamiltonian ground state problems \eqref{eq:FP1} and \eqref{eq:FP2} by small finite time evolution steps, which yields the thermodynamic limit version of the time evolution algorithm presented in Ref.~\onlinecite{TDVP_Uni}. This enables e.g.\ to efficiently study real time evolution of quantum states on systems with long range interactions in the thermodynamic limit.\cite{DQPT1,DQPT2}

We believe that the ideas presented in this paper should be relevant for other classes of tensor-network states as well. Specifically in the case of projected entangled pair states (PEPS) \cite{PEPS}, designed to capture ground states on two-dimensional quantum lattices, the further development of efficient variational algorithms -- as an alternative to ITEBD inspired approaches -- is still much desired.\cite{PEPS_var_Corboz,PEPS_var_LV} In particular, it motivates the search for (approximate) canonical forms for PEPS, which would enable a translation of the VUMPS algorithm to the two-dimensional setting. Equivalently, an adaptation of the new algorithm for continuous Matrix Product State (cMPS) \cite{cMPS} would allow for a much desired increase of efficiency for cMPS simulations, building on recent advances in that direction.\cite{CMPSGanahl2016}

\section*{acknowledgments}
The authors acknowledge inspiring and insightful discussions with M. Ganahl, M. Mari\"en and I.P. McCulloch. We thank B. Buyens for providing templates for drawing tensor network diagrams. We also thank J. Osorio Iregui for providing IDMRG data for infinite cylinder simulations. This project has received funding from the European Research Council (ERC) under the European Union’s Horizon 2020 research and innovation programme (grant agreement No 647905). The authors gratefully acknowledge support from the Austrian Science Fund (FWF): F4104 SFB ViCoM and F4014 SFB FoQuS (V.Z.-S. and F.V.), and GRW 1-N36 (M.F.). J.H. and L.V. are supporte  d by the Research Foundation Flanders (FWO). 

%


\begin{thebibliography}{78}%
\makeatletter
\providecommand \@ifxundefined [1]{%
 \@ifx{#1\undefined}
}%
\providecommand \@ifnum [1]{%
 \ifnum #1\expandafter \@firstoftwo
 \else \expandafter \@secondoftwo
 \fi
}%
\providecommand \@ifx [1]{%
 \ifx #1\expandafter \@firstoftwo
 \else \expandafter \@secondoftwo
 \fi
}%
\providecommand \natexlab [1]{#1}%
\providecommand \enquote  [1]{``#1''}%
\providecommand \bibnamefont  [1]{#1}%
\providecommand \bibfnamefont [1]{#1}%
\providecommand \citenamefont [1]{#1}%
\providecommand \href@noop [0]{\@secondoftwo}%
\providecommand \href [0]{\begingroup \@sanitize@url \@href}%
\providecommand \@href[1]{\@@startlink{#1}\@@href}%
\providecommand \@@href[1]{\endgroup#1\@@endlink}%
\providecommand \@sanitize@url [0]{\catcode `\\12\catcode `\$12\catcode
  `\&12\catcode `\#12\catcode `\^12\catcode `\_12\catcode `\%12\relax}%
\providecommand \@@startlink[1]{}%
\providecommand \@@endlink[0]{}%
\providecommand \url  [0]{\begingroup\@sanitize@url \@url }%
\providecommand \@url [1]{\endgroup\@href {#1}{\urlprefix }}%
\providecommand \urlprefix  [0]{URL }%
\providecommand \Eprint [0]{\href }%
\providecommand \doibase [0]{http://dx.doi.org/}%
\providecommand \selectlanguage [0]{\@gobble}%
\providecommand \bibinfo  [0]{\@secondoftwo}%
\providecommand \bibfield  [0]{\@secondoftwo}%
\providecommand \translation [1]{[#1]}%
\providecommand \BibitemOpen [0]{}%
\providecommand \bibitemStop [0]{}%
\providecommand \bibitemNoStop [0]{.\EOS\space}%
\providecommand \EOS [0]{\spacefactor3000\relax}%
\providecommand \BibitemShut  [1]{\csname bibitem#1\endcsname}%
\let\auto@bib@innerbib\@empty
\bibitem [{\citenamefont {Wilson}(1975)}]{RG_Wilson}%
  \BibitemOpen
  \bibfield  {author} {\bibinfo {author} {\bibfnamefont {K.~G.}\ \bibnamefont
  {Wilson}},\ }\href {http://link.aps.org/doi/10.1103/RevModPhys.47.773}
  {\bibfield  {journal} {\bibinfo  {journal} {Rev. Mod. Phys.}\ }\textbf
  {\bibinfo {volume} {47}},\ \bibinfo {pages} {773} (\bibinfo {year}
  {1975})}\BibitemShut {NoStop}%
\bibitem [{\citenamefont {Fisher}(1974)}]{RG_Fisher1}%
  \BibitemOpen
  \bibfield  {author} {\bibinfo {author} {\bibfnamefont {M.~E.}\ \bibnamefont
  {Fisher}},\ }\href {http://link.aps.org/doi/10.1103/RevModPhys.46.597}
  {\bibfield  {journal} {\bibinfo  {journal} {Rev. Mod. Phys.}\ }\textbf
  {\bibinfo {volume} {46}},\ \bibinfo {pages} {597} (\bibinfo {year}
  {1974})}\BibitemShut {NoStop}%
\bibitem [{\citenamefont {Fisher}(1998)}]{RG_Fisher2}%
  \BibitemOpen
  \bibfield  {author} {\bibinfo {author} {\bibfnamefont {M.~E.}\ \bibnamefont
  {Fisher}},\ }\href {http://link.aps.org/doi/10.1103/RevModPhys.70.653}
  {\bibfield  {journal} {\bibinfo  {journal} {Rev. Mod. Phys.}\ }\textbf
  {\bibinfo {volume} {70}},\ \bibinfo {pages} {653} (\bibinfo {year}
  {1998})}\BibitemShut {NoStop}%
\bibitem [{\citenamefont {Zinn-Justin}(1996)}]{RG_ZJ}%
  \BibitemOpen
  \bibfield  {author} {\bibinfo {author} {\bibfnamefont {J.}~\bibnamefont
  {Zinn-Justin}},\ }\href@noop {} {\emph {\bibinfo {title} {{Quantum Field
  Theory and Critical Phenomena}}}}\ (\bibinfo  {publisher} {Oxford University
  Press},\ \bibinfo {year} {1996})\BibitemShut {NoStop}%
\bibitem [{\citenamefont {White}(1992)}]{DMRG1}%
  \BibitemOpen
  \bibfield  {author} {\bibinfo {author} {\bibfnamefont {S.~R.}\ \bibnamefont
  {White}},\ }\href {http://link.aps.org/doi/10.1103/PhysRevLett.69.2863}
  {\bibfield  {journal} {\bibinfo  {journal} {Phys. Rev. Lett.}\ }\textbf
  {\bibinfo {volume} {69}},\ \bibinfo {pages} {2863} (\bibinfo {year}
  {1992})}\BibitemShut {NoStop}%
\bibitem [{\citenamefont {White}(1993)}]{DMRG2}%
  \BibitemOpen
  \bibfield  {author} {\bibinfo {author} {\bibfnamefont {S.~R.}\ \bibnamefont
  {White}},\ }\href {http://prb.aps.org/abstract/PRB/v48/i14/p10345_1}
  {\bibfield  {journal} {\bibinfo  {journal} {Phys. Rev. B}\ }\textbf {\bibinfo
  {volume} {48}},\ \bibinfo {pages} {10345} (\bibinfo {year}
  {1993})}\BibitemShut {NoStop}%
\bibitem [{\citenamefont {Fannes}\ \emph {et~al.}(1992)\citenamefont {Fannes},
  \citenamefont {Nachtergaele},\ and\ \citenamefont {Werner}}]{MPS1_FNW}%
  \BibitemOpen
  \bibfield  {author} {\bibinfo {author} {\bibfnamefont {M.}~\bibnamefont
  {Fannes}}, \bibinfo {author} {\bibfnamefont {B.}~\bibnamefont
  {Nachtergaele}}, \ and\ \bibinfo {author} {\bibfnamefont {R.}~\bibnamefont
  {Werner}},\ }\href
  {http://www.springerlink.com/content/787321531240h148/abstract} {\bibfield
  {journal} {\bibinfo  {journal} {Comm. Math. Phys.}\ }\textbf {\bibinfo
  {volume} {144}},\ \bibinfo {pages} {443} (\bibinfo {year}
  {1992})}\BibitemShut {NoStop}%
\bibitem [{\citenamefont {Kl{\"u}mper}\ \emph {et~al.}(1993)\citenamefont
  {Kl{\"u}mper}, \citenamefont {Schadschneider},\ and\ \citenamefont
  {Zittartz}}]{MPS2_KSZ}%
  \BibitemOpen
  \bibfield  {author} {\bibinfo {author} {\bibfnamefont {A.}~\bibnamefont
  {Kl{\"u}mper}}, \bibinfo {author} {\bibfnamefont {A.}~\bibnamefont
  {Schadschneider}}, \ and\ \bibinfo {author} {\bibfnamefont {J.}~\bibnamefont
  {Zittartz}},\ }\href {http://iopscience.iop.org/0295-5075/24/4/010}
  {\bibfield  {journal} {\bibinfo  {journal} {Europhys. Lett.}\ }\textbf
  {\bibinfo {volume} {24}},\ \bibinfo {pages} {293} (\bibinfo {year}
  {1993})}\BibitemShut {NoStop}%
\bibitem [{\citenamefont {{\"O}stlund}\ and\ \citenamefont
  {Rommer}(1995)}]{MPS3_OR}%
  \BibitemOpen
  \bibfield  {author} {\bibinfo {author} {\bibfnamefont {S.}~\bibnamefont
  {{\"O}stlund}}\ and\ \bibinfo {author} {\bibfnamefont {S.}~\bibnamefont
  {Rommer}},\ }\href {http://link.aps.org/doi/10.1103/PhysRevLett.75.3537}
  {\bibfield  {journal} {\bibinfo  {journal} {Phys. Rev. Lett.}\ }\textbf
  {\bibinfo {volume} {75}},\ \bibinfo {pages} {3537} (\bibinfo {year}
  {1995})}\BibitemShut {NoStop}%
\bibitem [{\citenamefont {Perez-Garcia}\ \emph {et~al.}(2007)\citenamefont
  {Perez-Garcia}, \citenamefont {Verstraete}, \citenamefont {Wolf},\ and\
  \citenamefont {Cirac}}]{MPS4_PVWC}%
  \BibitemOpen
  \bibfield  {author} {\bibinfo {author} {\bibfnamefont {D.}~\bibnamefont
  {Perez-Garcia}}, \bibinfo {author} {\bibfnamefont {F.}~\bibnamefont
  {Verstraete}}, \bibinfo {author} {\bibfnamefont {M.~M.}\ \bibnamefont
  {Wolf}}, \ and\ \bibinfo {author} {\bibfnamefont {J.~I.}\ \bibnamefont
  {Cirac}},\ }\href {http://www.rintonpress.com/journals/qiconline.html#v7n56}
  {\bibfield  {journal} {\bibinfo  {journal} {Quant. Inf. Comput.}\ }\textbf
  {\bibinfo {volume} {75}},\ \bibinfo {pages} {401} (\bibinfo {year}
  {2007})}\BibitemShut {NoStop}%
\bibitem [{\citenamefont {Verstraete}\ \emph {et~al.}(2008)\citenamefont
  {Verstraete}, \citenamefont {Murg},\ and\ \citenamefont {Cirac}}]{MPS5_VMC}%
  \BibitemOpen
  \bibfield  {author} {\bibinfo {author} {\bibfnamefont {F.}~\bibnamefont
  {Verstraete}}, \bibinfo {author} {\bibfnamefont {V.}~\bibnamefont {Murg}}, \
  and\ \bibinfo {author} {\bibfnamefont {J.~I.}\ \bibnamefont {Cirac}},\ }\href
  {http://www.tandfonline.com/doi/abs/10.1080/14789940801912366} {\bibfield
  {journal} {\bibinfo  {journal} {Adv. Phys.}\ }\textbf {\bibinfo {volume}
  {57}},\ \bibinfo {pages} {143} (\bibinfo {year} {2008})}\BibitemShut
  {NoStop}%
\bibitem [{\citenamefont {Schollw{\"o}ck}(2011)}]{MPS6_S}%
  \BibitemOpen
  \bibfield  {author} {\bibinfo {author} {\bibfnamefont {U.}~\bibnamefont
  {Schollw{\"o}ck}},\ }\href
  {http://www.sciencedirect.com/science/article/pii/S0003491610001752}
  {\bibfield  {journal} {\bibinfo  {journal} {Ann. Phys.}\ }\textbf {\bibinfo
  {volume} {326}},\ \bibinfo {pages} {96} (\bibinfo {year} {2011})}\BibitemShut
  {NoStop}%
\bibitem [{\citenamefont {Haegeman}\ \emph
  {et~al.}(2013{\natexlab{a}})\citenamefont {Haegeman}, \citenamefont
  {Osborne},\ and\ \citenamefont {Verstraete}}]{MPSTP}%
  \BibitemOpen
  \bibfield  {author} {\bibinfo {author} {\bibfnamefont {J.}~\bibnamefont
  {Haegeman}}, \bibinfo {author} {\bibfnamefont {T.~J.}\ \bibnamefont
  {Osborne}}, \ and\ \bibinfo {author} {\bibfnamefont {F.}~\bibnamefont
  {Verstraete}},\ }\href {http://link.aps.org/doi/10.1103/PhysRevB.88.075133}
  {\bibfield  {journal} {\bibinfo  {journal} {Phys. Rev. B}\ }\textbf {\bibinfo
  {volume} {88}},\ \bibinfo {pages} {075133} (\bibinfo {year}
  {2013}{\natexlab{a}})}\BibitemShut {NoStop}%
\bibitem [{\citenamefont {Or{\'u}s}(2014)}]{TN_Orus}%
  \BibitemOpen
  \bibfield  {author} {\bibinfo {author} {\bibfnamefont {R.}~\bibnamefont
  {Or{\'u}s}},\ }\href
  {http://www.sciencedirect.com/science/article/pii/S0003491614001596}
  {\bibfield  {journal} {\bibinfo  {journal} {Ann. Phys.}\ }\textbf {\bibinfo
  {volume} {349}},\ \bibinfo {pages} {117} (\bibinfo {year}
  {2014})}\BibitemShut {NoStop}%
\bibitem [{\citenamefont {Bridgeman}\ and\ \citenamefont
  {Chubb}(2016)}]{TN_BC}%
  \BibitemOpen
  \bibfield  {author} {\bibinfo {author} {\bibfnamefont {J.}~\bibnamefont
  {Bridgeman}}\ and\ \bibinfo {author} {\bibfnamefont {C.~T.}\ \bibnamefont
  {Chubb}},\ }\href@noop {} {\enquote {\bibinfo {title} {{Hand-waving and
  Interpretive Dance: An Introductory Course on Tensor Networks}},}\ }
  (\bibinfo {year} {2016}),\ \Eprint {http://arxiv.org/abs/1603.03039}
  {arXiv:1603.03039} \BibitemShut {NoStop}%
\bibitem [{\citenamefont {Hastings}(2007{\natexlab{a}})}]{AreaLaw_Hastings}%
  \BibitemOpen
  \bibfield  {author} {\bibinfo {author} {\bibfnamefont {M.}~\bibnamefont
  {Hastings}},\ }\href {http://iopscience.iop.org/1742-5468/2007/08/P08024}
  {\bibfield  {journal} {\bibinfo  {journal} {J. Stat. Mech. Theor. Exp.}\
  }\textbf {\bibinfo {volume} {2007}},\ \bibinfo {pages} {P08024} (\bibinfo
  {year} {2007}{\natexlab{a}})}\BibitemShut {NoStop}%
\bibitem [{\citenamefont {Eisert}\ \emph {et~al.}(2010)\citenamefont {Eisert},
  \citenamefont {Cramer},\ and\ \citenamefont {Plenio}}]{AreaLaw_Eisert}%
  \BibitemOpen
  \bibfield  {author} {\bibinfo {author} {\bibfnamefont {J.}~\bibnamefont
  {Eisert}}, \bibinfo {author} {\bibfnamefont {M.}~\bibnamefont {Cramer}}, \
  and\ \bibinfo {author} {\bibfnamefont {M.~B.}\ \bibnamefont {Plenio}},\
  }\href {http://link.aps.org/doi/10.1103/RevModPhys.82.277} {\bibfield
  {journal} {\bibinfo  {journal} {Rev. Mod. Phys.}\ }\textbf {\bibinfo {volume}
  {82}},\ \bibinfo {pages} {277} (\bibinfo {year} {2010})}\BibitemShut
  {NoStop}%
\bibitem [{\citenamefont {Verstraete}\ and\ \citenamefont
  {Cirac}(2006)}]{MPS_faithful}%
  \BibitemOpen
  \bibfield  {author} {\bibinfo {author} {\bibfnamefont {F.}~\bibnamefont
  {Verstraete}}\ and\ \bibinfo {author} {\bibfnamefont {J.~I.}\ \bibnamefont
  {Cirac}},\ }\href {http://link.aps.org/doi/10.1103/PhysRevB.73.094423}
  {\bibfield  {journal} {\bibinfo  {journal} {Phys. Rev. B}\ }\textbf {\bibinfo
  {volume} {73}},\ \bibinfo {pages} {094423} (\bibinfo {year}
  {2006})}\BibitemShut {NoStop}%
\bibitem [{\citenamefont {Hastings}(2007{\natexlab{b}})}]{MPS_Hastings}%
  \BibitemOpen
  \bibfield  {author} {\bibinfo {author} {\bibfnamefont {M.~B.}\ \bibnamefont
  {Hastings}},\ }\href {http://link.aps.org/doi/10.1103/PhysRevB.76.035114}
  {\bibfield  {journal} {\bibinfo  {journal} {Phys. Rev. B}\ }\textbf {\bibinfo
  {volume} {76}},\ \bibinfo {pages} {035114} (\bibinfo {year}
  {2007}{\natexlab{b}})}\BibitemShut {NoStop}%
\bibitem [{\citenamefont {Haegeman}\ \emph
  {et~al.}(2013{\natexlab{b}})\citenamefont {Haegeman}, \citenamefont
  {Michalakis}, \citenamefont {Nachtergaele}, \citenamefont {Osborne},
  \citenamefont {Schuch},\ and\ \citenamefont {Verstraete}}]{MPS_excitations}%
  \BibitemOpen
  \bibfield  {author} {\bibinfo {author} {\bibfnamefont {J.}~\bibnamefont
  {Haegeman}}, \bibinfo {author} {\bibfnamefont {S.}~\bibnamefont
  {Michalakis}}, \bibinfo {author} {\bibfnamefont {B.}~\bibnamefont
  {Nachtergaele}}, \bibinfo {author} {\bibfnamefont {T.~J.}\ \bibnamefont
  {Osborne}}, \bibinfo {author} {\bibfnamefont {N.}~\bibnamefont {Schuch}}, \
  and\ \bibinfo {author} {\bibfnamefont {F.}~\bibnamefont {Verstraete}},\
  }\href {http://link.aps.org/doi/10.1103/PhysRevLett.111.080401} {\bibfield
  {journal} {\bibinfo  {journal} {Phys. Rev. Lett.}\ }\textbf {\bibinfo
  {volume} {111}},\ \bibinfo {pages} {080401} (\bibinfo {year}
  {2013}{\natexlab{b}})}\BibitemShut {NoStop}%
\bibitem [{\citenamefont {Verstraete}\ \emph
  {et~al.}(2004{\natexlab{a}})\citenamefont {Verstraete}, \citenamefont
  {Porras},\ and\ \citenamefont {Cirac}}]{DMRG_variational}%
  \BibitemOpen
  \bibfield  {author} {\bibinfo {author} {\bibfnamefont {F.}~\bibnamefont
  {Verstraete}}, \bibinfo {author} {\bibfnamefont {D.}~\bibnamefont {Porras}},
  \ and\ \bibinfo {author} {\bibfnamefont {J.~I.}\ \bibnamefont {Cirac}},\
  }\href {http://link.aps.org/doi/10.1103/PhysRevLett.93.227205} {\bibfield
  {journal} {\bibinfo  {journal} {Phys. Rev. Lett.}\ }\textbf {\bibinfo
  {volume} {93}},\ \bibinfo {pages} {227205} (\bibinfo {year}
  {2004}{\natexlab{a}})}\BibitemShut {NoStop}%
\bibitem [{\citenamefont {White}(2005)}]{DMRG_SS}%
  \BibitemOpen
  \bibfield  {author} {\bibinfo {author} {\bibfnamefont {S.~R.}\ \bibnamefont
  {White}},\ }\href {http://link.aps.org/doi/10.1103/PhysRevB.72.180403}
  {\bibfield  {journal} {\bibinfo  {journal} {Phys. Rev. B}\ }\textbf {\bibinfo
  {volume} {72}},\ \bibinfo {pages} {180403} (\bibinfo {year}
  {2005})}\BibitemShut {NoStop}%
\bibitem [{\citenamefont {McCulloch}(2007)}]{DMRG_MC}%
  \BibitemOpen
  \bibfield  {author} {\bibinfo {author} {\bibfnamefont {I.~P.}\ \bibnamefont
  {McCulloch}},\ }\href
  {http://iopscience.iop.org/1742-5468/2007/10/P10014?fromSearchPage=true}
  {\bibfield  {journal} {\bibinfo  {journal} {J. Stat. Mech.}\ ,\ \bibinfo
  {pages} {P10014}} (\bibinfo {year} {2007})}\BibitemShut {NoStop}%
\bibitem [{\citenamefont {McCulloch}(2008)}]{IDMRG}%
  \BibitemOpen
  \bibfield  {author} {\bibinfo {author} {\bibfnamefont {I.~P.}\ \bibnamefont
  {McCulloch}},\ }\href@noop {} {\enquote {\bibinfo {title} {{Infinite size
  density matrix renormalization group, revisited}},}\ } (\bibinfo {year}
  {2008}),\ \Eprint {http://arxiv.org/abs/0804.2509} {arXiv:0804.2509}
  \BibitemShut {NoStop}%
\bibitem [{\citenamefont {Vidal}(2003)}]{TEBD}%
  \BibitemOpen
  \bibfield  {author} {\bibinfo {author} {\bibfnamefont {G.}~\bibnamefont
  {Vidal}},\ }\href {http://link.aps.org/doi/10.1103/PhysRevLett.91.147902}
  {\bibfield  {journal} {\bibinfo  {journal} {Phys. Rev. Lett.}\ }\textbf
  {\bibinfo {volume} {91}},\ \bibinfo {pages} {147902} (\bibinfo {year}
  {2003})}\BibitemShut {NoStop}%
\bibitem [{\citenamefont {Vidal}(2007)}]{ITEBD}%
  \BibitemOpen
  \bibfield  {author} {\bibinfo {author} {\bibfnamefont {G.}~\bibnamefont
  {Vidal}},\ }\href {http://link.aps.org/doi/10.1103/PhysRevLett.98.070201}
  {\bibfield  {journal} {\bibinfo  {journal} {Phys. Rev. Lett.}\ }\textbf
  {\bibinfo {volume} {98}},\ \bibinfo {pages} {070201} (\bibinfo {year}
  {2007})}\BibitemShut {NoStop}%
\bibitem [{\citenamefont {Haegeman}\ \emph {et~al.}(2011)\citenamefont
  {Haegeman}, \citenamefont {Cirac}, \citenamefont {Osborne}, \citenamefont
  {Pi\v{z}orn}, \citenamefont {Verschelde},\ and\ \citenamefont
  {Verstraete}}]{TDVP}%
  \BibitemOpen
  \bibfield  {author} {\bibinfo {author} {\bibfnamefont {J.}~\bibnamefont
  {Haegeman}}, \bibinfo {author} {\bibfnamefont {J.~I.}\ \bibnamefont {Cirac}},
  \bibinfo {author} {\bibfnamefont {T.~J.}\ \bibnamefont {Osborne}}, \bibinfo
  {author} {\bibfnamefont {I.}~\bibnamefont {Pi\v{z}orn}}, \bibinfo {author}
  {\bibfnamefont {H.}~\bibnamefont {Verschelde}}, \ and\ \bibinfo {author}
  {\bibfnamefont {F.}~\bibnamefont {Verstraete}},\ }\href
  {http://link.aps.org/doi/10.1103/PhysRevLett.107.070601} {\bibfield
  {journal} {\bibinfo  {journal} {Phys. Rev. Lett.}\ }\textbf {\bibinfo
  {volume} {107}},\ \bibinfo {pages} {070601} (\bibinfo {year}
  {2011})}\BibitemShut {NoStop}%
\bibitem [{\citenamefont {Haegeman}\ \emph {et~al.}(2016)\citenamefont
  {Haegeman}, \citenamefont {Lubich}, \citenamefont {Oseledets}, \citenamefont
  {Vandereycken},\ and\ \citenamefont {Verstraete}}]{TDVP_Uni}%
  \BibitemOpen
  \bibfield  {author} {\bibinfo {author} {\bibfnamefont {J.}~\bibnamefont
  {Haegeman}}, \bibinfo {author} {\bibfnamefont {C.}~\bibnamefont {Lubich}},
  \bibinfo {author} {\bibfnamefont {I.}~\bibnamefont {Oseledets}}, \bibinfo
  {author} {\bibfnamefont {B.}~\bibnamefont {Vandereycken}}, \ and\ \bibinfo
  {author} {\bibfnamefont {F.}~\bibnamefont {Verstraete}},\ }\href
  {http://link.aps.org/doi/10.1103/PhysRevB.94.165116} {\bibfield  {journal}
  {\bibinfo  {journal} {Phys. Rev. B}\ }\textbf {\bibinfo {volume} {94}},\
  \bibinfo {pages} {165116} (\bibinfo {year} {2016})}\BibitemShut {NoStop}%
\bibitem [{Note1()}]{Note1}%
  \BibitemOpen
  \bibinfo {note} {In many cases $h_{j,j+1}$ is sparse and the number $d_{h}$
  of non-zero elements is usually of the order $\protect \mathcal {O}(d^{2})$.
  The first two terms can then be applied in $\protect \mathcal
  {O}(d_{h}D^{3})$ operations.}\BibitemShut {Stop}%
\bibitem [{\citenamefont {Bhatia}(1997)}]{Bhatia}%
  \BibitemOpen
  \bibfield  {author} {\bibinfo {author} {\bibfnamefont {R.}~\bibnamefont
  {Bhatia}},\ }\href@noop {} {\emph {\bibinfo {title} {{Matrix Analysis}}}}\
  (\bibinfo  {publisher} {Springer-Verlag},\ \bibinfo {year}
  {1997})\BibitemShut {NoStop}%
\bibitem [{Note2()}]{Note2}%
  \BibitemOpen
  \bibinfo {note} {These values can be different and depend on the subtraction
  scheme for the divergent energy expectation value. If $h \to \protect
  \mathaccentV {tilde}07E{h}$ is performed everywhere, we have
  $E_{A_C}=E_{C}=0$. If, in the case of nearest neighbor interactions, we only
  substitute $h \to \protect \mathaccentV {tilde}07E{h}$ in the construction of
  $H_L$ and $H_R$, but not in the local terms, we will have
  $E_{A_C}=2E_{C}=2e$.}\BibitemShut {Stop}%
\bibitem [{Note3()}]{Note3}%
  \BibitemOpen
  \bibinfo {note} {Further approximations to comparable accuracy can be made
  within the construction of the effective Hamiltonians, e.g.\ when determining
  $H_L$ and $H_R$ to precision $\epsilon _{\protect \rm S}$. There, the
  approximations $\protect \mathaccentV {tilde}07E{R}=CC^{\dagger }$ and
  $\protect \mathaccentV {tilde}07E{L}=C^{\dagger }C$ for the true $L$ and $R$
  needed for some of these operations are good enough, if $\epsilon _{\protect
  \rm S}$ is roughly of the same order of magnitude as $\epsilon _{\protect \rm
  prec}$.}\BibitemShut {Stop}%
\bibitem [{\citenamefont {Phien}\ \emph {et~al.}(2015)\citenamefont {Phien},
  \citenamefont {McCulloch},\ and\ \citenamefont {Vidal}}]{ITEBD_Phien}%
  \BibitemOpen
  \bibfield  {author} {\bibinfo {author} {\bibfnamefont {H.~N.}\ \bibnamefont
  {Phien}}, \bibinfo {author} {\bibfnamefont {I.~P.}\ \bibnamefont
  {McCulloch}}, \ and\ \bibinfo {author} {\bibfnamefont {G.}~\bibnamefont
  {Vidal}},\ }\href {http://link.aps.org/doi/10.1103/PhysRevB.91.115137}
  {\bibfield  {journal} {\bibinfo  {journal} {Phys. Rev. B}\ }\textbf {\bibinfo
  {volume} {91}},\ \bibinfo {pages} {115137} (\bibinfo {year}
  {2015})}\BibitemShut {NoStop}%
\bibitem [{\citenamefont {Pirvu}\ \emph {et~al.}(2010)\citenamefont {Pirvu},
  \citenamefont {Murg}, \citenamefont {Cirac},\ and\ \citenamefont
  {Verstraete}}]{MPO3}%
  \BibitemOpen
  \bibfield  {author} {\bibinfo {author} {\bibfnamefont {B.}~\bibnamefont
  {Pirvu}}, \bibinfo {author} {\bibfnamefont {V.}~\bibnamefont {Murg}},
  \bibinfo {author} {\bibfnamefont {J.~I.}\ \bibnamefont {Cirac}}, \ and\
  \bibinfo {author} {\bibfnamefont {F.}~\bibnamefont {Verstraete}},\ }\href
  {http://iopscience.iop.org/1367-2630/12/2/025012} {\bibfield  {journal}
  {\bibinfo  {journal} {New J. Phys.}\ }\textbf {\bibinfo {volume} {12}},\
  \bibinfo {pages} {025012} (\bibinfo {year} {2010})}\BibitemShut {NoStop}%
\bibitem [{\citenamefont {Zaletel}\ \emph {et~al.}(2015)\citenamefont
  {Zaletel}, \citenamefont {Mong}, \citenamefont {Karrasch}, \citenamefont
  {Moore},\ and\ \citenamefont {Pollmann}}]{LRITEBD}%
  \BibitemOpen
  \bibfield  {author} {\bibinfo {author} {\bibfnamefont {M.~P.}\ \bibnamefont
  {Zaletel}}, \bibinfo {author} {\bibfnamefont {R.~S.~K.}\ \bibnamefont
  {Mong}}, \bibinfo {author} {\bibfnamefont {C.}~\bibnamefont {Karrasch}},
  \bibinfo {author} {\bibfnamefont {J.~E.}\ \bibnamefont {Moore}}, \ and\
  \bibinfo {author} {\bibfnamefont {F.}~\bibnamefont {Pollmann}},\ }\href
  {http://link.aps.org/doi/10.1103/PhysRevB.91.165112} {\bibfield  {journal}
  {\bibinfo  {journal} {Phys. Rev. B}\ }\textbf {\bibinfo {volume} {91}},\
  \bibinfo {pages} {165112} (\bibinfo {year} {2015})}\BibitemShut {NoStop}%
\bibitem [{\citenamefont {Hastings}(2011)}]{ITEBD_Hastings}%
  \BibitemOpen
  \bibfield  {author} {\bibinfo {author} {\bibfnamefont {M.~B.}\ \bibnamefont
  {Hastings}},\ }\href
  {http://scitation.aip.org/content/aip/journal/jmp/50/9/10.1063/1.3149556}
  {\bibfield  {journal} {\bibinfo  {journal} {J. Math. Phys.}\ }\textbf
  {\bibinfo {volume} {50}},\ \bibinfo {pages} {095207} (\bibinfo {year}
  {2011})}\BibitemShut {NoStop}%
\bibitem [{\citenamefont {Singh}\ \emph {et~al.}(2011)\citenamefont {Singh},
  \citenamefont {Pfeifer},\ and\ \citenamefont {Vidal}}]{Sukhi_symmetries}%
  \BibitemOpen
  \bibfield  {author} {\bibinfo {author} {\bibfnamefont {S.}~\bibnamefont
  {Singh}}, \bibinfo {author} {\bibfnamefont {R.~N.~C.}\ \bibnamefont
  {Pfeifer}}, \ and\ \bibinfo {author} {\bibfnamefont {G.}~\bibnamefont
  {Vidal}},\ }\href {http://link.aps.org/doi/10.1103/PhysRevB.83.115125}
  {\bibfield  {journal} {\bibinfo  {journal} {Phys. Rev. B}\ }\textbf {\bibinfo
  {volume} {83}},\ \bibinfo {pages} {115125} (\bibinfo {year}
  {2011})}\BibitemShut {NoStop}%
\bibitem [{\citenamefont {Pfeuty}(1970)}]{Pfeuty}%
  \BibitemOpen
  \bibfield  {author} {\bibinfo {author} {\bibfnamefont {P.}~\bibnamefont
  {Pfeuty}},\ }\href
  {http://www.sciencedirect.com/science/article/B6WB1-4DF54PD-25T/2/49df3093573535fd4cf7987fb9d8199c}
  {\bibfield  {journal} {\bibinfo  {journal} {Ann. Phys.}\ }\textbf {\bibinfo
  {volume} {57}},\ \bibinfo {pages} {79} (\bibinfo {year} {1970})}\BibitemShut
  {NoStop}%
\bibitem [{\citenamefont {Takahashi}(1999)}]{Takahashi}%
  \BibitemOpen
  \bibfield  {author} {\bibinfo {author} {\bibfnamefont {M.}~\bibnamefont
  {Takahashi}},\ }\href@noop {} {\emph {\bibinfo {title} {{Thermodynamics of
  one-dimensional solvable models}}}}\ (\bibinfo  {publisher} {Cambridge
  University Press},\ \bibinfo {year} {1999})\BibitemShut {NoStop}%
\bibitem [{\citenamefont {Lieb}\ and\ \citenamefont {Wu}(1968)}]{Lieb_Wu68}%
  \BibitemOpen
  \bibfield  {author} {\bibinfo {author} {\bibfnamefont {E.~H.}\ \bibnamefont
  {Lieb}}\ and\ \bibinfo {author} {\bibfnamefont {F.~Y.}\ \bibnamefont {Wu}},\
  }\href {http://link.aps.org/doi/10.1103/PhysRevLett.20.1445} {\bibfield
  {journal} {\bibinfo  {journal} {Phys. Rev. Lett.}\ }\textbf {\bibinfo
  {volume} {20}},\ \bibinfo {pages} {1445} (\bibinfo {year}
  {1968})}\BibitemShut {NoStop}%
\bibitem [{\citenamefont {Essler}\ \emph {et~al.}(2005)\citenamefont {Essler},
  \citenamefont {Frahm}, \citenamefont {G{\"o}hmann}, \citenamefont
  {Kl{\"u}mper},\ and\ \citenamefont {Korepin}}]{Essler}%
  \BibitemOpen
  \bibfield  {author} {\bibinfo {author} {\bibfnamefont {F.~H.~L.}\
  \bibnamefont {Essler}}, \bibinfo {author} {\bibfnamefont {H.}~\bibnamefont
  {Frahm}}, \bibinfo {author} {\bibfnamefont {F.}~\bibnamefont {G{\"o}hmann}},
  \bibinfo {author} {\bibfnamefont {A.}~\bibnamefont {Kl{\"u}mper}}, \ and\
  \bibinfo {author} {\bibfnamefont {V.}~\bibnamefont {Korepin}},\ }\href@noop
  {} {\emph {\bibinfo {title} {{The One-Dimensional Hubbard Model}}}}\
  (\bibinfo  {publisher} {Cambridge University Press},\ \bibinfo {year}
  {2005})\BibitemShut {NoStop}%
\bibitem [{\citenamefont {Haldane}(1988)}]{Haldane88}%
  \BibitemOpen
  \bibfield  {author} {\bibinfo {author} {\bibfnamefont {F.~D.~M.}\
  \bibnamefont {Haldane}},\ }\href
  {http://link.aps.org/doi/10.1103/PhysRevLett.60.635} {\bibfield  {journal}
  {\bibinfo  {journal} {Phys. Rev. Lett.}\ }\textbf {\bibinfo {volume} {60}},\
  \bibinfo {pages} {635} (\bibinfo {year} {1988})}\BibitemShut {NoStop}%
\bibitem [{\citenamefont {Shastry}(1988)}]{Shastry88}%
  \BibitemOpen
  \bibfield  {author} {\bibinfo {author} {\bibfnamefont {B.~S.}\ \bibnamefont
  {Shastry}},\ }\href {http://link.aps.org/doi/10.1103/PhysRevLett.60.639}
  {\bibfield  {journal} {\bibinfo  {journal} {Phys. Rev. Lett.}\ }\textbf
  {\bibinfo {volume} {60}},\ \bibinfo {pages} {639} (\bibinfo {year}
  {1988})}\BibitemShut {NoStop}%
\bibitem [{\citenamefont {{Osorio Iregui}}\ \emph {et~al.}(2017)\citenamefont
  {{Osorio Iregui}}, \citenamefont {Troyer},\ and\ \citenamefont
  {Corboz}}]{Osorio17}%
  \BibitemOpen
  \bibfield  {author} {\bibinfo {author} {\bibfnamefont {J.}~\bibnamefont
  {{Osorio Iregui}}}, \bibinfo {author} {\bibfnamefont {M.}~\bibnamefont
  {Troyer}}, \ and\ \bibinfo {author} {\bibfnamefont {P.}~\bibnamefont
  {Corboz}},\ }\href@noop {} {\bibfield  {journal} {\bibinfo  {journal} {Phys.
  Rev. B}\ }\textbf {\bibinfo {volume} {96}},\ \bibinfo {pages} {115113}
  (\bibinfo {year} {2017})}\BibitemShut {NoStop}%
\bibitem [{\citenamefont {Haegeman}\ \emph {et~al.}(2012)\citenamefont
  {Haegeman}, \citenamefont {Pirvu}, \citenamefont {Weir}, \citenamefont
  {Cirac}, \citenamefont {Osborne}, \citenamefont {Verschelde},\ and\
  \citenamefont {Verstraete}}]{haegeman2012variational}%
  \BibitemOpen
  \bibfield  {author} {\bibinfo {author} {\bibfnamefont {J.}~\bibnamefont
  {Haegeman}}, \bibinfo {author} {\bibfnamefont {B.}~\bibnamefont {Pirvu}},
  \bibinfo {author} {\bibfnamefont {D.~J.}\ \bibnamefont {Weir}}, \bibinfo
  {author} {\bibfnamefont {J.~I.}\ \bibnamefont {Cirac}}, \bibinfo {author}
  {\bibfnamefont {T.~J.}\ \bibnamefont {Osborne}}, \bibinfo {author}
  {\bibfnamefont {H.}~\bibnamefont {Verschelde}}, \ and\ \bibinfo {author}
  {\bibfnamefont {F.}~\bibnamefont {Verstraete}},\ }\href
  {http://link.aps.org/doi/10.1103/PhysRevB.85.100408} {\bibfield  {journal}
  {\bibinfo  {journal} {Phys. Rev. B}\ }\textbf {\bibinfo {volume} {85}},\
  \bibinfo {pages} {100408} (\bibinfo {year} {2012})}\BibitemShut {NoStop}%
\bibitem [{\citenamefont {Vanderstraeten}\ \emph {et~al.}(2014)\citenamefont
  {Vanderstraeten}, \citenamefont {Haegeman}, \citenamefont {Osborne},\ and\
  \citenamefont {Verstraete}}]{vanderstraeten2014s}%
  \BibitemOpen
  \bibfield  {author} {\bibinfo {author} {\bibfnamefont {L.}~\bibnamefont
  {Vanderstraeten}}, \bibinfo {author} {\bibfnamefont {J.}~\bibnamefont
  {Haegeman}}, \bibinfo {author} {\bibfnamefont {T.~J.}\ \bibnamefont
  {Osborne}}, \ and\ \bibinfo {author} {\bibfnamefont {F.}~\bibnamefont
  {Verstraete}},\ }\href
  {http://link.aps.org/doi/10.1103/PhysRevLett.112.257202} {\bibfield
  {journal} {\bibinfo  {journal} {Phys. Rev. Lett.}\ }\textbf {\bibinfo
  {volume} {112}},\ \bibinfo {pages} {257202} (\bibinfo {year}
  {2014})}\BibitemShut {NoStop}%
\bibitem [{\citenamefont {Vanderstraeten}\ \emph {et~al.}(2015)\citenamefont
  {Vanderstraeten}, \citenamefont {Verstraete},\ and\ \citenamefont
  {Haegeman}}]{vanderstraeten2015scattering}%
  \BibitemOpen
  \bibfield  {author} {\bibinfo {author} {\bibfnamefont {L.}~\bibnamefont
  {Vanderstraeten}}, \bibinfo {author} {\bibfnamefont {F.}~\bibnamefont
  {Verstraete}}, \ and\ \bibinfo {author} {\bibfnamefont {J.}~\bibnamefont
  {Haegeman}},\ }\href {http://link.aps.org/doi/10.1103/PhysRevB.92.125136}
  {\bibfield  {journal} {\bibinfo  {journal} {Phys. Rev. B}\ }\textbf {\bibinfo
  {volume} {92}},\ \bibinfo {pages} {125136} (\bibinfo {year}
  {2015})}\BibitemShut {NoStop}%
\bibitem [{\citenamefont {Zauner-Stauber}\ \emph {et~al.}(tion)\citenamefont
  {Zauner-Stauber}, \citenamefont {McCulloch},\ and\ \citenamefont
  {Verstraete}}]{QN_implementation}%
  \BibitemOpen
  \bibfield  {author} {\bibinfo {author} {\bibfnamefont {V.}~\bibnamefont
  {Zauner-Stauber}}, \bibinfo {author} {\bibfnamefont {I.~P.}\ \bibnamefont
  {McCulloch}}, \ and\ \bibinfo {author} {\bibfnamefont {F.}~\bibnamefont
  {Verstraete}},\ }\href@noop {} {} (\bibinfo {year} {in
  preparation})\BibitemShut {NoStop}%
\bibitem [{\citenamefont {Tagliacozzo}\ \emph {et~al.}(2008)\citenamefont
  {Tagliacozzo}, \citenamefont {de~Oliveira}, \citenamefont {Iblisdir},\ and\
  \citenamefont {Latorre}}]{FES_Luca}%
  \BibitemOpen
  \bibfield  {author} {\bibinfo {author} {\bibfnamefont {L.}~\bibnamefont
  {Tagliacozzo}}, \bibinfo {author} {\bibfnamefont {T.~R.}\ \bibnamefont
  {de~Oliveira}}, \bibinfo {author} {\bibfnamefont {S.}~\bibnamefont
  {Iblisdir}}, \ and\ \bibinfo {author} {\bibfnamefont {J.~I.}\ \bibnamefont
  {Latorre}},\ }\href {\doibase 10.1103/PhysRevB.78.024410} {\bibfield
  {journal} {\bibinfo  {journal} {Phys. Rev. B}\ }\textbf {\bibinfo {volume}
  {78}},\ \bibinfo {pages} {024410} (\bibinfo {year} {2008})}\BibitemShut
  {NoStop}%
\bibitem [{\citenamefont {Pollmann}\ \emph {et~al.}(2009)\citenamefont
  {Pollmann}, \citenamefont {Mukerjee}, \citenamefont {Turner},\ and\
  \citenamefont {Moore}}]{FES_Pollmann}%
  \BibitemOpen
  \bibfield  {author} {\bibinfo {author} {\bibfnamefont {F.}~\bibnamefont
  {Pollmann}}, \bibinfo {author} {\bibfnamefont {S.}~\bibnamefont {Mukerjee}},
  \bibinfo {author} {\bibfnamefont {A.~M.}\ \bibnamefont {Turner}}, \ and\
  \bibinfo {author} {\bibfnamefont {J.~E.}\ \bibnamefont {Moore}},\ }\href
  {\doibase 10.1103/PhysRevLett.102.255701} {\bibfield  {journal} {\bibinfo
  {journal} {Phys. Rev. Lett.}\ }\textbf {\bibinfo {volume} {102}},\ \bibinfo
  {pages} {255701} (\bibinfo {year} {2009})}\BibitemShut {NoStop}%
\bibitem [{\citenamefont {Stojevic}\ \emph {et~al.}(2015)\citenamefont
  {Stojevic}, \citenamefont {Haegeman}, \citenamefont {McCulloch},
  \citenamefont {Tagliacozzo},\ and\ \citenamefont {Verstraete}}]{FES_Vid}%
  \BibitemOpen
  \bibfield  {author} {\bibinfo {author} {\bibfnamefont {V.}~\bibnamefont
  {Stojevic}}, \bibinfo {author} {\bibfnamefont {J.}~\bibnamefont {Haegeman}},
  \bibinfo {author} {\bibfnamefont {I.~P.}\ \bibnamefont {McCulloch}}, \bibinfo
  {author} {\bibfnamefont {L.}~\bibnamefont {Tagliacozzo}}, \ and\ \bibinfo
  {author} {\bibfnamefont {F.}~\bibnamefont {Verstraete}},\ }\href {\doibase
  10.1103/PhysRevB.91.035120} {\bibfield  {journal} {\bibinfo  {journal} {Phys.
  Rev. B}\ }\textbf {\bibinfo {volume} {91}},\ \bibinfo {pages} {035120}
  (\bibinfo {year} {2015})}\BibitemShut {NoStop}%
\bibitem [{\citenamefont {White}\ and\ \citenamefont
  {Huse}(1993)}]{White_Huse93}%
  \BibitemOpen
  \bibfield  {author} {\bibinfo {author} {\bibfnamefont {S.~R.}\ \bibnamefont
  {White}}\ and\ \bibinfo {author} {\bibfnamefont {D.~A.}\ \bibnamefont
  {Huse}},\ }\href {http://link.aps.org/doi/10.1103/PhysRevB.48.3844}
  {\bibfield  {journal} {\bibinfo  {journal} {Phys. Rev. B}\ }\textbf {\bibinfo
  {volume} {48}},\ \bibinfo {pages} {3844} (\bibinfo {year}
  {1993})}\BibitemShut {NoStop}%
\bibitem [{\citenamefont {Legeza}\ and\ \citenamefont
  {F{\'a}th}(1996)}]{Legeza_Fath96}%
  \BibitemOpen
  \bibfield  {author} {\bibinfo {author} {\bibfnamefont {{\"O}.}~\bibnamefont
  {Legeza}}\ and\ \bibinfo {author} {\bibfnamefont {G.}~\bibnamefont
  {F{\'a}th}},\ }\href {http://link.aps.org/doi/10.1103/PhysRevB.53.14349}
  {\bibfield  {journal} {\bibinfo  {journal} {Phys. Rev. B}\ }\textbf {\bibinfo
  {volume} {53}},\ \bibinfo {pages} {14349} (\bibinfo {year}
  {1996})}\BibitemShut {NoStop}%
\bibitem [{Note4()}]{Note4}%
  \BibitemOpen
  \bibinfo {note} {The fact that ITEBD usually yields better variational states
  than IDMRG for the same bond dimension has already been observed in
  Ref.~\protect \rev@citealpnum {IDMRG}.}\BibitemShut {Stop}%
\bibitem [{\citenamefont {{Osorio Iregui}}(tion)}]{OsorioPriv}%
  \BibitemOpen
  \bibfield  {author} {\bibinfo {author} {\bibfnamefont {J.}~\bibnamefont
  {{Osorio Iregui}}},\ }\href@noop {} {} (\bibinfo {year} {private
  communication})\BibitemShut {NoStop}%
\bibitem [{Note5()}]{Note5}%
  \BibitemOpen
  \bibinfo {note} {IDMRG simulations have been performed without a rotation by
  $\pi $ of every second spin and with a $N=2W$ unit cell to accommodate an
  antiferromagnetically ordered ground state.\cite {OsorioPriv}}\BibitemShut
  {NoStop}%
\bibitem [{\citenamefont {Halimeh}\ and\ \citenamefont
  {Zauner-Stauber}(2016)}]{DQPT1}%
  \BibitemOpen
  \bibfield  {author} {\bibinfo {author} {\bibfnamefont {J.~C.}\ \bibnamefont
  {Halimeh}}\ and\ \bibinfo {author} {\bibfnamefont {V.}~\bibnamefont
  {Zauner-Stauber}},\ }\href@noop {} {\enquote {\bibinfo {title} {{Enriching
  the dynamical phase diagram of spin chains with long-range interactions}},}\
  } (\bibinfo {year} {2016}),\ \Eprint {http://arxiv.org/abs/1610.02019}
  {arXiv:1610.02019} \BibitemShut {NoStop}%
\bibitem [{\citenamefont {Halimeh}\ \emph {et~al.}(2017)\citenamefont
  {Halimeh}, \citenamefont {Zauner-Stauber}, \citenamefont {McCulloch},
  \citenamefont {de~Vega}, \citenamefont {Schollw{\"o}ck},\ and\ \citenamefont
  {Kastner}}]{DQPT2}%
  \BibitemOpen
  \bibfield  {author} {\bibinfo {author} {\bibfnamefont {J.~C.}\ \bibnamefont
  {Halimeh}}, \bibinfo {author} {\bibfnamefont {V.}~\bibnamefont
  {Zauner-Stauber}}, \bibinfo {author} {\bibfnamefont {I.~P.}\ \bibnamefont
  {McCulloch}}, \bibinfo {author} {\bibfnamefont {I.}~\bibnamefont {de~Vega}},
  \bibinfo {author} {\bibfnamefont {U.}~\bibnamefont {Schollw{\"o}ck}}, \ and\
  \bibinfo {author} {\bibfnamefont {M.}~\bibnamefont {Kastner}},\ }\href
  {http://link.aps.org/doi/10.1103/PhysRevB.95.024302} {\bibfield  {journal}
  {\bibinfo  {journal} {Phys. Rev. B}\ }\textbf {\bibinfo {volume} {95}},\
  \bibinfo {pages} {024302} (\bibinfo {year} {2017})}\BibitemShut {NoStop}%
\bibitem [{\citenamefont {Verstraete}\ and\ \citenamefont
  {Cirac}(2004)}]{PEPS}%
  \BibitemOpen
  \bibfield  {author} {\bibinfo {author} {\bibfnamefont {F.}~\bibnamefont
  {Verstraete}}\ and\ \bibinfo {author} {\bibfnamefont {J.~I.}\ \bibnamefont
  {Cirac}},\ }\href@noop {} {\enquote {\bibinfo {title} {{Renormalization
  algorithms for Quantum-Many Body Systems in two and higher dimensions}},}\ }
  (\bibinfo {year} {2004}),\ \Eprint {http://arxiv.org/abs/cond-mat/0407066}
  {arXiv:cond-mat/0407066} \BibitemShut {NoStop}%
\bibitem [{\citenamefont {Corboz}(2016)}]{PEPS_var_Corboz}%
  \BibitemOpen
  \bibfield  {author} {\bibinfo {author} {\bibfnamefont {P.}~\bibnamefont
  {Corboz}},\ }\href {http://link.aps.org/doi/10.1103/PhysRevB.94.035133}
  {\bibfield  {journal} {\bibinfo  {journal} {Phys. Rev. B}\ }\textbf {\bibinfo
  {volume} {94}},\ \bibinfo {pages} {035133} (\bibinfo {year}
  {2016})}\BibitemShut {NoStop}%
\bibitem [{\citenamefont {Vanderstraeten}\ \emph {et~al.}(2016)\citenamefont
  {Vanderstraeten}, \citenamefont {Haegeman}, \citenamefont {Corboz},\ and\
  \citenamefont {Verstraete}}]{PEPS_var_LV}%
  \BibitemOpen
  \bibfield  {author} {\bibinfo {author} {\bibfnamefont {L.}~\bibnamefont
  {Vanderstraeten}}, \bibinfo {author} {\bibfnamefont {J.}~\bibnamefont
  {Haegeman}}, \bibinfo {author} {\bibfnamefont {P.}~\bibnamefont {Corboz}}, \
  and\ \bibinfo {author} {\bibfnamefont {F.}~\bibnamefont {Verstraete}},\
  }\href {http://link.aps.org/doi/10.1103/PhysRevB.94.155123} {\bibfield
  {journal} {\bibinfo  {journal} {Phys. Rev. B}\ }\textbf {\bibinfo {volume}
  {94}},\ \bibinfo {pages} {155123} (\bibinfo {year} {2016})}\BibitemShut
  {NoStop}%
\bibitem [{\citenamefont {Verstraete}\ and\ \citenamefont
  {Cirac}(2010)}]{cMPS}%
  \BibitemOpen
  \bibfield  {author} {\bibinfo {author} {\bibfnamefont {F.}~\bibnamefont
  {Verstraete}}\ and\ \bibinfo {author} {\bibfnamefont {J.~I.}\ \bibnamefont
  {Cirac}},\ }\href@noop {} {\bibfield  {journal} {\bibinfo  {journal} {Phys.
  Rev. Lett.}\ }\textbf {\bibinfo {volume} {104}},\ \bibinfo {pages} {190405}
  (\bibinfo {year} {2010})}\BibitemShut {NoStop}%
\bibitem [{\citenamefont {Ganahl}\ \emph {et~al.}(2017)\citenamefont {Ganahl},
  \citenamefont {Rinc{\'o}n},\ and\ \citenamefont {Vidal}}]{CMPSGanahl2016}%
  \BibitemOpen
  \bibfield  {author} {\bibinfo {author} {\bibfnamefont {M.}~\bibnamefont
  {Ganahl}}, \bibinfo {author} {\bibfnamefont {J.}~\bibnamefont {Rinc{\'o}n}},
  \ and\ \bibinfo {author} {\bibfnamefont {G.}~\bibnamefont {Vidal}},\
  }\href@noop {} {\bibfield  {journal} {\bibinfo  {journal} {Phys. Rev. Lett.}\
  }\textbf {\bibinfo {volume} {118}},\ \bibinfo {pages} {220402} (\bibinfo
  {year} {2017})}\BibitemShut {NoStop}%
\bibitem [{\citenamefont {Haegeman}\ and\ \citenamefont
  {Verstraete}(2016)}]{MPO_review}%
  \BibitemOpen
  \bibfield  {author} {\bibinfo {author} {\bibfnamefont {J.}~\bibnamefont
  {Haegeman}}\ and\ \bibinfo {author} {\bibfnamefont {F.}~\bibnamefont
  {Verstraete}},\ }\href@noop {} {\enquote {\bibinfo {title} {{Diagonalizing
  transfer matrices and matrix product operators: a medley of exact and
  computational methods}},}\ } (\bibinfo {year} {2016}),\ \Eprint
  {http://arxiv.org/abs/1611.08519} {arXiv:1611.08519} \BibitemShut {NoStop}%
\bibitem [{\citenamefont {Haegeman}\ \emph {et~al.}(2014)\citenamefont
  {Haegeman}, \citenamefont {Mari{\"e}n}, \citenamefont {Osborne},\ and\
  \citenamefont {Verstraete}}]{haegeman2014geometry}%
  \BibitemOpen
  \bibfield  {author} {\bibinfo {author} {\bibfnamefont {J.}~\bibnamefont
  {Haegeman}}, \bibinfo {author} {\bibfnamefont {M.}~\bibnamefont
  {Mari{\"e}n}}, \bibinfo {author} {\bibfnamefont {T.}~\bibnamefont {Osborne}},
  \ and\ \bibinfo {author} {\bibfnamefont {F.}~\bibnamefont {Verstraete}},\
  }\href
  {http://scitation.aip.org/content/aip/journal/jmp/55/2/10.1063/1.4862851}
  {\bibfield  {journal} {\bibinfo  {journal} {J. Math. Phys.}\ }\textbf
  {\bibinfo {volume} {55}},\ \bibinfo {pages} {021902} (\bibinfo {year}
  {2014})}\BibitemShut {NoStop}%
\bibitem [{\citenamefont {Hubig}\ \emph {et~al.}(2015)\citenamefont {Hubig},
  \citenamefont {McCulloch}, \citenamefont {Schollw{\"o}ck},\ and\
  \citenamefont {Wolf}}]{bonddim2}%
  \BibitemOpen
  \bibfield  {author} {\bibinfo {author} {\bibfnamefont {C.}~\bibnamefont
  {Hubig}}, \bibinfo {author} {\bibfnamefont {I.~P.}\ \bibnamefont
  {McCulloch}}, \bibinfo {author} {\bibfnamefont {U.}~\bibnamefont
  {Schollw{\"o}ck}}, \ and\ \bibinfo {author} {\bibfnamefont {F.~A.}\
  \bibnamefont {Wolf}},\ }\href
  {http://link.aps.org/doi/10.1103/PhysRevB.91.155115} {\bibfield  {journal}
  {\bibinfo  {journal} {Phys. Rev. B}\ }\textbf {\bibinfo {volume} {91}},\
  \bibinfo {pages} {155115} (\bibinfo {year} {2015})}\BibitemShut {NoStop}%
\bibitem [{Note6()}]{Note6}%
  \BibitemOpen
  \bibinfo {note} {For fermionic Hamiltonians with long range interactions, we
  can employ a Jordan-Wigner transformation to spin operators, introducing a
  string operator counting the number of fermions between $o_{j}$ and
  $o_{j+n}$; geometric sums of transfer matrices in \protect \textup {\hbox
  {\mathsurround \z@ \protect \normalfont (\ignorespaces \ref
  {eq:edens_LR}\unskip \@@italiccorr )}} then turn into geometric sums of
  (string) operator transfer matrices.}\BibitemShut {Stop}%
\bibitem [{\citenamefont {Ruelle}(1968)}]{Ruelle68}%
  \BibitemOpen
  \bibfield  {author} {\bibinfo {author} {\bibfnamefont {D.}~\bibnamefont
  {Ruelle}},\ }\href {http://link.springer.com/article/10.1007/BF01654281}
  {\bibfield  {journal} {\bibinfo  {journal} {Comm. Math. Phys.}\ }\textbf
  {\bibinfo {volume} {9}},\ \bibinfo {pages} {267} (\bibinfo {year}
  {1968})}\BibitemShut {NoStop}%
\bibitem [{\citenamefont {Dyson}(1969)}]{Dyson69}%
  \BibitemOpen
  \bibfield  {author} {\bibinfo {author} {\bibfnamefont {F.~J.}\ \bibnamefont
  {Dyson}},\ }\href {http://link.springer.com/article/10.1007/BF01645907}
  {\bibfield  {journal} {\bibinfo  {journal} {Comm. Math. Phys.}\ }\textbf
  {\bibinfo {volume} {12}},\ \bibinfo {pages} {91} (\bibinfo {year}
  {1969})}\BibitemShut {NoStop}%
\bibitem [{\citenamefont {Cardy}(1981)}]{Cardy81}%
  \BibitemOpen
  \bibfield  {author} {\bibinfo {author} {\bibfnamefont {J.~L.}\ \bibnamefont
  {Cardy}},\ }\href {http://stacks.iop.org/0305-4470/14/i=6/a=017} {\bibfield
  {journal} {\bibinfo  {journal} {J. Phys. A: Math. Gen.}\ }\textbf {\bibinfo
  {volume} {144}},\ \bibinfo {pages} {1407} (\bibinfo {year}
  {1981})}\BibitemShut {NoStop}%
\bibitem [{\citenamefont {Verstraete}\ \emph
  {et~al.}(2004{\natexlab{b}})\citenamefont {Verstraete}, \citenamefont
  {Garcia-Ripoll},\ and\ \citenamefont {Cirac}}]{MPO1}%
  \BibitemOpen
  \bibfield  {author} {\bibinfo {author} {\bibfnamefont {F.}~\bibnamefont
  {Verstraete}}, \bibinfo {author} {\bibfnamefont {J.~J.}\ \bibnamefont
  {Garcia-Ripoll}}, \ and\ \bibinfo {author} {\bibfnamefont {J.~I.}\
  \bibnamefont {Cirac}},\ }\href
  {http://link.aps.org/doi/10.1103/PhysRevLett.93.207204} {\bibfield  {journal}
  {\bibinfo  {journal} {Phys. Rev. Lett.}\ }\textbf {\bibinfo {volume} {93}},\
  \bibinfo {pages} {207204} (\bibinfo {year} {2004}{\natexlab{b}})}\BibitemShut
  {NoStop}%
\bibitem [{\citenamefont {Crosswhite}\ and\ \citenamefont
  {Bacon}(2008)}]{MPO4}%
  \BibitemOpen
  \bibfield  {author} {\bibinfo {author} {\bibfnamefont {G.~M.}\ \bibnamefont
  {Crosswhite}}\ and\ \bibinfo {author} {\bibfnamefont {D.}~\bibnamefont
  {Bacon}},\ }\href {http://link.aps.org/doi/10.1103/PhysRevA.78.012356}
  {\bibfield  {journal} {\bibinfo  {journal} {Phys. Rev. A}\ }\textbf {\bibinfo
  {volume} {78}},\ \bibinfo {pages} {012356} (\bibinfo {year}
  {2008})}\BibitemShut {NoStop}%
\bibitem [{\citenamefont {Crosswhite}\ \emph {et~al.}(2008)\citenamefont
  {Crosswhite}, \citenamefont {Doherty},\ and\ \citenamefont
  {Vidal}}]{MPOlong1}%
  \BibitemOpen
  \bibfield  {author} {\bibinfo {author} {\bibfnamefont {G.~M.}\ \bibnamefont
  {Crosswhite}}, \bibinfo {author} {\bibfnamefont {A.~C.}\ \bibnamefont
  {Doherty}}, \ and\ \bibinfo {author} {\bibfnamefont {G.}~\bibnamefont
  {Vidal}},\ }\href {http://link.aps.org/doi/10.1103/PhysRevB.78.035116}
  {\bibfield  {journal} {\bibinfo  {journal} {Phys. Rev. B}\ }\textbf {\bibinfo
  {volume} {78}},\ \bibinfo {pages} {035116} (\bibinfo {year}
  {2008})}\BibitemShut {NoStop}%
\bibitem [{\citenamefont {Fr{\"o}wis}\ \emph {et~al.}(2010)\citenamefont
  {Fr{\"o}wis}, \citenamefont {Nebendahl},\ and\ \citenamefont
  {D{\"u}r}}]{MPOlong2}%
  \BibitemOpen
  \bibfield  {author} {\bibinfo {author} {\bibfnamefont {F.}~\bibnamefont
  {Fr{\"o}wis}}, \bibinfo {author} {\bibfnamefont {V.}~\bibnamefont
  {Nebendahl}}, \ and\ \bibinfo {author} {\bibfnamefont {W.}~\bibnamefont
  {D{\"u}r}},\ }\href {http://link.aps.org/doi/10.1103/PhysRevA.81.062337}
  {\bibfield  {journal} {\bibinfo  {journal} {Phys. Rev. A}\ }\textbf {\bibinfo
  {volume} {81}},\ \bibinfo {pages} {062337} (\bibinfo {year}
  {2010})}\BibitemShut {NoStop}%
\bibitem [{\citenamefont {Michel}\ and\ \citenamefont
  {McCulloch}(2010)}]{MPOinf}%
  \BibitemOpen
  \bibfield  {author} {\bibinfo {author} {\bibfnamefont {L.}~\bibnamefont
  {Michel}}\ and\ \bibinfo {author} {\bibfnamefont {I.~P.}\ \bibnamefont
  {McCulloch}},\ }\href@noop {} {\enquote {\bibinfo {title} {{Schur Forms of
  Matrix Product Operators in the Infinite Limit}},}\ } (\bibinfo {year}
  {2010}),\ \Eprint {http://arxiv.org/abs/1008.4667} {arXiv:1008.4667}
  \BibitemShut {NoStop}%
\bibitem [{Note7()}]{Note7}%
  \BibitemOpen
  \bibinfo {note} {The generalization to nontrivial operators on the diagonal,
  such as e.g. Pauli-$Z$ operators in the case of long range fermion hopping,
  is straight forward.}\BibitemShut {Stop}%
\bibitem [{\citenamefont {van~der Vorst}(1992)}]{BICGSTAB}%
  \BibitemOpen
  \bibfield  {author} {\bibinfo {author} {\bibfnamefont {H.}~\bibnamefont
  {van~der Vorst}},\ }\href {http://link.aip.org/link/?SCE/13/631/1} {\bibfield
   {journal} {\bibinfo  {journal} {SIAM J. Sci. Stat. Comput.}\ }\textbf
  {\bibinfo {volume} {13}},\ \bibinfo {pages} {631} (\bibinfo {year}
  {1992})}\BibitemShut {NoStop}%
\bibitem [{\citenamefont {Saad}\ and\ \citenamefont {Schultz}(1986)}]{GMRES}%
  \BibitemOpen
  \bibfield  {author} {\bibinfo {author} {\bibfnamefont {Y.}~\bibnamefont
  {Saad}}\ and\ \bibinfo {author} {\bibfnamefont {M.~H.}\ \bibnamefont
  {Schultz}},\ }\href {http://epubs.siam.org/doi/abs/10.1137/0907058}
  {\bibfield  {journal} {\bibinfo  {journal} {SIAM J. Sci. Stat. Comput.}\
  }\textbf {\bibinfo {volume} {75}},\ \bibinfo {pages} {856} (\bibinfo {year}
  {1986})}\BibitemShut {NoStop}%
\end{thebibliography}

\clearpage
\appendix

\section{Theoretical background}
\label{sec:theory}

In this Appendix we reiterate definitions and concepts needed for the algorithm presented in \Sec{sec:algorithm} in more detail, and motivate the VUMPS algorithm from a variational perspective.

\subsection{Variational principle on manifolds}

The variational principle in quantum mechanics characterizes the ground state of a given Hamiltonian as the state $\ket{\Psi}$ which minimizes the normalized energy expectation value
\begin{equation*}
E=\frac{\braket{\Psi|H|\Psi}}{\braket{\Psi|\Psi}}.
\end{equation*}
If (typically for computational reasons) we only have access to a subset of Hilbert space, the variational principle still gives a way to find an approximation to the true ground state, namely by solving the minimization problem within the restricted set. If this subset is a linear subspace spanned by a number of basis vectors $\{\ket{i},i=1,\ldots,N\}$, we obtain a generalized eigenvalue problem
\begin{equation*}
\braket{i|H|j} c_{j} = E \braket{i|j} c_{j}
\end{equation*}
for the expansion coefficients $c_i$ in $\ket{\Psi}=\sum_{i=1}^{N} c_i \ket{i}$. This is known as the Rayleigh-Ritz method, and by orthonormalizing the basis it clearly amounts to projecting the full time-independent Schr\"{o}dinger equation into the variational subspace.

If, more generally, we have a variational ansatz $\ket{\Psi(\bm{A})}$ which depends analytically on a number of complex parameters, as encoded in the complex vector $\bm{A}$, a variational minimum $\ket{\Psi(\bm{A}^\ast)}$ is characterized by a vanishing gradient of the energy expectation value, i.e.
\begin{equation}
\braket{\partial_{\overline{\imath}} \Psi(\bar{\bm{A}}^\ast)| H - E(\bar{\bm{A}}^\ast,\bm{A}^\ast) |\Psi(\bm{A}^\ast}=0,\label{eq:varmingalerkin}
\end{equation}
with $\bar{\bm{A}}$ the (formally independent) complex conjugate of $\bm{A}$, $\partial_i$ and $\partial_{\bar{\imath}}$ the complex derivatives with respect to the $i$'th component of $\bm{A}$ and $\bar{\bm{A}}$ and
\begin{equation*}
E(\bar{\bm{A}},\bm{A}) = \frac{\braket{\Psi(\bar{\bm{A}})| H |\Psi(\bm{A})}}{\braket{\Psi(\bar{\bm{A}})|\Psi(\bm{A})}}.
\end{equation*}
Eq.~\eqref{eq:varmingalerkin} can be interpreted as a Galerkin condition: It forces the residual $(H-E)\ket{\Psi}$ of the full Schr\"{o}dinger eigenvalue equation -- which does not have an exact solution in the variational subset -- to be orthogonal to the space spanned by the states $\ket{\partial_i \Psi(\bm{A}^\ast)}$. If the variational subset is a manifold, these states can be interpreted as a basis for the tangent space of the manifold at the point of the variational optimum. Hence, geometrically, the residual has to be orthogonal to the manifold (and thus to its tangent space) at the point of the variational optimum. Interpreting Eq.~\eqref{eq:varmingalerkin} as a Galerkin condition on the ground state eigenvalue problem is useful because it can be generalized to other eigenvalue problems which do not necessarily have a variational characterization (and thus no gradient), as e.g.\ when the operator is non-Hermitian. Indeed, a similar approach as is developed here was described for finding fixed points of transfer matrices, encoded as matrix product operators, in Ref.~\onlinecite{MPO_review}.

However, before discussing Eq.~\eqref{eq:varmingalerkin} in the context of MPS, let us conclude this section by relating it to the time-dependent variational principle (TDVP).\cite{TDVP} Geometrically, the TDVP also amounts to an orthogonal projection of the equation of motion (the time-dependent Schr\"{o}dinger equation) onto the tangent space of the variational manifold. In the case of imaginary time evolution, it can be written as
\begin{equation}
g_{\bar{\imath},j}(\overline{\bm{A}},\bm{A}) \frac{\mathrm{d}\ }{\mathrm{d} t} A^j =
- \braket{\partial_{\bar{\imath}} \Psi(\overline{\bm{A}}) | H - E(\overline{\bm{A}},\bm{A}) | \Psi(\bm{A})}
\label{eq:imagTDVP}
\end{equation}
where
\begin{equation*}
g_{\bar{\imath},j}(\overline{\bm{A}},\bm{A}) = \braket{\partial_{\bm{\imath}} \Psi(\overline{\bm{A}})|\partial_j \Psi(\bm{A})}	
\end{equation*}
is the Gram matrix of the tangent vectors and thus the metric of the manifold. The right hand side of Eq.~\eqref{eq:imagTDVP} is again the gradient of the objective function, and the TDVP will thus converge when it reaches a variational optimum $\bm{A}^\ast$ where Eq.~\eqref{eq:varmingalerkin} is satisfied. However, the metric $g_{\bar{\imath},j}$ in the left hand side shows that the TDVP equation is not a normal gradient flow, but rather a proper covariant gradient flow that takes the geometry of the manifold and its embedding into the Hilbert space into account. We can thus also associate a quantum state with the gradient, which is given by
\begin{equation}
\ket{\partial_i \Psi} g^{i,\bar{\jmath}}\braket{\partial_{\bar{\jmath}} \Psi | (H - E) | \Psi} = \P_{T_{\ket{\Psi}} \mathcal{M}} (H-E) \ket{\Psi}
\end{equation}
where we have omitted the arguments $\bm{A}$ and $\bar{\bm{A}}$, $g^{i,\bar{\jmath}}$ is the inverse of the metric and
\begin{equation}
\P_{T_{\ket{\Psi}} \mathcal{M}} = \ket{\partial_i \Psi} g^{i,\bar{\jmath}}\bra{\partial_{\bar{\jmath}}\Psi}
\end{equation}
is the projector onto the tangent space at the point $\ket{\Psi}$ in the variational manifold $\mathcal{M}$. This latter expression is only valid when the variational parameters are proper coordinates for the manifold (i.e.\ a bijective mapping). While this is not the case for MPS because of gauge freedom (see below), the geometrical interpretation for the Galerkin condition
\begin{equation} \label{eq:galerkin2}
\P_{T_{\ket{\Psi}} \mathcal{M}} (H-E) \ket{\Psi} = 0
\end{equation}
remains valid; the correct expression for the MPS tangent-space projector will be discussed in more detail in the following section. Independent of whether a variational algorithm is based on a gradient flow, the Hilbert space norm of the gradient $\lVert \P_{T_{\ket{\Psi}} \mathcal{M}} H \ket{\Psi}\rVert$ provides an objective measure for the convergence of the state towards the variational optimum. Note that this is different from the standard Euclidean norm of the naive gradient vector with components $\braket{\partial_{\bar{\jmath}} \Psi | H - E | \Psi} $.

\subsection{Manifold of Uniform MPS and its tangent space}
\label{sec:umps}
We have already introduced the set of uniform MPS and some of its properties in Sec~\ref{sec:umps_short}. For completeness, we restate the definition as
\begin{equation}
\label{eq:TH_MPS1}
\ket{\Psi(A)}=\sum_{\bm{s}}\vec{v}_{L}^{\dagger}\Big(\prod_{n\in\ints}A^{s_{n}}\Big)\vec{v}_{R}\ket{\bm{s}}.
\end{equation}
Such states can be rigorously defined in the thermodynamic limit and correspond to the so-called purely generated finitely correlated states of Ref.~\onlinecite{MPS1_FNW}. However, for all practical purposes, we can interpret this state as a large but finite MPS, which happens to be uniformly parameterized (and therefore translation invariant) in the bulk. We have now introduced boundary vectors $\bm{v}_L,\bm{v}_R$ living at $\pm \infty$. They represent just one way to close the matrix product near the boundaries, but any other behavior is equally fine and won't affect the bulk properties, provided that the MPS tensor $A$ has the property of injectivity.\cite{MPS4_PVWC} This condition is generically fulfilled and is in one-to-one correspondence with the MPS transfer matrix 
\begin{equation}
T=\sum_{s}\bar{A}^{s}\otimes A^{s}=
\begin{tikzpicture}[baseline = (X.base),every node/.style={scale=0.750},scale=.55]
\draw[rounded corners] (1,1.5) rectangle (2,.5);
\draw[rounded corners] (1,-1.5) rectangle (2,-.5);
\draw (1.5,1) node {$A$}; \draw (1.5,-1) node (X) {$\bar{A}$};
\draw (1.5,.5) -- (1.5,-.5); 
\draw (1,1) -- (.5,1); \draw (2,1) -- (2.5,1);
\draw (1,-1) -- (.5,-1); \draw (2,-1) -- (2.5,-1);
\draw (1.5,0) node (X) {$\phantom{X}$};
\end{tikzpicture}
\label{eq:MPSTM}
\end{equation} 
having a unique eigenvalue of largest magnitude, with corresponding eigenvectors that can be reinterpreted as full rank positive $D \times D$ matrices. A proper normalization of the state is obtained by rescaling the tensor $A$ such that this largest magnitude eigenvalue is $1$.

We can then use gauge invariance to transform the tensor $A$ into the left and right canonical representations \eqref{eq:gauges} discussed in the main text. These representations are themselves related via the bond matrix $C$ as $A_L^s C = C A_R^s$, which allows to write the state in the mixed canonical representation \eqref{eq:psi_mixed} familiar from DMRG. We can also make contact with the representation commonly used in TEBD\cite{TEBD,ITEBD} by additional unitary gauge transforms. We make $C$ diagonal by considering the SVD $C=U\lambda V^{\dagger}$ and gauge transform $\tilde{A}^{s}_{L}= U^{\dagger}\AL U$, $\tilde{A}_{R}^{s}= V^{\dagger}\AR V$. The singular values $\lambda$ are then the Schmidt values of a bipartition of the state and we can obtain $\Gamma^{s}=\lambda^{-1}\tilde{A}^{s}_{L}=\tilde{A}^{s}_{R}\lambda^{-1}$, or equivalently
\begin{equation}
 \tilde{A}^{s}_{C} = \tilde{A}^{s}_{L} \lambda = \lambda \tilde{A}^{s}_{R} = \lambda \Gamma^{s} \lambda,
\end{equation} 
and $L=R=\lambda^{2}$.

It was proven in Ref.~\onlinecite{haegeman2014geometry} that the set of injective uMPS constitute a (complex) manifold. We now construct the tangent space projector $\P_{\ket{\Psi(A)}} \equiv \P_{T_{\ket{\Psi(A)}}\mathcal{M}}$ for a uMPS $\ket{\Psi(A)}$ in the thermodynamic limit.

Let us first discuss generic tangent vectors $\ket{\Phi}$ in this tangent space. To define them via the partial derivatives of $\ket{\Psi(A)}$, we require that the latter is represented using a uniform parameterization or gauge choice of the tensor $A$ throughout. By applying the chain rule, the derivative will give rise to a uniform superposition of states where a single $A$ tensor is replaced by a new tensor $B$.\cite{MPSTP,haegeman2014geometry} But it is clear that we can afterwards change gauges again and  absorb the gauge factors in the tensor $B$ that parameterizes the tangent vector, so as to obtain the most general tangent vector representation
\begin{align}
 \ket{\Phi(B)} &= \sum_{n\in\ints}\sum_{\bm{s}} (\ldots A_{L}^{s_{n-2}}A_{L}^{s_{n-1}}B^{s_{n}}A_{R}^{s_{n+1}}A_{R}^{s_{n+2}})\ket{\bm{s}}\notag\\
  &= \sum_{n\in\ints}\sum_{s_{n},\alpha,\beta} B^{s_{n}}_{(\alpha,\beta)}\ket{\Psi_{A_{C}}^{(\alpha,s_{n},\beta)}(n)} \label{eq:TH_phiB}\\
  &= \sum_{n\in\ints} \dots
 \begin{tikzpicture}[baseline = (X.base),every node/.style={scale=0.75},scale=.5]
\mpsT{-4.5}{0}{A_L} \mpsT{-2.5}{0}{A_L} \mpsT{-0.5}{0}{B} \mpsT{1.5}{0}{A_R} \mpsT{3.5}{0}{A_R}
\draw (-4.5,-1.5) node {$\dots$}; \draw (-2.5,-1.5) node {$s_{n-1}$}; \draw (-0.5,-1.5) node {$s_n$}; 
\draw (1.5,-1.5) node {$s_{n+1}$}; \draw (3.5,-1.5) node {$\dots$};
\lineH{-5.5}{-5}{0} \lineH{-4}{-3}{0} \lineH{-2}{-1}{0} \lineH{0}{1}{0} \lineH{2}{3}{0} \lineH{4}{4.5}{0} 
\draw (-4.5,-.5) -- (-4.5,-1) ; \lineV{-.5}{-1}{-2.5} \lineV{-.5}{-1}{-0.5} \lineV{-.5}{-1}{1.5} \lineV{-.5}{-1}{3.5} 
\dobase{0} 
\end{tikzpicture}\dots \notag
\end{align} 
The multiplicative gauge freedom of the MPS translates into an additive gauge freedom in the tangent space, i.e.\ a transformation $B^{s}\to B^{s} + A_{L}^{s}X - XA_{R}^{s}$ with $X\in\mathbb{C}^{D\times D}$ leaves $\ket{\Phi(B)}$ invariant, as can readily be verified by explicit substitution. We can exploit these gauge degrees of freedom to impose e.g.\ the left tangent space gauge
\begin{equation}
\sum_s {A^{s}_{L}}^{\dagger} B^s 
= 
\begin{tikzpicture}[baseline = (X.base),every node/.style={scale=0.750},scale=.55]
\draw (1,-1.5) edge[out=180,in=180] (1,1.5);
\draw[rounded corners] (1,2) rectangle (2,1);
\draw[rounded corners] (1,-1) rectangle (2,-2);
\draw (1.5,1) -- (1.5,-1);
\draw (1.5,1.5) node {$B$};
\draw (1.5,-1.5) node {$\bar{A}_L$};
\draw (2,1.5) -- (2.5,1.5); \draw (2,-1.5) -- (2.5,-1.5);
\end{tikzpicture} 
= 0 
\label{eq:leftgaugefixingcondition} 
\end{equation} 
or the right tangent space gauge
\begin{equation}
\sum_s B^s {A^{s}_{R}}^{\dagger}  
= 
\begin{tikzpicture}[baseline = (X.base),every node/.style={scale=0.750},scale=.55]
\draw (2,-1.5) edge[out=0,in=0] (2,1.5);
\draw[rounded corners] (1,2) rectangle (2,1);
\draw[rounded corners] (1,-1) rectangle (2,-2);
\draw (1.5,1) -- (1.5,-1);
\draw (1.5,1.5) node {$B$};
\draw (1.5,-1.5) node {$\bar{A}_R$};
\draw (0.5,1.5) -- (1,1.5); 
\draw (0.5,-1.5) -- (1,-1.5);
\end{tikzpicture} 
=0.\label{eq:rightgaugefixingcondition}
\end{equation}
Strictly speaking, these conditions can only be imposed for tangent vectors $\ket{\Phi(B)}\perp \ket{\Psi(A)}$. This is no restriction as we can always evaluate the contribution in the direction of $\ket{\Psi(A)}$ separately. Under either of these two gauge constraints, we indeed have $\braket{\Psi(A)|\Phi(B)}=0$, but more importantly, the overlap between two tangent vectors simplifies to
\begin{equation} \label{eq:tangentspacenormalization}
\braket{\Phi(B_2)|\Phi(B_1)} = |\mathbb{Z}| \sum_s \Tr \left({B_2^s}^\dag B_1^s\right).
\end{equation}
This corresponds to an Euclidean inner product for the $B$ tensors and thus to an orthonormal basis for the tangent space. The diverging factor $|\mathbb{Z}|$ arises because a tangent vector contains a sum over all lattice sites, i.e.\ its norm is extensive. Fortunately, this diverging factor will drop out in all computations.

Given Eq.~\eqref{eq:galerkin2} we need to derive the explicit form of the projector onto the tangent space, a derivation that was written down in Ref.~\onlinecite{TDVP_Uni} for the case of finite MPS. The tangent vector $\ket{\Phi(B)} = \P_{\ket{\Psi(A)}} \ket{\Xi}$ resulting from the orthogonal projection of a general translation invariant state $\ket{\Xi}$ onto the tangent space can be readily found by solving the minimization problem
\begin{equation*}
\min_B \left\lVert \ket{\Xi} - \ket{\Phi(B)} \right\rVert^2,
\end{equation*}
or, equivalently,
\begin{equation*}
\min_B \left( \braket{\Phi(B)|\Phi(B)} - \braket{\Xi|\Phi(B)} - \braket{\Phi(B)|\Xi} \right).
\end{equation*}
In order to use Eq.~\eqref{eq:tangentspacenormalization} for the first term, we however need to impose the constraint in Eq.~\eqref{eq:leftgaugefixingcondition} or \eqref{eq:rightgaugefixingcondition}. In the former case, this will add a term $\Tr[\Lambda \sum_s (A_L^s)^\dagger B^s ] + \Tr[\bar{\Lambda} \sum_s (B^s)^\dagger A_L^s]$ to the objective function, with $\Lambda$ and $\bar{\Lambda}$ corresponding Lagrange multipliers. The solution is readily obtained by demanding $\partial_{\bar{B}} (\dots) = 0$, where Eq.~\eqref{eq:tangentspacenormalization} simply results in $\partial_{\bar{B}}\braket{\Phi(B)|\Phi(B)} = |\mathbb{Z}| B$. 
The overlap between a tangent vector and $\ket{\Xi}$ is given by
\begin{multline*}
\braket{\Phi(B)|\Xi} = | \mathbb{Z} | \\ \times \dots \begin{tikzpicture}[baseline = (X.base),every node/.style={scale=0.750},scale=.55]
\draw (1,3) -- (10,3); \draw (1,4) -- (10,4);
\draw (5.5,3.5) node {$\Xi$};
\draw (0.5,1.5) -- (1,1.5); 
\draw[rounded corners] (1,2) rectangle (2,1);
\draw (1.5,1.5) node (X) {$\bar{A}_L$};
\draw (2,1.5) -- (3,1.5); 
\draw[rounded corners] (3,2) rectangle (4,1);
\draw (3.5,1.5) node {$\bar{A}_L$};
\draw (4,1.5) -- (5,1.5);
\draw[rounded corners] (5,2) rectangle (6,1);
\draw (5.5,1.5) node {$\bar{B}$};
\draw (6,1.5) -- (7,1.5);
\draw[rounded corners] (7,2) rectangle (8,1);
\draw (7.5,1.5) node {$\bar{A}_R$};
\draw (8,1.5) -- (9,1.5); 
\draw[rounded corners] (9,2) rectangle (10,1);
\draw (9.5,1.5) node {$\bar{A}_R$};
\draw (10,1.5) -- (10.5,1.5);
\draw (1.5,2) -- (1.5,3); \draw (3.5,2) -- (3.5,3); \draw (5.5,2) -- (5.5,3);
\draw (7.5,2) -- (7.5,3); \draw (9.5,2) -- (9.5,3);
\draw (4.5,2.5) node (X) {$\phantom{X}$};
\end{tikzpicture} \dots,
\end{multline*}
so that its derivative $\partial_{\bar{B}}\braket{\Phi(B)|\Xi}$ is easily obtained by omitting the tensor $\bar{B}$ from the diagram and interpreting the open legs as defining the indices of a new tensor. Without the Lagrange multiplier, we would simply obtain
\begin{align*}
\begin{tikzpicture}[baseline = (X.base),every node/.style={scale=0.750},scale=.55]
\draw (0,-0.5) -- (0.5,-0.5);
\draw[rounded corners] (0.5,0) rectangle (1.5,-1);  
\draw (1,-0.5) node (X) {$B$};
\draw (1,-1) -- (1,-1.5);
\draw (1.5,-0.5) -- (2,-0.5);
\end{tikzpicture}
 =  \dots
\begin{tikzpicture}[baseline = (X.base),every node/.style={scale=0.750},scale=.55]
\draw (1,3) -- (10,3); \draw (1,4) -- (10,4);
\draw (5.5,3.5) node {$\Xi$};
\draw (0.5,1.5) -- (1,1.5); 
\draw[rounded corners] (1,2) rectangle (2,1);
\draw (1.5,1.5) node (X) {$\bar{A}_L$};
\draw (2,1.5) -- (3,1.5); 
\draw[rounded corners] (3,2) rectangle (4,1);
\draw (3.5,1.5) node {$\bar{A}_L$};
\draw (4,1.5) -- (4.5,1.5);
\draw (6.5,1.5) -- (7,1.5);
\draw[rounded corners] (7,2) rectangle (8,1);
\draw (7.5,1.5) node {$\bar{A}_R$};
\draw (8,1.5) -- (9,1.5); 
\draw[rounded corners] (9,2) rectangle (10,1);
\draw (9.5,1.5) node {$\bar{A}_R$};
\draw (10,1.5) -- (10.5,1.5);
\draw (1.5,2) -- (1.5,3); \draw (3.5,2) -- (3.5,3); \draw (5.5,-1) -- (5.5,3);
\draw (7.5,2) -- (7.5,3); \draw (9.5,2) -- (9.5,3);
\draw (4.5,-.5) edge[out=0,in=0] (4.5,1.5); \draw (6.5,-.5) edge[out=180,in=180] (6.5,1.5);
\draw (4.5,2.5) node (X) {$\phantom{X}$};
\end{tikzpicture} \dots \; .
\end{align*}
With the additional constraint, the solution is still straightforward. It can be easily verified that the correct value of the Lagrange multiplier is such that the additional term acts as a projection
\begin{equation}
B^s \rightarrow B^s - A_L^s \left[\sum_t {A_L^t}^\dag B^t\right] 
\end{equation} 
or similarly
\begin{equation}
B^s \rightarrow B^s -  \left[\sum_t B^t{A_R^t}^\dag\right] A_R^s
\end{equation} 
if we would have chosen the right gauge of Eq.~\eqref{eq:rightgaugefixingcondition}. 
\begin{widetext}
By inserting the solution for $B$ back into Eq.~\eqref{eq:TH_phiB}, we can read off the tangent space projector. While the value of $B$ depends on the gauge condition, the resulting projector is of course gauge independent and given by
\begin{equation*}
\P_{\ket{\Psi(A)}} = \sum_{n\in\ints}
\dots \begin{tikzpicture}[baseline = (X.base),every node/.style={scale=0.750},scale=.55]
\draw (5,.5) node (X) {$\phantom{X}$};
\draw (0.5,1.5) -- (1,1.5); 
\draw[rounded corners] (1,2) rectangle (2,1);
\draw (1.5,1.5) node {$\bar{A}_L$};
\draw (2,1.5) -- (3,1.5); 
\draw[rounded corners] (3,2) rectangle (4,1);
\draw (3.5,1.5) node {$\bar{A}_L$};
\draw[rounded corners] (7,2) rectangle (8,1);
\draw (7.5,1.5) node {$\bar{A}_R$};
\draw (8,1.5) -- (9,1.5); 
\draw[rounded corners] (9,2) rectangle (10,1);
\draw (9.5,1.5) node {$\bar{A}_R$};
\draw (10,1.5) -- (10.5,1.5);
\draw (1.5,2) -- (1.5,2.5); \draw (3.5,2) -- (3.5,2.5); \draw (5.5,-1.5) -- (5.5,2.5);
\draw (7.5,2) -- (7.5,2.5); \draw (9.5,2) -- (9.5,2.5);
\draw (4,-.5) edge[out=0,in=0] (4,1.5); \draw (7,-.5) edge[out=180,in=180] (7,1.5);
\draw (.5,-.5) -- (1,-.5); 
\draw[rounded corners] (1,0) rectangle (2,-1);
\draw (1.5,-.5) node {$A_L$};
\draw (2,-.5) -- (3,-.5); 
\draw[rounded corners] (3,0) rectangle (4,-1);
\draw (3.5,-.5) node {$A_L$};
\draw[rounded corners] (7,0) rectangle (8,-1);
\draw (7.5,-.5) node {$A_R$};
\draw (8,-.5) -- (9,-.5);
\draw[rounded corners] (9,0) rectangle (10,-1);
\draw (9.5,-.5) node {$A_R$};
\draw (10,-.5) -- (10.5,-.5); 
\draw (1.5,-1) -- (1.5,-1.5); \draw (3.5,-1) -- (3.5,-1.5);
\draw (7.5,-1) -- (7.5,-1.5); \draw (9.5,-1) -- (9.5,-1.5);
\draw (1.5,-2) node {$s_{n-2}$}; \draw (3.5,-2) node {$s_{n-1}$}; \draw (5.5,-2) node {$s_n$}; \draw (7.5,-2) node {$s_{n+1}$}; \draw (9.5,-2) node {$s_{n+2}$};
\end{tikzpicture}
\dots 
\quad
-
\quad
 \dots
\begin{tikzpicture}[baseline = (X.base),every node/.style={scale=0.750},scale=.55]
\draw (5,.5) node (X) {$\phantom{X}$};
\draw (0.5,1.5) -- (1,1.5); 
\draw[rounded corners] (1,2) rectangle (2,1);
\draw (1.5,1.5) node {$\bar{A}_L$};
\draw (2,1.5) -- (3,1.5); 
\draw[rounded corners] (3,2) rectangle (4,1);
\draw (3.5,1.5) node {$\bar{A}_L$};
\draw (4,1.5) -- (5,1.5);
\draw[rounded corners] (5,2) rectangle (6,1);
\draw (5.5,1.5) node {$\bar{A}_L$};
\draw[rounded corners] (8,2) rectangle (9,1);
\draw (8.5,1.5) node {$\bar{A}_R$};
\draw (9,1.5) -- (10,1.5); 
\draw[rounded corners] (10,2) rectangle (11,1);
\draw (10.5,1.5) node {$\bar{A}_R$};
\draw (11,1.5) -- (11.5,1.5);
\draw (1.5,2) -- (1.5,2.5); \draw (3.5,2) -- (3.5,2.5); \draw (5.5,2) -- (5.5,2.5);
\draw (8.5,2) -- (8.5,2.5); \draw (10.5,2) -- (10.5,2.5);
\draw (6,-.5) edge[out=0,in=0] (6,1.5); \draw (8,-.5) edge[out=180,in=180] (8,1.5);
\draw (.5,-.5) -- (1,-.5); 
\draw[rounded corners] (1,0) rectangle (2,-1);
\draw (1.5,-.5) node {$A_L$};
\draw (2,-.5) -- (3,-.5); 
\draw[rounded corners] (3,0) rectangle (4,-1);
\draw (3.5,-.5) node {$A_L$};
\draw (4,-.5) -- (5,-.5);
\draw[rounded corners] (5,0) rectangle (6,-1);
\draw (5.5,-.5) node {$A_L$};
\draw[rounded corners] (8,0) rectangle (9,-1);
\draw (8.5,-.5) node {$A_R$};
\draw (9,-.5) -- (10,-.5);
\draw[rounded corners] (10,0) rectangle (11,-1);
\draw (10.5,-.5) node {$A_R$};
\draw (11,-.5) -- (11.5,-.5); 
\draw (1.5,-1) -- (1.5,-1.5); \draw (3.5,-1) -- (3.5,-1.5); \draw (5.5,-1) -- (5.5,-1.5);
\draw (8.5,-1) -- (8.5,-1.5); \draw (10.5,-1) -- (10.5,-1.5);
\draw (1.5,-2) node {$s_{n-2}$}; \draw (3.5,-2) node {$s_{n-1}$}; \draw (5.5,-2) node {$s_n$}; \draw (8.5,-2) node {$s_{n+1}$}; \draw (10.5,-2) node {$s_{n+2}$};
\end{tikzpicture}\;\dots .
\end{equation*}
\end{widetext}
We can represent the tangent space projector as
\begin{equation}
\P_{\ket{\Psi(A)}} = \sum_{n\in\ints} P_{A_{C}}(n) - P_{C}(n),\label{eq:TH_PT}
\end{equation}
by defining the partial projectors
\begin{subequations}
\label{eq:TH_projectors}
\begin{align}
P_{A_{C}}(n) =& P_{L}(n-1)\otimes \unity_{n}\otimes P_{R}(n+1),\label{eq:TH_PAC}\\
P_{C}(n) =&\P_{L}(n)\otimes\P_{R}(n+1)\label{eq:TH_PC},\\
P_{L}(n) =& \sum_{\alpha}\ket{\Psi_{L}^{\alpha}(n)}\bra{\Psi_{L}^{\alpha}(n)},\label{eq:TH_PL}\\
P_{R}(n) =& \sum_{\alpha}\ket{\Psi_{R}^{\alpha}(n)}\bra{\Psi_{R}^{\alpha}(n)}.\label{eq:TH_PR}
\end{align}
\end{subequations}
We can verify that $\P_{\ket{\Psi(A)}}^{2}=\P_{\ket{\Psi(A)}}$ by using
\begin{subequations}
\label{eq:TH_Pfuse}
\begin{align}
P_{L}(m)\,P_{L}(n)&= P_{L}(\max(m,n)),\label{eq:TH_PLfuse}\\
P_{R}(m)\,P_{R}(n)&= P_{R}(\min(m,n)).\label{eq:TH_PRfuse}
\end{align} 
\end{subequations}

\subsection{Gradient and Effective Hamiltonians}
\label{sec:grad_effH}

As discussed in the beginning of this section, a variational optimum can be characterized geometrically as
\begin{equation}
\P_{\ket{\Psi(A)}} (H - E) \ket{\Psi(A)} = 0\label{eq:TH_galerkinmps}
\end{equation}
where $E = \braket{\Psi(A)|H|\Psi(A)}$ (unit normalization is assumed). Since the Galerkin condition is automatically ensured in the direction of the MPS itself, the only non-trivial information of Eq.~\eqref{eq:TH_galerkinmps} is thus contained in the part of the tangent space orthogonal to $\ket{\Psi(A)}$. This is convenient, as $\P_{\ket{\Psi(A)}}$ was actually constructed as the projector onto the part of tangent space orthogonal to $\ket{\Psi(A)}$ in the first place. While this implies that the $E$ subtraction does not contribute, it is convenient to keep it around, as it ensures that the individual terms in the final expression are finite in the thermodynamic limit.

Applying the tangent space projection as in the previous section to the state $\ket{\Xi}=H\ket{\Psi(A)}$ gives rise to a tangent vector of the form Eq.~\eqref{eq:TH_phiB}, with
\begin{equation}
 B^{s} =  {A}_{C}^{\prime s} - {C^{\prime}} A_{R}^{s} \quad\text{or}\quad B^{s}= {A}_{C}^{\prime s} - A_{L}^{s}{C^{\prime}}
\label{eq:TH_HpsiB}
\end{equation}
where ${A}_{C}^{\prime}$ originates from applying $P_{A_C}(n)$ and $C^{\prime}$ from applying $P_C(n)$. By writing $\ket{\Psi(A)}$ itself in a compatible gauge for every individual term, we obtain the diagrammatic expressions
\begin{align*}
\begin{tikzpicture}[baseline = (X.base),every node/.style={scale=0.750},scale=.55]
\draw (0,-0.5) -- (0.5,-0.5);
\draw[rounded corners] (0.5,0) rectangle (1.5,-1);  
\draw (1,-0.5) node (X) {$A^{\prime}_C$};
\draw (1,-1) -- (1,-1.5);
\draw (1.5,-0.5) -- (2,-0.5);
\end{tikzpicture}  = \dots \begin{tikzpicture}[baseline = (X.base),every node/.style={scale=0.750},scale=.55]
\mpsT{-4.5}{1.5}{A_L} \mpsT{-2.5}{1.5}{A_L} \mpsT{-0.5}{1.5}{A_C} \mpsT{1.5}{1.5}{A_R} \mpsT{3.5}{1.5}{A_R} 
\mpsT{-4.5}{-1.5}{\bar{A}_L} \mpsT{-2.5}{-1.5}{\bar{A}_L}  \mpsT{1.5}{-1.5}{\bar{A}_R} \mpsT{3.5}{-1.5}{\bar{A}_R}
\lineH{-5.5}{-5}{+1.5} \lineH{-4}{-3}{+1.5} \lineH{-2}{-1}{1.5} \lineH{0}{1}{+1.5} \lineH{2}{3}{+1.5} \lineH{4}{4.5}{+1.5} 
\lineH{-5.5}{-5}{-1.5} \lineH{-4}{-3}{-1.5} \lineH{-2}{-1}{-1.5} \lineH{0}{1}{-1.5} \lineH{2}{3}{-1.5} \lineH{4}{4.5}{-1.5} 
\lineV{-1}{-.5}{-4.5} \lineV{-1}{-.5}{-2.5} \lineV{.5}{1}{-4.5} \lineV{.5}{1}{-2.5} 
\lineV{.5}{1}{-0.5} \lineV{-3}{-.5}{-0.5}
\lineV{-1}{-0.5}{1.5} \lineV{0.5}{1}{1.5}  \lineV{-1}{-.5}{3.5} \lineV{.5}{1}{3.5}
\lineH{-5.5}{4.5}{0.5}
\draw (-0.5,0) node (X) {$H$};
\lineH{-5.5}{4.5}{-0.5}
\draw (-1,-1.5) edge[out=0,in=90] (-0.75,-2);
\draw (-0.75,-2) edge[out=-90,in=0] (-1,-2.5);
\draw (-1,-2.5) -- (-1.5,-2.5);
\draw (-0,-1.5) edge[out=180,in=90] (-0.25,-2);
\draw (-0.25,-2) edge[out=-90,in=180] (-0,-2.5);
\draw (-0,-2.5) -- (.5,-2.5);
\end{tikzpicture} \dots
\end{align*}
and
\begin{align*}
\begin{tikzpicture}[baseline = (X.base),every node/.style={scale=0.750},scale=.55]
\draw (-.5,1.5) circle (.5); \draw (-.5,1.5) node (X) {${C^{\prime}}$};
\draw (-1.5,1.5) -- (-1,1.5); \draw (0,1.5) -- (0.5,1.5); 
\end{tikzpicture} &=
\dots \begin{tikzpicture}[baseline = (X.base),every node/.style={scale=0.750},scale=.55]
\mpsT{-4.5}{1.5}{A_L} \mpsT{-2.5}{1.5}{A_L} 
\draw (-.5,1.5) circle (.5); \draw (-.5,1.5) node {$C$};
\mpsT{1.5}{1.5}{A_R} \mpsT{3.5}{1.5}{A_R} 
\mpsT{-4.5}{-1.5}{\bar{A}_L} \mpsT{-2.5}{-1.5}{\bar{A}_L}  \mpsT{1.5}{-1.5}{\bar{A}_R} \mpsT{3.5}{-1.5}{\bar{A}_R}
\lineH{-5.5}{-5}{+1.5} \lineH{-4}{-3}{+1.5} \lineH{-2}{-1}{1.5} \lineH{0}{1}{+1.5} \lineH{2}{3}{+1.5} \lineH{4}{4.5}{+1.5} 
\lineH{-5.5}{-5}{-1.5} \lineH{-4}{-3}{-1.5} \lineH{-2}{-1}{-1.5} \lineH{0}{1}{-1.5} \lineH{2}{3}{-1.5} \lineH{4}{4.5}{-1.5} 
\lineV{-1}{-.5}{-4.5} \lineV{-1}{-.5}{-2.5} \lineV{.5}{1}{-4.5} \lineV{.5}{1}{-2.5} 
\lineV{-1}{-0.5}{1.5} \lineV{0.5}{1}{1.5}  \lineV{-1}{-.5}{3.5} \lineV{.5}{1}{3.5}
\lineH{-5.5}{4.5}{0.5}
\draw (-0.5,0) node (X) {$H$};
\lineH{-5.5}{4.5}{-0.5}
\draw (-1,-1.5) edge[out=0,in=90] (-0.75,-2);
\draw (-0.75,-2) edge[out=-90,in=0] (-1,-2.5);
\draw (-1,-2.5) -- (-1.5,-2.5);
\draw (-0,-1.5) edge[out=180,in=90] (-0.25,-2);
\draw (-0.25,-2) edge[out=-90,in=180] (-0,-2.5);
\draw (-0,-2.5) -- (.5,-2.5);
\end{tikzpicture} \dots 
\end{align*}
We can thus obtain ${A}_{C}^{\prime}$ and ${C^{\prime}}$ by acting with the effective Hamiltonians $H_{A_{C}}$ and $H_{C}$ introduced in \eqref{eq:HAC} and \eqref{eq:HC} in the main text onto $A_{C}$ and $C$
\begin{align}
 \bm{A}_{C}^{\prime} &= \bm{H}_{A_{C}}\, \bm{A}_{C}^{\prime}, & \bm{C}^{\prime} &= \bm{H}_{C}\,\bm{C}.\label{eq:TH_ApplyHeff}
\end{align}

Even without subtracting the energy, the two choices of $B$ (which are related by the additive gauge transform with $X=C^{\prime}$) will be finite in the thermodynamic limit. However, the individual tensors ${A}^{\prime}_{C}$ and ${C^{\prime}}$ will have a divergent contribution proportional to $A_C$ and $C$, respectively. Indeed, as discussed in the main text for the case of nearest neighbor interactions, the effective Hamiltonians $H_{A_C}$ and $H_{C}$ have a divergent contribution corresponding to the total energy times the identity operator. It is thus by subtracting $H\to \tilde{H} = H-E$ (or  $h\to\tilde{h}=h-e$ for the local terms) that these divergences are canceled. \App{sec:effH_construct} provides a detailed description of the construction of the effective Hamiltonian for other types of interactions and illustrates explicitly that the diverging contributions cancel exactly.

A variational extremum is characterized by $\ket{\Phi(B)}=0$, which leads to $B=0$ for either gauge choice, as these choices completely fix the gauge freedom. This gives rise to the following simultaneous conditions:
\begin{align}
{A}_{C}^{\prime s} = A_L^s C^{\prime} = C^{\prime} A_R^s \label{eq:TH_FP} \\
A_{C}^{s}=A_{L}^{s}C=CA_{R}^{s} \label{eq:TH_FP3}.
\end{align}
However, because the gauge transformation that relates $A_L$ and $A_R$ is unique up to a factor (for injective MPS), $C$ and $C^{\prime}$ have to be proportional, and we have actually obtained the eigenvalue equations
\begin{align}
\bm{A}_{C}^{\prime} &= \bm{H}_{A_C}\, \bm{A}_C = E_{A_{C}}\, \bm{A}_C \label{eq:TH_FP1}\\
\bm{C}^{\prime} &= \bm{H}_{C}\, \bm{C} = E_{C} \,\bm{C}. \label{eq:TH_FP2}
\end{align}

As we are looking for a variational minimum, the eigenvalues $E_{A_{C}}$ and $E_{C}$ should be the lowest eigenvalues of the effective Hamiltonians $H_{A_{C}}$ and $H_{C}$. Depending on how we have regularized the divergent contributions, these eigenvalues might be different. If we have completely subtracted the energy expectation value from every term, we then have $E_C=E_{A_C} = 0$.

\subsection{Convergence and error measures}
\label{sec:converr} 
While neither \algorithmname\ nor IDMRG or ITEBD directly use the gradient itself, the Hilbert space norm of $\ket{\Phi(B)}$ can be used as an objective convergence measure to indicate how far the current state is from the variational optimum. For either choice of $B$, we obtain $\lVert \ket{\Phi(B)}\rVert = \sqrt{N} \lVert B\rVert$, with $N$ the diverging number of sites and $\lVert B\rVert$ the 2-norm of the tensor $B$. Its square is given by
\begin{equation}
\begin{split}
\lVert B \rVert^{2} &= \sum_{s,\alpha,\beta} |B^s_{\alpha,\beta}|^2\\
&= \sum_s \lVert {A}_{C}^{\prime s} - A_L^s {C^{\prime}}\rVert^2\\
&= \sum_s \lVert {A}_{C}^{\prime s} - {C^{\prime}} A_R^s\rVert^2\\
&= \lVert {A}_{C}^{\prime}\rVert^2 - \lVert {C^{\prime}}\rVert^2
\end{split}
\label{eq:gradnorm}
\end{equation} 
where the equalities follow from $C^{\prime} = \sum_s (A_L^s)^\dagger A_{C}^{\prime s} = \sum_s {A}_{C}^{\prime s} (A_R^s)^\dagger$. Note that none of these expressions are well suited for numerically evaluating the norm close to convergence, as they involve subtracting quantities that are almost equal, especially when the state is close to convergence.

An alternative strategy for evaluating $\lVert B \rVert$ is by using the matrix notation for tensors \eqref{eq:vectorizations}
to write $\mathcal{B}^{[\ell]} = \mathcal{A}^{\prime[\ell]}_C - \mathcal{A}_L C^{\prime}$. Since $\mathcal{A}_{L}$ is an isometry, we can extend it to a $dD \times dD$ unitary matrix $U=\begin{bmatrix}\mathcal{A}_L &\mathcal{N}_L\end{bmatrix}$, where $\mathcal{N}_L$ contains an orthonormal basis for the $(d-1)D$-dimensional null space of $\mathcal{A}_L^\dagger$, i.e.\  $\mathcal{A}_L^\dagger \mathcal{N}_L = 0$. As the 2-norm is unitarily invariant, we can write
\begin{align*}
\lVert B\rVert = \lVert \mathcal{B}^{[\ell]}\rVert = \lVert U^\dagger \mathcal{B}^{[\ell]}\rVert = \lVert \mathcal{N}_L^\dagger \mathcal{B}^{[\ell]} \rVert = \lVert\mathcal{N}_L^\dagger \mathcal{A}^{\prime[\ell]}_C \rVert.
\end{align*}
The second equality follows from $\mathcal{A}_L^\dagger \mathcal{B}^{[\ell]}=0$ and the third from the null space property of $\mathcal{N}_L$. We then obtain $\lVert B\rVert$ as the Frobenius norm of a single matrix, which can be calculated accurately as a sum of strictly positive numbers. For further reference, we reshape $\mathcal{N}_L$ into a $D \times d \times (d-1)D$ tensor $N_L$ and similarly introduce a $(d-1)D \times d \times D$ tensor $N_R$ via the defining relations
\begin{align}
\sum_s (N_L^s)^\dagger A_L^s &=0,&\sum_s (N_L^s)^\dagger N_L^s = \openone,\\
\sum_s A_R^s (N_R^s)^\dagger &=0,&\sum_s N_R^s (N_R^s)^\dagger = \openone.
\end{align}

While the norm of the gradient provides a measure for the quality of approaching the variational minimum, it does not provide any information about the quality of the (u)MPS approximation to the true ground state itself. In the context of (two-site) DMRG schemes, a popular measure is the truncation error, as it is naturally accessible throughout the algorithm (see e.g.\ Ref.~\onlinecite{White_Huse93,Legeza_Fath96,MPS6_S}). But also within \algorithmname\ we can compute this quantity by first writing the state $\ket{\Psi(A)}$ in a mixed canonical form with a two-site center block
\begin{align*}
\ket{\Psi(A)} = \sum_{n,\alpha,\beta,s_n,s_{n+1}} (A_{2C})_{\alpha,\beta}^{s_ns_{n+1}} \ket{\Psi_L^\alpha} \ket{s_n}\ket{s_{n+1}} \ket{\Psi_R^\beta}.
\end{align*}
The two-cite center tensor $A_{2C}^{ss'} = A_L^s A_C^{s'} = A_C^s A_R^{s'} = A_L^s C A_R^{s'}$ (known as the two-site wave function $\psi^{ss'}$ in standard DMRG) has an associated effective Hamiltonian $H_{A_{2C}}$. We can compute its lowest eigenvector $\tilde{A}_{2C}$ and compute its singular value decomposition (by first reshaping it to a $dD \times dD$ matrix) $\tilde{A}_{2C}^{ss'} = U^s S V^{s'}$. The truncation error then corresponds to the discarded weight 
\begin{equation}
\epsilon_{\rho} = \sum_{k=D+1}^{dD} S_k^2\label{eq:truncation_error}
\end{equation}
when truncating the inner bond dimension of this two-site tensor to its original value $D$.

A more generic measure for the error in the variational approximation is given by the energy variance $\braket{\Psi(A)|(H-E)^2|\Psi(A)} = \lVert (H-E)\ket{\Psi(A)}\rVert^2$. This quantity is also used in the context of e.g.\ variational Monte Carlo and various other methods. We can systematically decompose $(H-E)\ket{\Psi(A)}$ into various parts: the projection onto $\ket{\Psi(A)}$ is automatically zero by the definition of $E$. The projection onto the tangent space is zero when we are at the variational minimum. Next, we can project $(H-E)\ket{\Psi(A)}$ onto the space of all 2-site variations, which is given by states of the form
\begin{equation}
\ket{\Phi^{(2)}(B_2)}=\sum_{\bm{s}}\sum_{n\in\ints}(\ldots A^{s_{n-1}}_{L}B_2^{s_{n}s_{n+1}}A^{s_{n+2}}_{R}\ldots)\ket{\bm{s}}
\end{equation}
In the case of nearest neighbor Hamiltonians, this space captures $(H-E)\ket{\Psi(A)}$ completely, namely by choosing $B_2^{st} = \braket{st|\tilde{h}|s't'} A_{2C}^{s't'}$ with $A_{2C}$ the two-site center tensor defined in the previous paragraph and $\tilde{h}$ the local terms of the Hamiltonian, with the current expectation value subtracted, i.e.\ $H - E = \sum_{n} \tilde{h}_{n,n+1}$. However, there is again an additive representation redundancy (gauge freedom) $B_2^{st} \to B_2^{st} + A_L^s X^t - X^s A_R^t$ which enables us to choose representations $B$ satisfying e.g.\ a left gauge condition $\sum_{s} (A_L^s)^\dagger B_2^{st} = 0$ ($\forall t$). The advantage of this representation is again that it facilitates the calculation of the norm, as $\lVert \ket{\Phi^{(2)}(B_2)}\rVert^2 = N \lVert B_2\rVert^2$ with $N$ the diverging number of sites. The projection of $(H-E)\ket{\Psi(A)}$ onto this space can be worked out similarly as for the tangent space, and leads to the general result (for any Hamiltonian)
\begin{align}
B_2^{st} &= A_{2C}^{\prime st} - A_L^s A_{C}^{\prime t}& \text{or}&& B_2^{st} &= A_{2C}^{\prime st} - A_{C}^{\prime s}A_R^t
\end{align}
with $\bm{A}_{2C}^{\prime} = \bm{H}_{A_{2C}} \bm{A}_{2C}$ a single application of the two-site effective Hamiltonian. Using $A_L^s (A_L^t)^\dagger + N_L^s (N_L^t)^\dagger = \delta_{s,t} \openone$, we can rewrite the first form of $B_2$ as
\begin{align*}
B_2^{st} &= N_L^s \sum_{s'} (N_L^{s'})^\dagger A_{2C}^{\prime s't}.
\end{align*}
We now also apply $(A_R^s)^\dagger A_R^t + (N_R^s)^\dagger N_R^t  = \delta_{s,t} \openone$ to the right hand side and recognize $A_C^{\prime s} = A_{2C}^{\prime st} (A_R^t)^\dagger$. But since $\sum_{s} (N_L^s)^\dagger A_C^{\prime s}=0$ at the variational minimum, we obtain at the variational minimum
\begin{align*}
B_2^{st} &= N_L^s \left[ \sum_{s't'} (N_L^{s'})^\dagger A_{2C}^{\prime s't'} (N_R^{t'})^\dagger\right] N_R^t
\end{align*}
and in particular
\begin{align}
\lVert B_2 \rVert = \lVert \sum_{s't'} (N_L^{s'})^\dagger A_{2C}^{\prime s't'} (N_R^{t'})^\dagger \rVert.
\end{align}

We can also relate $\lVert B_2 \rVert^2$ to the truncation error defined in the previous paragraph. For the truncation error arising in the context of two-site DMRG schemes, the lowest eigenvector $\tilde{A}_{2C}$ of the two-site effective Hamiltonian is used. When the DMRG algorithm has converged, the rank $D$ approximation of $\tilde{A}_{2C}$ should again be the original two-site center tensor $A_{2C}^{st} = A_L^s C A_R^t$. But this means that we can construct $N_L$ and $N_R$ exactly from the singular vectors corresponding to the $(d-1) D$ singular values that were truncated away, and thus that the truncation error is given by $\epsilon_{\rho}=\lVert \sum_{st} (N_L^s)^\dagger \tilde{A}_{2C}^{st} (N_R^t)^\dagger\rVert^2$. This definition is close to $\lVert B_2 \rVert^2$, except that the latter uses the tensor $A_{2C}^{\prime}$ arising from applying the two-site effective Hamiltonian once. As $A_{2C}$ and $\tilde{A}_{2C}$ are anyway close, we can think of $A_{2C}^{\prime}$ as providing the leading order correction from $A_{2C}$ to $\tilde{A}_{2C}$ in the sense of a Krylov scheme. Indeed, in the first iteration of the Lanczos method, the eigenvector $\tilde{A}_{2C}$ would be approximated in the form $\alpha A_{2C} + \beta A_{2C}^{\prime}$. Since the first term drops out when projecting onto $N_L$ and $N_R$, the DMRG truncation error and $\lVert B_2 \rVert^2$ will be of the same order of magnitude.

Note, however, that $\lVert B_2 \rVert^2$ only captures the full energy variance (per site) for nearest neighbor Hamiltonians, whose action on $\ket{\Psi(A)}$ is completely contained within the space of two-site variations as noticed above. In that case, we can see that the only term that survives in $A_{2C}^{\prime}$ after projection onto $N_L$ and $N_R$ is the local term, where $\tilde{h}$ acts on the two-site center tensor. We can thus also write
\begin{align}
\lVert B_2 \rVert = \lVert \sum_{s't'st} \braket{s't'|\tilde{h}|st} (N_L^{s'})^\dagger A_{2C}^{st} (N_R^{t'})^\dagger \rVert.
\end{align}
We can also relate this to the truncation step in (I)TEBD, where we would apply $\exp(-\Delta t\ \tilde{h})$ to every two-site block of the state. The resulting truncation would lead to a discarded weight of the order $\Delta t^2 \lVert B_2\rVert^2$.

The considerations regarding the projection of $(H-E)\ket{\Psi(A)}$ onto the space of two-site variations can also be used to devise a scheme for expanding the bond dimension of the uMPS. This approach is presented in the next section.

\section{Dynamic Control of the Bond Dimension}
\label{sec:bonddim}
A characteristic feature of two-site implementations of conventional MPS methods -- such as e.g.\ (I)TEBD or (I)DMRG -- is that the bond dimension $D$ of the MPS is automatically increased in every iteration and has to be \textit{truncated} in order to remain at a finite maximum bond dimension. This truncation step lies at the basis of why such schemes for finding ground states will never truly converge to the variational minimum up to machine precision, as observed in the results. Indeed, even in finite size simulations, two-site DMRG is used to initialize the state and one-site DMRG to obtain final convergence. However, the truncation step in two-site methods has the advantage that the bond dimension can be dynamically increased (or decreased) according to some quality constraint, such as the magnitude of the smallest Schmidt-value or the discarded weight. Especially in the presence of symmetry, this is important to automatically obtain the correct symmetry sectors within the virtual MPS space.

The  \algorithmname\ algorithm presented in the main text is variational from the start and therefore works at fixed bond dimension $D$, i.e.\ it is a one-site scheme in DMRG terminology. Alternative subspace expansion strategies for dynamically increasing the bond dimension in such one-site schemes have been proposed.\cite{DMRG_SS,bonddim2} These methods use information from acting with the global Hamiltonian onto the current state to either add a tiny perturbation to the current MPS or to generate a larger basis in which the effective eigenvalue problem is solved. 

\begin{widetext}
We have developed a similar subspace expansion technique that works for a uMPS in the thermodynamic limit. It is based on projecting the full action of the Hamiltonian $(H-E)\ket{\Psi(A)}$ onto the space of two-site variations, as developed in the previous section. There we have found the representation
\begin{align*}
B_{2}^{st} &= N_L^s \left[ \sum_{s't'} (N_L^{s'})^\dagger A_{C}^{\prime s't'} (A_R^{t'})^\dagger \right] A_R^t + N_L^s \left[ \sum_{s't'} (N_L^{s'})^\dagger A_{2C}^{\prime s't'} (N_R^{t'})^\dagger \right] N_R^t\\
&= A_{C}^{\prime s} A_R^t - A_L^s C^{\prime} A_R^t + N_L^s \left[ \sum_{s't'} (N_L^{s'})^\dagger A_{2C}^{\prime s't'} (N_R^{t'})^\dagger \right] N_R^t
\end{align*}
Even when we have not yet reached the variational minimum, the first term (on line 1) or the first two terms (on line 2) are captured in the tangent space, and only the last term (on either line) contains a new search direction. To capture it completely, we would need to expand the bond dimension from value $D$ to $d D$. If we want to expand to a new dimension $\tilde{D} = D+\Delta D$, we can use a singular value decomposition to compute the rank $\Delta D$ approximation of $\sum_{s't'} (N_L^{s'})^\dagger A_{2C}^{\prime s't'} (N_R^{t'})^\dagger = U S V$. By keeping only the largest $\Delta D$ singular values, $U$ and $V$ are left and right isometries of size $(d-1)D \times \Delta D$ and $\Delta D \times (d-1)D$ respectively. As remarked in the previous section, in the case of nearest neighbor interactions, the projection of $A_{2C}^{\prime}$ onto $N_L$ and $N_R$ does not require the full two-site effective Hamiltonian but reduces to the local term.
\end{widetext}

We do not directly update the current MPS, but rather write it in an en expanded basis in a mixed canonical form with matrices
\begin{align*}
\tilde{A}^{s}_{L}&=\begin{bmatrix}A^{s}_{L}&N^{s}_{L}U\\ 0 & 0\end{bmatrix},&
\tilde{A}^{s}_{R}&=\begin{bmatrix}A^{s}_{R} & 0\\ V^{\dagger}N^{s}_{R} & 0 \end{bmatrix},&
\tilde{C}&=\begin{bmatrix} C & 0 \\ 0 & 0 \end{bmatrix}.
\end{align*} 
With these initial tensors, we can now start a new iteration of \algorithmname. Note that we can straightforwardly update the environments used to construct the effective Hamiltonians into this expanded basis, which is necessary if we want to use them as initial guess.

\section{Explicit Construction of Effective Hamiltonians}
\label{sec:effH_construct}
In this section we describe how to efficiently apply the effective Hamiltonians $H_{A_{C}}$ and $H_{C}$ onto the center site tensor $A^{s}_{C}$ and bond matrix $C$ and how the necessary individual terms are explicitly constructed. Such a procedure is needed for solving the effective eigenvalue problems \eqref{eq:FP1} and \eqref{eq:FP2} by means of an iterative eigensolver. The case of systems with nearest neighbor interaction has already been discussed in \Sec{sec:effH_NN}. In the following we consider the cases of Hamiltonians with long range interactions in \Sec{sec:effH_LR} and general Hamiltonians given in terms of Matrix Product Operators (MPOs) in \Sec{sec:effH_MPO}.

\subsection{Long Range Interactions}
Consider Hamiltonians with long range interactions of the form $H=\sum_{j\in\ints}h_{j}$, where $h_{j}$ is itself an infinite sum 
\begin{align}
h_{j} = \sum_{n>0} f(n)\,o_{j}o_{j+n}
\label{eq:hj_LR}
\end{align}
and operators $o_{i}$ act on a single-site $i$ and commute when acting on different sites $[o_{i},o_{j}]=0,\,i\neq j$.
Without loss of generality, we restrict to a single pair of (bounded) operators $o$, which commute when acting on different sites $[o_{i},o_{j}]=0,\,i\neq j$ \footnote{For fermionic Hamiltonians with long range interactions, we can employ a Jordan-Wigner transformation to spin operators, introducing a string operator counting the number of fermions between $o_{j}$ and $o_{j+n}$; geometric sums of transfer matrices in \eqref{eq:edens_LR} then turn into geometric sums of (string) operator transfer matrices.}. The generalization to Hamiltonians containing several terms of that form is straight forward. Furthermore, we assume distance functions $f(n)$ that are bounded in the sense of $\sum_{n>0}\lvert f(n)\rvert <\infty$, such that $\lVert h_{j} \rVert<\infty$, and that can be well approximated by a sum of $K$ exponentials, i.e.
\begin{equation}
f(n)\approx\sum_{k=1}^{K}c_{k} \lambda_{k}^{n-1},
\label{eq:sumofexp}
\end{equation} 
with $\lvert \lambda_{k}\rvert<1$ and $n>0$. In practice, for an infinite system we fit $f(n)$ with a suitable number of $K$ exponentials over a distance $N$ large enough, such that $f(N)$ and the largest residuals are below some desired threshold.

Examples of Hamiltonians that fall in this class are the transverse field Ising (TFI) model or XXZ model with power-law interactions,\cite{Ruelle68,Dyson69,Cardy81} as well as the famous Haldane-Shastry model,\cite{Haldane88,Shastry88} for which the ground state is exactly known.

Similar to the case of nearest neighbor interactions in \Sec{sec:effH_NN}, the effective Hamiltonians factorize into a number of terms which can all be applied efficiently. For $H_{A_{C}}$ these are five terms, out of which four are already familiar from the case of nearest neighbor interactions. Two of these are the left and right block Hamiltonians $H_{L}$ and $H_{R}$ with infinitely many local contributions from $h_{j}$ acting on sites strictly left or right of the current center site, and the other two are the terms containing interactions between the center site and the left and right block respectively, i.e.\ where $h_{j}$ partially acts on $A_{C}$. For long range interactions we have one additional term, containing infinitely many interaction terms between the left and the right block only without involving the center site, i.e.\ where $o_{j}$ acts to the left of the current center site, and $o_{j+n}$ acts to the right.

To construct all these terms we start by defining the operator transfer matrices
\begin{align}
 T_{L}^{[o]}&=\sum_{st}o_{st}\bar{A}_{L}^{s}\otimes A_{L}^{t}&
  T_{R}^{[o]}&=\sum_{st}o_{st}\bar{A}_{R}^{s}\otimes A_{R}^{t}.
\end{align}
The current energy density expectation value $e=\braket{\Psi(A)|h|\Psi(A)}$ can thus be written as
\begin{equation}
\begin{split}
e &=\rbra{\unity}T^{[o]}_{L}\left[\sum_{n>0}f(n)(T_{L})^{n-1}\right]T^{[o]}_{L}\rket{R}\\
&=\rbra{L}T^{[o]}_{R}\left[\sum_{n>0}f(n)(T_{R})^{n-1}\right]T^{[o]}_{R}\rket{\unity},
\end{split}
\label{eq:edens_LR}
\end{equation} 
or using \eqref{eq:sumofexp}
\begin{equation}
\begin{split}
e &=\sum_{k}c_{k}\rbra{\unity}T^{[o]}_{L}\left[\sum_{n\geq0}(\lambda_{k}T_{L})^{n}\right]T^{[o]}_{L}\rket{R}\\
&=\sum_{k}c_{k}\rbra{L}T^{[o]}_{L}\left[\sum_{n\geq0}(\lambda_{k}T_{R})^{n}\right]T^{[o]}_{L}\rket{\unity}.
\end{split}
\label{eq:edens_LR_approx}
\end{equation} 
Since $\lvert\lambda_{k}\rvert<1$ the geometric series converge and we can perform them explicitly. We proceed by defining
\begin{equation}
\begin{split}
 \rbra{O^{[k]}_{L}}&=\rbra{\unity}T^{[o]}_{L}\left[\unity - \lambda_{k}T_{L}\right]^{-1}\\
\rket{O^{[k]}_{R}}&= \left[\unity - \lambda_{k}T_{R}\right]^{-1}T^{[o]}_{L}\rket{\unity}.
\end{split}
\label{eq:GSconv}
\end{equation}
These terms can again either be calculated recursively by explicitly evaluating the geometric sums term by term until convergence, or more efficiently by iteratively solving the following systems of linear equations
\begin{equation}
\begin{split}
 \rbra{O^{[k]}_{L}}\left[\unity - \lambda_{k}T_{L}\right]&=\rbra{\unity}T^{[o]}_{L}\\
\left[\unity - \lambda_{k}T_{R}\right]\rket{O^{[k]}_{R}}&= T^{[o]}_{L}\rket{\unity}
\end{split}
\label{eq:GSconv_LSE}
\end{equation}
using iterative methods. 

We represent these terms by the diagrams
\begin{align*}
\drawMatrixLeft{$O_L^{[k]}$}&= 
\begin{tikzpicture}[baseline = (X.base),every node/.style={scale=0.750},scale=.55]
\draw (1,-1.5) edge[out=180,in=180] (1,1.5);
\draw[rounded corners] (1,2) rectangle (2,1);
\draw[rounded corners] (1,-1) rectangle (2,-2);
\draw (1.5,0) circle (.5);
\draw (1.5,0) node {$O$};
\draw (1.5,1.5) node {$A_L$};
\draw (1.5,-1.5) node {$\bar{A}_L$};
\draw (1.5,1) -- (1.5,.5);
\draw (1.5,-1) -- (1.5,-.5);
\draw (2,1.5) -- (3,1.5); \draw (2,-1.5) -- (3,-1.5);
\draw[rounded corners] (3,2) rectangle (6,-2);
\draw (4.5,0) node {$[\unity-\lambda_k T_L]^{-1}$};
\draw (6,1.5) -- (6.5,1.5);  \draw (6,-1.5) -- (6.5,-1.5);
\end{tikzpicture} \\
\drawMatrixRight{$O_R^{[k]}$} &= 
\begin{tikzpicture}[baseline = (X.base),every node/.style={scale=0.75},scale=.55]
\draw (2.5,1.5) -- (3,1.5); \draw (2.5,-1.5) -- (3,-1.5);
\draw[rounded corners] (3,2) rectangle (6,-2);
\draw (4.5,0) node {$[\unity-\lambda_k T_R]^{-1}$};
\draw (6,1.5) -- (7,1.5);  \draw (6,-1.5) -- (7,-1.5);
\draw[rounded corners] (7,2) rectangle (8,1);
\draw[rounded corners] (7,-1) rectangle (8,-2);
\draw (7.5,1.5) node {$A_R$};
\draw (7.5,-1.5) node {$\bar{A}_R$};
\draw (7.5,0) node {$O$};
\draw (7.5,1) -- (7.5,.5);
\draw (7.5,-1) -- (7.5,-.5);
\draw (7.5,0) circle (.5);
\draw (8,+1.5) edge[out = 0, in =0] (8, -1.5);
\end{tikzpicture}
\end{align*}
and collect all such terms into single left and right environment contributions
\begin{align}
\rbra{O_{L}}&= \sum_{k}c_{k}\rbra{O^{[k]}_{L}}&
\rket{O_{R}}&= \sum_{k}c_{k}\rket{O^{[k]}_{T}}
\label{eq:O_LR}
\end{align}
and further
\begin{align}
\rbra{h_{L}}&= \rbra{O_{L}}T^{[o]}_{L}&
\rket{h_{R}}&= T^{[o]}_{R}\rket{O_{R}}.
\label{eq:hl_LR}
\end{align}
We can then write for the energy density
\begin{equation}
e =\rbraket{h_{L}}{R}=\rbraket{L}{h_{R}}.
\end{equation}
Comparing with \eqref{eq:edens_LR} we have thus defined
\begin{equation}
\begin{split}
 \rbra{h_{L}}&= \rbra{\unity}T^{[o]}_{L}\left[\sum_{n>0}f(n)(T_{L})^{n-1}\right]T^{[o]}_{L}\\
 \rket{h_{R}}&= T^{[o]}_{R}\left[\sum_{n>0}f(n)(T_{R})^{n-1}\right]T^{[o]}_{R}\rket{\unity}.
\end{split}
\label{eq:hlr_LR}
\end{equation} 

With these definitions at hand we can write the left and right block Hamiltonians as
\begin{align}
 \rbra{H_{L}}&=\rbra{h_{L}}\sum_{n=0}^{\infty}[T_{L}]^{n}&
  \rket{H_{R}}&=\sum_{n=0}^{\infty}[T_{L}]^{n}\rket{h_{R}}.
  \label{eq:InfGS_LR}
\end{align}
These equations are exactly the same as Eq.~\eqref{eq:InfGS_NN} for the case of nearest neighbor interactions, but with different $\rbra{h_{L}}$ and $\rket{h_{R}}$. We can thus evaluate the geometric sums recursively or by solving a linear system iteratively, as explained in \Sec{sec:effH_NN}. Note that we again start by applying an energy shift $\rbra{h_L} \to \rbra{\tilde{h}_L} = \rbra{h_L} - e \rket{R} \rbra{\openone}$ and similar for $\rket{h_R}$, such that $(\tilde{h}_{L}|R)=(L|\tilde{h}_{R})=0$.

\begin{widetext}
We are now ready to formulate the action of $H_{A_{C}}$ onto $A^{s}_{C}$ as
\begin{equation}
 \begin{split}
 {A}^{\prime s}_{C}&=H_{L}A^{s}_{C} + A^{s}_{C}H_{R} + O_{L}\left[\sum_{t}o^{s}_{t}A^{t}_{C}\right] + \left[\sum_{t}o^{s}_{t}A^{t}_{C}\right] O_{R}
+\sum_{k}c_{k}\lambda_{k}\,O^{[k]}_{L}A^{s}_{C}\,O^{[k]}_{R}
\\
\begin{tikzpicture}[baseline = (X.base),every node/.style={scale=0.750},scale=.55]
\draw[rounded corners] (1,-.5) rectangle (2,.5);
\draw (0.5,0) -- (1,0); \draw (2,0) -- (2.5,0); \draw (1.5,-.5) -- (1.5,-1);
\draw (1.5,0) node {${A}^{\prime}_C$};
\end{tikzpicture} &= 
\begin{tikzpicture}[baseline = (X.base),every node/.style={scale=0.750},scale=.55]
\draw (0,0) circle (.5);
\draw (0,0) node (X) {$H_L$};
\draw (0,0.5) edge[out=90,in=180] (1,1.5);
\draw (0,-0.5) edge[out=270,in=180] (1,-1.5);
\draw[rounded corners] (1,2) rectangle (2,1);
\draw (1.5,1.5) node {$A_C$};
\draw (1,-1.5) edge[out=0,in=90] (1.25,-2);
\draw (1.25,-2) edge[out=-90,in=0] (1,-2.5);
\draw (1,-2.5) -- (.5,-2.5);
\draw (2,-1.5) edge[out=180,in=90] (1.75,-2);
\draw (1.75,-2) edge[out=-90,in=180] (2,-2.5);
\draw (2,-2.5) -- (2.5,-2.5);
\draw (2,+1.5) edge[out = 0, in =0] (2, -1.5);
\draw (1.5,-3) -- (1.5,1);
\end{tikzpicture}
+
\begin{tikzpicture}[baseline = (X.base),every node/.style={scale=0.750},scale=.55]
\draw (1,-1.5) edge[out=180,in=180] (1,1.5);
\draw[rounded corners] (1,2) rectangle (2,1);
\draw (1.5,1.5) node {$A_C$};
\draw (1,-1.5) edge[out=0,in=90] (1.25,-2);
\draw (1.25,-2) edge[out=-90,in=0] (1,-2.5);
\draw (1,-2.5) -- (0.5,-2.5);
\draw (2,-1.5) edge[out=180,in=90] (1.75,-2);
\draw (1.75,-2) edge[out=-90,in=180] (2,-2.5);
\draw (2,-2.5) -- (2.5,-2.5);
\draw (2,1.5) edge[out=0,in=90] (3,0.5);
\draw (3,0) circle (.5);
\draw (3,-.5) edge[out = -90, in = 0] (2, -1.5);
\draw (3,0) node {$H_R$};
\draw (1.5,-3) -- (1.5,1);
\end{tikzpicture}
+
\begin{tikzpicture}[baseline = (X.base),every node/.style={scale=0.750},scale=.55]
\draw (0,0) circle (.5); \draw (0,0) node {$O_L$};
\draw (1,-1.5) edge[out=180,in=270] (0,-0.5);
\draw (0,0.5) edge[out=90,in=180] (1,1.5);
\draw[rounded corners] (1,2) rectangle (2,1);
\draw (1.5,0) circle (.5);
\draw (1.5,0) node {$o$};
\draw (1.5,1.5) node {$A_C$};
\draw (1.5,1) -- (1.5,.5);
\draw (1.5,-3) -- (1.5,-.5);
\draw (2,+1.5) edge[out = 0, in =0] (2, -1.5);
\draw (1,-1.5) edge[out=0,in=90] (1.25,-2);
\draw (1.25,-2) edge[out=-90,in=0] (1,-2.5);
\draw (1,-2.5) -- (0.5,-2.5); \draw (2,-2.5) -- (2.5,-2.5);
\draw (2,-1.5) edge[out=180,in=90] (1.75,-2);
\draw (1.75,-2) edge[out=-90,in=180] (2,-2.5);
\end{tikzpicture}
+
\begin{tikzpicture}[baseline = (X.base),every node/.style={scale=0.750},scale=.55]
\draw (1,-1.5) edge[out=180,in=180] (1,1.5);
\draw[rounded corners] (1,2) rectangle (2,1);
\draw (1.5,0) circle (.5);
\draw (1.5,0) node {$o$};
\draw (1.5,1.5) node {$A_C$};
\draw (1.5,1) -- (1.5,.5);
\draw (1.5,-3) -- (1.5,-.5);
\draw (2,+1.5) edge[out=0,in=90] (3,0.5);
\draw (3,-0.5) edge[out=270,in=0] (2,-1.5);
\draw (3,0) circle (.5); \draw (3,0) node {$O_R$};
\draw (1,-1.5) edge[out=0,in=90] (1.25,-2);
\draw (1.25,-2) edge[out=-90,in=0] (1,-2.5);
\draw (1,-2.5) -- (0.5,-2.5); \draw (2,-2.5) -- (2.5,-2.5);
\draw (2,-1.5) edge[out=180,in=90] (1.75,-2);
\draw (1.75,-2) edge[out=-90,in=180] (2,-2.5);
\end{tikzpicture}
+  \sum_k c_k\lambda_k\;
\begin{tikzpicture}[baseline = (X.base),every node/.style={scale=0.750},scale=.55]
\draw (0,0) circle (.5); \draw (0,0) node {$O_L^{[k]}$};
\draw (1,-1.5) edge[out=180,in=270] (0,-0.5);
\draw (0,0.5) edge[out=90,in=180] (1,1.5);
\draw[rounded corners] (1,2) rectangle (2,1);
\draw (1.5,1.5) node {$A_C$};
\draw (1.5,-3) -- (1.5,1);
\draw (2,+1.5) edge[out=0,in=90] (3,0.5);
\draw (3,-0.5) edge[out=270,in=0] (2,-1.5);
\draw (3,0) circle (.5); \draw (3,0) node {$O_R^{[k]}$};
\draw (1,-1.5) edge[out=0,in=90] (1.25,-2);
\draw (1.25,-2) edge[out=-90,in=0] (1,-2.5);
\draw (1,-2.5) -- (0.5,-2.5); \draw (2,-2.5) -- (2.5,-2.5);
\draw (2,-1.5) edge[out=180,in=90] (1.75,-2);
\draw (1.75,-2) edge[out=-90,in=180] (2,-2.5);
\end{tikzpicture} 
 \end{split}
 \label{eq:HontoAC_LR}
\end{equation}
The additional factor of $\lambda_{k}$ in the sum in the last term arises due to $A^{s}_{C}$ adding an additional site between the left and right operators $o$. 
Similarly, the action of $H_{C}$ onto $C$ becomes
\begin{equation}
 \begin{split}
 {C^{\prime}} &= H_{L}C + C H_{R} + \sum_{k}c_{k}O^{[k]}_{L}CO^{[k]}_{R}
\\
 \begin{tikzpicture}[baseline = (X.base),every node/.style={scale=0.750},scale=.55]
\draw (1.5,0) circle (.5);
\draw (0.5,0) -- (1,0); \draw (2,0) -- (2.5,0);
\draw (1.5,0) node {${C^{\prime}}$};
\end{tikzpicture} &=  
\begin{tikzpicture}[baseline = (X.base),every node/.style={scale=0.750},scale=.55]
\draw (0,0) circle (.5);
\draw (0,0) node (X) {$H_L$};
\draw (0,0.5) edge[out=90,in=180] (1,1.5);
\draw (0,-0.5) edge[out=270,in=180] (1,-1.5);
\draw (1.5,1.5) circle (.5);
\draw (1.5,1.5) node {$C$};
\draw (1,-1.5) edge[out=0,in=90] (1.25,-2);
\draw (1.25,-2) edge[out=-90,in=0] (1,-2.5);
\draw (1,-2.5) -- (0,-2.5);
\draw (2,-1.5) edge[out=180,in=90] (1.75,-2);
\draw (1.75,-2) edge[out=-90,in=180] (2,-2.5);
\draw (2,-2.5) -- (3,-2.5);
\draw (2,+1.5) edge[out = 0, in =0] (2, -1.5);
\end{tikzpicture}
+
\begin{tikzpicture}[baseline = (X.base),every node/.style={scale=0.750},scale=.55]
\draw (1,-1.5) edge[out=180,in=180] (1,1.5);
\draw (1.5,1.5) circle (.5);
\draw (1.5,1.5) node {$C$};
\draw (1,-1.5) edge[out=0,in=90] (1.25,-2);
\draw (1.25,-2) edge[out=-90,in=0] (1,-2.5);
\draw (1,-2.5) -- (0,-2.5);
\draw (2,-1.5) edge[out=180,in=90] (1.75,-2);
\draw (1.75,-2) edge[out=-90,in=180] (2,-2.5);
\draw (2,-2.5) -- (3,-2.5);
\draw (2,1.5) edge[out=0,in=90] (3,0.5);
\draw (3,0) circle (.5);
\draw (3,-.5) edge[out = -90, in = 0] (2, -1.5);
\draw (3,0) node {$H_R$};
\end{tikzpicture} 
+ \sum_k c_k \;
\begin{tikzpicture}[baseline = (X.base),every node/.style={scale=0.750},scale=.55]
\draw (0,0) circle (.5);
\draw (0,0) node (X) {$O_L^{[k]}$};
\draw (0,0.5) edge[out=90,in=180] (1,1.5);
\draw (0,-0.5) edge[out=270,in=180] (1,-1.5);
\draw (1.5,1.5) circle (.5);
\draw (1.5,1.5) node {$C$};
\draw (1,-1.5) edge[out=0,in=90] (1.25,-2);
\draw (1.25,-2) edge[out=-90,in=0] (1,-2.5);
\draw (1,-2.5) -- (0,-2.5);
\draw (2,-1.5) edge[out=180,in=90] (1.75,-2);
\draw (1.75,-2) edge[out=-90,in=180] (2,-2.5);
\draw (2,-2.5) -- (3,-2.5);
\draw (2,1.5) edge[out=0,in=90] (3,0.5);
\draw (3,0) circle (.5);
\draw (3,-.5) edge[out = -90, in = 0] (2, -1.5);
\draw (3,0) node {$O_R^{[k]}$};
\end{tikzpicture}
 \end{split}
 \label{eq:HontoC_LR}
\end{equation}
\end{widetext}

\label{sec:effH_LR}
\begin{table*}[t]
\begin{minipage}{\linewidth}
\begin{algorithm}[H]
  \caption{Explicit terms of effective Hamiltonians with long range interactions and their application onto a state}
  \label{alg:Heff_LR}
   \begin{algorithmic}[1]
   \Require operator $o$ defining \eqref{eq:hj_LR}, parameters $c_{k}$ and $\lambda_{k}$ defining \eqref{eq:sumofexp}, current uMPS tensors $A_{L}$, $A_{R}$ in left and right gauge, left dominant eigenvector $\rbra{L}$ of $T_{R}$, right dominant eigenvector $\rket{R}$ of $T_{L}$, desired precision $\epsilon_{\rm S}$ for terms involving infinite geometric sums
  \Ensure Explicit terms of effective Hamiltonians $H_{A_{C}}$ and $H_{C}$, updated ${A}^{\prime}_{C}$ and $C^{\prime}$
  
  \Function{HeffTerms}{$H=\{o,\{c_{k}\},\{\lambda_{k}\}\}$,$A_{L}$,$A_{R}$,$L$,$R$,$\epsilon_{\rm S}$} \Comment{Calculates explicit terms of effective Hamiltonians}
   \State Calculate $O^{[k]}_{L}$ and $O^{[k]}_{R}$ by iteratively solving \eqref{eq:GSconv_LSE} for each $\lambda_{k}$ to machine precision
   \State Calculate single environment contributions $O_{L}$ and $O_{R}$ from \eqref{eq:O_LR} and $h_{L}$ and $h_{R}$ from \eqref{eq:hl_LR}
    \State Calculate $H_{L}$ and $H_{R}$ by iteratively solving \eqref{eq:HLR_NN_iterative} or (preferably) \eqref{eq:HLR_NN}, to precision $\epsilon_{\rm S}$
   \State $H_{A_{C}}\gets\{o,\{c_{k}\},\{\lambda_{k}\},\{O^{k}_{L}\},\{O^{k}_{R}\},O_{L},O_{R},H_{L},H_{R}\}$
   \State $H_{C}\gets\{\{c_{k}\},\{O^{k}_{L}\},\{O^{k}_{R}\},H_{L},H_{R}\}$
    \State \Return $H_{A_{C}},H_{C}$
  \EndFunction
  
   \Function{ApplyHAC}{$A_{C}$,$H_{A_{C}}$} 
   \Comment{Terms of $H_{A_{C}}$ from \Call{HeffTerms}{$H$,$A_{L}$,$A_{R}$,$L$,$R$,$\epsilon_{\rm S}$}}
   \State Calculate updated ${A}^{\prime}_{C}$ from \eqref{eq:HontoAC_LR}
   \State \Return ${A}^{\prime}_{C}$
   \EndFunction
   
      \Function{ApplyHC}{$C$,$H_{C}$} 
      \Comment{Terms of $H_{C}$ from \Call{HeffTerms}{$H$,$A_{L}$,$A_{R}$,$L$,$R$,$\epsilon_{\rm S}$}}
   \State Calculate updated ${C^{\prime}}$ from \eqref{eq:HontoC_LR}
   \State \Return ${C^{\prime}}$
   \EndFunction
   \end{algorithmic}
\end{algorithm}
\end{minipage}
\caption{
Pseudocode for obtaining the explicit terms of the effective Hamiltonians $H_{A_{C}}$ and $H_{C}$ for systems with with long range interactions and their applications onto a state.
}
\label{tab:Heff_LR}
\end{table*}

In \eqref{eq:HontoAC_LR} the first two terms can be applied in $\O(dD^{3})$, the second two in $\O(d^{2}D^{2}) + \O(dD^{3})$ and the last term in $\O(KdD^{3})$ operations, and in \eqref{eq:HontoC_LR} the first two terms in $\O(D^{3})$ and the last term in $\O(KD^{3})$ operations.
In general we have to perform $2(K+1)$ iterative inversions involving $\O(D^{3})$ operations and collect $K$ terms to arrive at the necessary terms for \eqref{eq:HontoAC_LR} and \eqref{eq:HontoC_LR}, where the solutions from the previous iteration can be used as starting vectors to speed up convergence.

If there are additional simple single or nearest neighbor two-site terms present in the Hamiltonian, appropriate terms as described in \Sec{sec:effH_NN} can be added. For a pseudocode summary for obtaining the necessary explicit terms of $H_{A_{C}}$ and $H_{C}$ for Hamiltonians with long range interactions, and their applications onto a state, required for solving the effective eigenvalue problems using an iterative eigensolver, see \Tab{tab:Heff_LR}.

\subsection{General Hamiltonians given in terms of MPOs}
\label{sec:effH_MPO}

Consider the Hamiltonian $H$ given in terms of an infinite Matrix Product Operator (MPO) \cite{MPO1,DMRG_MC,MPO3,MPO4,MPOlong1,MPOlong2,MPOinf} with 4-index MPO elements 
$W^{ab}_{ss'}$ with $a,b=1,\ldots,d_{W}$ and $s,s'=1,\ldots,d$ and we call $d_{W}$ the MPO bond dimension.
In terms of the operator valued matrices $\hat{W}^{ab}=\sum_{ss'}W^{ab}_{ss'}\ket{s}\bra{s'}$ the Hamiltonian can then be written as
\begin{align*}
 H &= \hat{w}_{L}\left[\prod_{j\in\ints}\hat{W}_{[j]}\right]\hat{w}_{R} \\
 &= \dots \begin{tikzpicture}[baseline = (X.base),every node/.style={scale=0.550},scale=.55]
\draw (0.5,1.5) -- (1,1.5); 
\draw[rounded corners] (1,2) rectangle (2,1);
\draw (1.5,1.5) node (X) {$W$};
\draw (2,1.5) -- (3,1.5); 
\draw[rounded corners] (3,2) rectangle (4,1);
\draw (3.5,1.5) node {$W$};
\draw (4,1.5) -- (5,1.5);
\draw[rounded corners] (5,2) rectangle (6,1);
\draw (5.5,1.5) node {$W$};
\draw (6,1.5) -- (7,1.5); 
\draw[rounded corners] (7,2) rectangle (8,1);
\draw (7.5,1.5) node {$W$};
\draw (8,1.5) -- (9,1.5); 
\draw[rounded corners] (9,2) rectangle (10,1);
\draw (9.5,1.5) node {$W$};
\draw (10,1.5) -- (10.5,1.5);
\draw (1.5,2) -- (1.5,2.5); \draw (3.5,2) -- (3.5,2.5); \draw (5.5,2) -- (5.5,2.5);
\draw (7.5,2) -- (7.5,2.5); \draw (9.5,2) -- (9.5,2.5);
\draw (1.5,1) -- (1.5,.5); \draw (3.5,1) -- (3.5,.5); \draw (5.5,1) -- (5.5,.5);
\draw (7.5,1) -- (7.5,.5); \draw (9.5,1) -- (9.5,.5);
\end{tikzpicture} \dots
\end{align*}
where $\hat{W}_{[j]}$ contains operators acting on site $j$ only and $\hat{w}_{L}$ and $\hat{w}_{R}$ are operator valued boundary vectors.

An example for such an MPO decomposition for the Transverse Field Ising (TFI) Hamiltonian with exponentially decaying long range interaction 
\begin{equation*}
H_{\rm TFI}=-J\sum_{j}\sum_{n>0}\lambda^{n-1}X_{j}X_{j+n}-h\sum_{j}Z_{j}
\end{equation*} 
with $\lambda<1$ is given by
\begin{equation}
\begin{split}
 \hat{W}&=
 \begin{bmatrix}
\unity&0&0\\
-JX&\lambda\unity&0\\
-hZ&X&\unity
\end{bmatrix}\\
 \hat{w}_{L}&=
 \begin{bmatrix} -hZ& X&\unity\end{bmatrix}\\
  \hat{w}_{R}&=
  \begin{bmatrix}\unity & -JX&-hZ \end{bmatrix}^{T},
\end{split}
\label{eq:TFI_MPO}
\end{equation} 
where $X$ and $Z$ are Pauli matrices. For the TFI Hamiltonian we thus have $d_{W}=3$ and the limit $\lambda=0$ corresponds to the nearest neighbor interaction case.

In order to efficiently apply the effective Hamiltonians $H_{A_{C}}$ and $H_{C}$, it is necessary to determine the left and right (quasi) fixed points $L^{[W]}_{a}$ and $R^{[W]}_{a}$ of the MPO transfer matrices
\begin{equation}
 {T^{[W]}_{L/R}}^{ab}=\sum_{ss'} W^{ab}_{s's}\bar{A}_{L/R}^{s'}\otimes A_{L/R}^{s},
 \label{eq:MPOTM}
\end{equation} 
where -- similar to MPS tensors -- $L^{[W]}_{a}$ and $R^{[W]}_{a}$ are collections of $d_{W}$ matrices of dimension $D\times D$, with $a=1,\ldots,d_{W}$. These two objects are in fact the thermodynamic limit versions of the objects defined in Eq.~(190) and (191) in Ref.~\onlinecite{MPS6_S}.

Typically, MPO representations $\hat{W}^{ab}$ of (quasi)local Hamiltonians (such as e.g.\ Eq.~\eqref{eq:TFI_MPO}) are of Schur form,\cite{MPOinf} such that the MPO transfer matrix contains Jordan blocks and that the dominant eigenvalue is one and of twofold algebraic degeneracy. Such MPO transfer matrices therefore technically do not have well defined fixed points. We can however find quasi fixed points $L^{[W]}_{a}$ and $R^{[W]}_{a}$, that are fixed points up to a term contributing to the energy density expectation value in one of the $d_{W}$ elements of $L^{[W]}_{a}$ and $R^{[W]}_{a}$. An application of ${T^{[W]}_{L/R}}^{ab}$ onto both quasi fixed points will therefore accumulate an additional term contributing to the extensive global energy expectation value. Similar to the terms in Eq.~\eqref{eq:InfGS_NN} or Eq.~\eqref{eq:InfGS_LR} in the previous cases involving infinite geometric sums, we can however safely discard these diverging contributions, which is equivalent to setting the energy expectation values of the semi-infinite left and right half of the system to zero (see below). 

In the following we briefly reiterate the procedure of Ref.~\onlinecite{MPOinf} to systematically determine $L^{[W]}_{a}$ and $R^{[W]}_{a}$ from given $\hat{W}^{ab}$ and $A^{s}_{L/R}$. The obtained solutions will of course contain the results of \Sec{sec:effH_NN} and \Sec{sec:effH_LR} as special cases.

Without loss of generality we assume $\hat{W}^{ab}$ to be of lower triangular form, i.e.\ $\hat{W}^{ab}=0,\forall b>a$. Furthermore, we assume the typical case of any nonzero diagonal elements being proportional to the identity, i.e.\ $\hat{W}^{aa}=\lambda_{a}\unity$, where $\lambda_{a}\leq1$ and usually $\lambda_{1}=\lambda_{d_{W}}=1$, as is e.g.\ the case in \eqref{eq:TFI_MPO}. 
\footnote{The generalization to nontrivial operators on the diagonal, such as e.g. Pauli-$Z$ operators in the case of long range fermion hopping, is straight forward.}
By defining the result of the action of the MPO transfer matrix as
\begin{align}
\rbra{{Y_{L}}_{a}}&= \sum_{b>a}\rbra{L^{[W]}_{b}}{T^{[W]}_{L}}^{ba}\label{eq:Y_MPO_L}\\
\rket{{Y_{R}}_{a}}&= \sum_{b<a}{T^{[W]}_{R}}^{ab}\rket{R^{[W]}_{b}}\label{eq:Y_MPO_R},
\end{align}
the system of fixed point equations can be written as
\begin{align}
\rbra{L^{[W]}_{a}} &= \rbra{L^{[W]}_{a}} {T^{[W]}_{L}}^{aa} + \rbra{{Y_{L}}_{a}}\label{eq:MPO_FPL}\\
\rket{R^{[W]}_{a}} &= {T^{[W]}_{R}}^{aa} \rket{R^{[W]}_{a}}  + \rket{{Y_{R}}_{a}}\label{eq:MPO_FPR}.
\end{align}
Notice that due to the lower triangular structure of $\hat{W}^{ab}$, the terms $\rbra{{Y_{L}}_{a}}$  and $\rket{{Y_{R}}_{a}}$ only contain contributions from $\rbra{L^{[W]}_{b>a}}$ and $\rket{R^{[W]}_{b<a}}$ and we can solve 
\eqref{eq:MPO_FPL} and \eqref{eq:MPO_FPR} 
recursively, starting with $a=d_{W}$ for $L^{[W]}_{a}$ and with $a=1$ for $R^{[W]}_{a}$, which initially amounts to $\rbra{L^{[W]}_{d_{W}}}=\rbra{\unity}$ and $\rket{R^{[W]}_{1}}=\rket{\unity}$. Terms with  ${T^{[W]}_{L/R}}^{aa}=0$ are particularly simple and simply reduce to the identification $\rbra{L^{[W]}_{a}}= \rbra{{Y_{L}}_{a}}$ and $\rket{R^{[W]}_{a}}= \rket{{Y_{R}}_{a}}$.

Terms with ${T^{[W]}_{L/R}}^{aa}=\lambda_{a}T_{L/R}$ where $\lambda_{a}<1$, now result in solutions of the form 
\begin{align}
\rbra{L^{[W]}_{a}} &= \rbra{{Y_{L}}_{a}}[\unity - \lambda_{a}T_{L}]^{-1} \label{eq:MPO_FP_T2_L}\\
\rket{R^{[W]}_{a}} &= [\unity - \lambda_{a}T_{R}]^{-1}\rket{{Y_{R}}_{a}},\label{eq:MPO_FP_T2_R}
\end{align} 
equivalent to terms such as \eqref{eq:GSconv} stemming from infinite geometric sums of (weighted) MPS transfer matrices.

Equivalently, terms with ${T^{[W]}_{L/R}}^{aa}=T_{L/R}$ then result in relations of the form
\begin{align}
\rbra{L^{[W]}_{a}}[\unity - T_{L}] &= \rbra{{Y_{L}}_{a}}\label{eq:MPO_FP_T1_L}\\
[\unity - T_{R}]\rket{R^{[W]}_{a}} &= \rket{{Y_{R}}_{a}},\label{eq:MPO_FP_T1_R}
\end{align}
which in general do not have a formal solution, since the left hand sides of these equations live in the subspace orthogonal to the dominant eigenspaces of $T_{L/R}$, while the right hand sides generally do have contributions in the dominant eigenspace. We can however discard these contributions by projecting onto the complementary subspace, and then obtain $\rbra{L^{[W]}_{a}}$ and $\rket{R^{[W]}_{a}}$ by solving the systems of equations (see also \App{sec:invertE})
\begin{subequations}
\label{eq:MPO_FP_T3}
\begin{align}
 \rbra{L^{[W]}_{a}}[\unity - T_{L} + \rket{R}\rbra{\unity}] &= \rbra{{Y_{L}}_{a}} -\rbraket{{Y_{L}}_{a}}{R}\rbra{\unity} \label{eq:MPO_FP_T3_L}\\
 [\unity - T_{R} + \rket{\unity}\rbra{L}]\rket{R^{[W]}_{a}} &= \rket{{Y_{R}}_{a}}- \rket{\unity}\rbraket{L}{{Y_{R}}_{a}} \label{eq:MPO_FP_T3_R}
\end{align}
\end{subequations}
We have encountered exactly the same type of equations in \eqref{eq:HLR_NN} when evaluating infinite geometric sums of transfer matrices, after a constant shift in energy to remove diverging terms. The MPO formalism thus automatically yields these contributions in a form where the sums have already been explicitly performed.

\begin{table*}[p]
\begin{minipage}{\linewidth}
\begin{algorithm}[H]
  \caption{Explicit terms of effective Hamiltonians in MPO form and their application onto a state}
  \label{alg:Heff_MPO}
   \begin{algorithmic}[1]
   \Require MPO $\hat{W}$ defining the Hamiltonian, current uMPS tensors $A_{L}$, $A_{R}$ in left and right gauge, left dominant eigenvector $\rbra{L}$ of $T_{R}$, right dominant eigenvector $\rket{R}$ of $T_{L}$, desired precision $\epsilon_{\rm S}$ for iterative solution of linear system of equations
  \Ensure Explicit terms of effective Hamiltonians $H_{A_{C}}$ and $H_{C}$, updated ${A}^{\prime}_{C}$ and $C^{\prime}$
  
  \Function{HeffTerms}{$H=\hat{W}$,$A_{L}$,$A_{R}$,$L$,$R$,$\epsilon_{\rm S}$} \Comment{Calculates explicit terms of effective Hamiltonians}
  \State $L^{[W]}\gets\,$\Call{CalcLW}{$\hat{W}$,$A_{L}$,$R$,$\epsilon_{\rm S}$}
  \State $R^{[W]}\gets\,$\Call{CalcRW}{$\hat{W}$,$A_{R}$,$L$,$\epsilon_{\rm S}$}
   \State $H_{A_{C}}\gets\{\hat{W},L^{[W]},R^{[W]}\}$
   \State $H_{C}\gets\{L^{[W]},R^{[W]}\}$
   \State \Return $H_{A_{C}},H_{C}$
  \EndFunction
  
  \Function{CalcLW($\hat{W}$,$A_{L}$,$R$,$\epsilon_{S}$)}{}
  \Comment{Calculates left quasi fixed point of MPO transfer matrix $T^{[W]}_{L}$}
  \State $\rbra{L^{[W]}_{d_{w}}}\gets \rbra{\unity}$
  \For{$a=d_{w}-1,\ldots,1$}
  \State Calculate $\rbra{{Y_{L}}_{a}}$ from \eqref{eq:Y_MPO_L}
  \If{${T^{[W]}_{L}}^{aa}==\lambda_{a}T_{L}$}
  \State Calculate $\rbra{L^{[W]}_{a}}$ by iteratively solving \eqref{eq:MPO_FP_T2_L} to machine precision
  \ElsIf{${T^{[W]}_{L}}^{aa}==T_{L}$}
  \State Calculate $\rbra{L^{[W]}_{a}}$ by iteratively solving \eqref{eq:MPO_FP_T3_L} to precision $\epsilon_{S}$
  \ElsIf{${T^{[W]}_{L}}^{aa}==0$}
  \State $\rbra{L^{[W]}_{a}}\gets \rbra{{Y_{L}}_{a}}$
  \EndIf 
  \EndFor
  \State \Return $L^{[W]}$.
  \EndFunction
  
  \Function{CalcRW}{$\hat{W}$,$A_{R}$,$L$,$\epsilon_{\rm S}$}
  \Comment{Calculate right quasi fixed point of MPO transfer matrix $T^{[W]}_{R}$}
  \State $\rket{R^{[W]}_{1}}\gets \rket{\unity}$
  \For{$a=2,\ldots,d_{w}$}
  \State Calculate $\rket{{Y_{R}}_{a}}$ from \eqref{eq:Y_MPO_R}
  \If{${T^{[W]}_{R}}^{aa}==\lambda_{a}T_{R}$}
  \State Calculate $\rket{R^{[W]}_{a}}$ by iteratively solving \eqref{eq:MPO_FP_T2_R} to machine precision
  \ElsIf{${T^{[W]}_{R}}^{aa}==T_{R}$}
  \State Calculate $\rket{R^{[W]}_{a}}$ by iteratively solving \eqref{eq:MPO_FP_T3_R} to precision $\epsilon_{S}$
  \ElsIf{${T^{[W]}_{R}}^{aa}==0$}
  \State $\rket{R^{[W]}_{a}}\gets \rket{{Y_{R}}_{a}}$
  \EndIf 
  \EndFor
  \State \Return $R^{[W]}$.
  \EndFunction
  
   \Function{ApplyHAC}{$A_{C}$,$H_{A_{C}}$}
      \Comment{Terms of $H_{A_{C}}$ from \Call{HeffTerms}{$H$,$A_{L}$,$A_{R}$,$L$,$R$,$\epsilon_{\rm S}$}}
   \State Calculate updated ${A}^{\prime}_{C}$ from \eqref{eq:HontoAC_MPO}
   \State \Return ${A}^{\prime}_{C}$
   \EndFunction
   
      \Function{ApplyHC}{$C$,$H_{A_{C}}$}
      \Comment{Terms of $H_{C}$ from \Call{HeffTerms}{$H$,$A_{L}$,$A_{R}$,$L$,$R$,$\epsilon_{\rm S}$}}
   \State Calculate updated ${C^{\prime}}$ from \eqref{eq:HontoC_MPO}
   \State \Return ${C^{\prime}}$
   \EndFunction
   \end{algorithmic}
\end{algorithm}
\end{minipage}
\caption{
Pseudocode for obtaining the explicit terms of the effective Hamiltonians $H_{A_{C}}$ and $H_{C}$ for general Hamiltonians in MPO form and their applications onto a state.
}
\label{tab:Heff_MPO}
\end{table*}

Such a situation typically occurs only for the final terms in the recursive solution of the fixed point equations, i.e.\ for $\rbra{L^{[W]}_{1}}$ and $\rket{R^{[W]}_{d_{W}}}$. A concrete evaluation (see below) of the discarded terms in these cases shows that they correspond to contributions to the energy density expectation value, i.e.\ discarding these terms is equivalent to a constant shift in energy, such that the energy density is zero and we have $(L^{[W]}_{1}|R)=(L|R^{[W]}_{d_{W}})=0$. After applying $T^{[W]}_{L/R}$ once onto the quasi fixed points we thus have for the first element of $L^{[W]}$ and the last element of $R^{[W]}$
\begin{equation}
\begin{split}
  \rbra{{Y_{L}}_{1}} &= \rbra{L^{[W]}_{1}} +  \rbraket{{Y_{L}}_{1}}{R}\rbra{\unity}\\
\rket{{Y_{R}}_{d_{W}}} &= \rket{R^{[W]}_{d_{W}}} + \rket{\unity}\rbraket{L}{{Y_{R}}_{d_{w}}},
 \end{split}
\end{equation}
i.e.\ the fixed point relations only hold up to an additive diagonal correction for these elements. These corrections correspond to the energy density expectation value
\begin{equation}
  e =\rbraket{{Y_{L}}_{1}}{R}= \rbraket{L}{{Y_{R}}_{d_{w}}}
\end{equation} 
and they can in fact be used for its evaluation. 

As a concrete example, for the long range TFI Hamiltonian given by MPO \eqref{eq:TFI_MPO} we obtain
\begin{align}
 \rbra{L^{[W]}_{1}}[\unity - T_{L}]&= -h\,\rbra{\unity}T^{Z}_{L} -J\,\rbra{\unity}T^{X}_{L}[\unity - \lambda T_{L}]^{-1}T^{X}_{L}\notag\\
 \rbra{L^{[W]}_{2}}&=\rbra{\unity}T^{X}_{L}[\unity - \lambda T_{L}]^{-1}\notag\\
 \rbra{L^{[W]}_{3}}&=\rbra{\unity}\notag
\end{align}
and
\begin{align}
 \rket{R^{[W]}_{1}} &= \rket{\unity}\notag\\
 \rket{R^{[W]}_{2}} &= -J[\unity-\lambda T_{R}]^{-1}T^{X}_{R}\rket{\unity}\notag\\
[\unity-T_{R}] \rket{R^{[W]}_{3}} &= -h\,T^{Z}_{R}\rket{\unity} -J\, T^{X}_{R}[\unity-\lambda T_{R}]^{-1}T^{X}_{R}\rket{\unity}.\notag
\end{align}

Having determined the left and right quasi fixed points of the MPO transfer matrices, it is now particularly easy to calculate the action of the effective Hamiltonians $H_{A_{C}}$ onto $A^{s}_{C}$ as
\begin{equation}
\begin{split}
 {A}^{\prime s}_{C} &= \sum_{abt}W^{ab}_{st}\,L^{[W]}_{a}\,A^{t}_{C}\,R^{[W]}_{b}\\
 \begin{tikzpicture}[baseline = (X.base),every node/.style={scale=0.750},scale=.55]
\draw[rounded corners] (1,1) rectangle (2,2);
\draw (0.5,1.5) -- (1,1.5); \draw (2,1.5) -- (2.5,1.5); \draw (1.5,1) -- (1.5,0.5);
\draw (1.5,1.5) node {${A}^{\prime}_C$};
\end{tikzpicture} 
 &= \;
 \begin{tikzpicture}[baseline = (X.base),every node/.style={scale=0.750},scale=.55]
\draw[rounded corners] (3,4) rectangle (4,-1);
\draw (3.5,1.5) node {$L^{[W]}$};
\draw[rounded corners] (7,4) rectangle (8,-1);
\draw (7.5,1.5) node {$R^{[W]}$};
\draw (4,1.5) -- (5,1.5);
\draw[rounded corners] (5,2) rectangle (6,1);
\draw (5.5,1.5) node {$W$};
\draw (6,1.5) -- (7,1.5); 
\draw (4,3.5) -- (5,3.5);
\draw[rounded corners] (5,4) rectangle (6,3);
\draw (5.5,3.5) node {$A_C$};
\draw (6,3.5) -- (7,3.5); 
\draw (5.5,2) -- (5.5,3); \draw (5.5,1) -- (5.5,-2);
\draw (4,-0.5) -- (5,-0.5); edge\draw (5,-0.5) edge[out=0,in=0] (5,-1.5); \draw (4.5,-1.5) -- (5,-1.5);
\draw (6,-0.5) -- (7,-0.5); \draw (6,-0.5) edge[out=180,in=180] (6,-1.5); \draw (6,-1.5) -- (6.5,-1.5);
\draw (5,1.5) node (X) {$\phantom{X}$};
\end{tikzpicture} 
\end{split}
\label{eq:HontoAC_MPO}
\end{equation} 
and equivalently the action of $H_{C}$ onto $C$ as
\begin{equation}
\begin{split}
 {C^{\prime}} &= \sum_{a}L^{[W]}_{a}\,C\,R^{[W]}_{a}\\
  \begin{tikzpicture}[baseline = (X.base),every node/.style={scale=0.750},scale=.55]
\draw (1.5,1.5) circle (.5);
\draw (0.5,1.5) -- (1,1.5); \draw (2,1.5) -- (2.5,1.5);
\draw (1.5,1.5) node {${C^{\prime}}$};
\end{tikzpicture} &= \; 
 \begin{tikzpicture}[baseline = (X.base),every node/.style={scale=0.750},scale=.55]
\draw[rounded corners] (3,4) rectangle (4,-1);
\draw (3.5,1.5) node {$L^{[W]}$};
\draw[rounded corners] (7,4) rectangle (8,-1);
\draw (7.5,1.5) node {$R^{[W]}$};
\draw (4,1.5) -- (7,1.5);
\draw (4,3.5) -- (5,3.5);
\draw (5.5,3.5) circle (.5);
\draw (5.5,3.5) node {$C$};
\draw (6,3.5) -- (7,3.5); 
\draw (4,-0.5) -- (5,-0.5); edge\draw (5,-0.5) edge[out=0,in=0] (5,-1.5); \draw (4.5,-1.5) -- (5,-1.5);
\draw (6,-0.5) -- (7,-0.5); \draw (6,-0.5) edge[out=180,in=180] (6,-1.5); \draw (6,-1.5) -- (6.5,-1.5);
\draw (5,1.5) node (X) {$\phantom{X}$};,
\end{tikzpicture} 
\end{split}
\label{eq:HontoC_MPO}
\end{equation} 
which can be performed in $\O(dd_{W}D^{3}) + \O(d^{2}d_{W}^{2}D^{2})$, respective $\O(d_{W}D^{3})$ operations. In total we also have to perform an iterative inversion for each diagonal element of $\hat{W}$.

This framework is very flexible, general and powerful, once a routine for determining quasi fixed points of general MPO transfer matrices has been implemented. The effective Hamiltonians of \Sec{sec:effH_NN} and \Sec{sec:effH_LR} are contained within as special cases. A pseudocode summary for obtaining the necessary explicit terms of $H_{A_{C}}$ and $H_{C}$ for Hamiltonians given in terms of an MPO, and their applications onto a state, required for solving the effective eigenvalue problems using an iterative eigensolver, is presented in \Tab{tab:Heff_MPO}.

\section{Geometric Sums of Transfer Matrices}
\label{sec:invertE}
We wish to evaluate terms involving infinite geometric sums of the form
\begin{align}
\rbra{y} &= \rbra{x} \sum_{n=0}^{\infty}T^{n}&
\rket{y} &=  \sum_{n=0}^{\infty}T^{n}\rket{x}. \label{eq:def}
\end{align} 
Such expressions typically arise in situations where one sums up contributions of successive applications of $T$ onto some fixed virtual boundary vector $x$, with the initial contribution being the boundary vector $x$ itself. 
This is reflected in the above expression by summing from $n=0$ and using the definition $T^{0}=\unity$. 

We assume a spectral decomposition of the transfer matrix given by
\begin{equation}
 T=\sum_{j=0}^{D^{2}-1}\lambda_{j}\rproj{j},
\end{equation} 
where the left and right eigenvectors are mutually orthonormal, i.e.\ $(j|k)=\delta_{jk}$. Note that $T$ is in general not hermitian and thus $\rbra{j}\neq\rket{j}^{\dagger}$.

For a generic injective normalized state, $T$ has a unique eigenvalue of largest magnitude given by $\lambda_{0}=1$, whereas all other eigenvalues are contained in the unit circle ($\lvert \lambda_{j>0}\rvert < 1$).

We divide into dominant and complementary subspaces and get for powers of $T$
\begin{equation}
 T^{n}=\PDS + \sum_{j=1}^{D^{2}-1}\lambda_{j}^{n}\rproj{j}.
\end{equation}

We can safely perform the geometric sum for all eigenvalues $|\lambda_{j>0}|<1$, while $\lambda_{0}=1$ contributes a formally diverging term
\begin{align}
 \sum_{n=0}^{\infty}T^{n}&= \sum_{n=0}^{\infty}\PDS + \sum_{j=1}^{D^{2}-1} \sum_{n=0}^{\infty}\lambda_{j}^{n}\rproj{j}\\
 &=|\mathbb{N}|\PDS + \sum_{j=1}^{D^{2}-1} (1-\lambda_{j})^{-1}\rproj{j}.\label{eq:TMpow}
\end{align} 
The interpretation of this diverging contribution depends on the situation. By using the projectors 
\begin{align}
 P&=\PDS & Q&=\unity-\PDS
\end{align}
onto the dominant and complementary subspaces we define the projected transfer matrix
\begin{equation}
 \T=\sum_{j=1}^{D^{2}-1}\lambda_{j}\rket{j}\rbra{j}=QT=TQ=T-P.
 \label{eq:PTM}
\end{equation} 

We realize that the spectral decomposition of $(\unity - \T)^{-1}$ has a component of $\PDS$
\begin{equation}
 (\unity - \T)^{-1} = \PDS + \sum_{j=1}^{D^{2}-1} (1-\lambda_{j})^{-1}\rproj{j}
\end{equation} 
and therefore identify the second term in \eqref{eq:TMpow} as
\begin{equation}
 \sum_{j=1}^{D^{2}-1} (1-\lambda_{j})^{-1}\rproj{j} = Q(\unity - \T)^{-1}Q.
\end{equation} 
For the geometric sum we then obtain
\begin{equation}
 \sum_{n=0}^{\infty}T^{n} = |\mathbb{N}|\PDS + Q(\unity - \T)^{-1}Q
\end{equation} 
with a diverging contribution from $P$. Plugging into \eqref{eq:def} we finally get
\begin{equation}
\begin{split}
 \rbra{y} &= |\mathbb{N}|\, (x|0)\,\rbra{0} + \rbra{x} Q(\unity - \T)^{-1}\\
 \rket{y} &= |\mathbb{N}|\, \rket{0}\,(0|x) + (\unity - \T)^{-1}Q\rket{x}.
\end{split}
\end{equation} 

Usually it is not necessary to calculate the full matrix expression of $\sum_{n}T^{n}$, but to just act with it onto some $\rbra{x}$ or $\rket{x}$. The diverging contributions can typically be safely discarded, as they correspond to a constant (albeit infinite) offset of some extensive observable (e.g.\ the Hamiltonian). The action of the finite remaining part can be calculated efficiently by iteratively solving the linear system of equations of the type $A\vec{y}=\vec{x}$ or $\vec{y}^{\dagger}A=\vec{x}^{\dagger}$
\begin{equation}
\begin{split}
\rbra{y} (\unity-\T) &= \rbra{x}Q\\
 (\unity-\T)\rket{y} &= Q\rket{x}
\end{split}
\end{equation} 
with inhomogeneities $\vec{x}=Q\rket{x}$ and $\vec{x}^{\dagger}=\rbra{x}Q$. One can then efficiently compute $\rket{y}$ and $\rbra{y}$ by employing an iterative Krylov subspace method such as \texttt{bicgstab}\cite{BICGSTAB} or \texttt{gmres}.\cite{GMRES} For such methods only the implementation of a (left or right) action of $(\unity-\T)$ onto a vector is necessary, which can be done efficiently with $\O(dD^{3})$ operations. If the transfer matrix is in left or right canonical form, we recover the linear systems in Eqs.~\eqref{eq:HLR_NN} and \eqref{eq:MPO_FP_T3}
\begin{equation}
\begin{split}
\rbra{y} [\unity-T_{L}+\rket{R}\rbra{\unity}] &= \rbra{x} - (x|R)\rbra{\unity}\\
 [\unity-T_{R}+\rket{\unity}\rbra{L}]\rket{y} &= \rket{x} - \rket{\unity}(L|x).
\end{split}
\end{equation} 

\end{document}